\documentclass{CSML}

\def\dOi{13(2:12)2017}
\lmcsheading%
{\dOi}
{1--61}
{}
{}
{Aug.~\phantom01, 2015}
{Jun.~30, 2017}
{}

\usepackage[nomargin,inline,index,%
  status=final 
]{fixme}
\fxusetheme{colorsig}
\FXRegisterAuthor{fxTC}{anfxTC}{\underline{\textbf{TC}}}
\FXRegisterAuthor{fxMD}{anfxMD}{\underline{\textbf{MD}}}
\FXRegisterAuthor{fxNY}{anfxNY}{\underline{\textbf{NY}}}
\FXRegisterAuthor{fxAS}{anfxAS}{\underline{\textbf{AS}}}

\usepackage[all]{xy}
\usepackage{mathpartir}
\usepackage{mathbbol}
\usepackage{stmaryrd,amssymb}
\usepackage{empheq}
\usepackage[mathcal]{euscript}
\usepackage{color}
\usepackage{ifpdf}
\usepackage{upgreek} 

\usepackage{hyperref}\hypersetup{hidelinks}
\hypersetup{final} 

\usepackage{breakurl}
\usepackage{etoolbox}
\usepackage{xspace}
\usepackage{mathtools}
\usepackage{keyval}
\usepackage{amsmath}
\usepackage{nicefrac} 

\usepackage{cmll} 

\usepackage{xspace}

\makeatletter

\let\proof\@undefined
\let\endproof\@undefined
\makeatother
\usepackage{amsthm}

\newcommand{\etal}{\emph{et al.}\@\xspace}%
\newcommand{\ie}{i.e.\@\xspace}%

\DeclareMathAlphabet{\mathbfsf}{\encodingdefault}{\sfdefault}{bx}{sl}

\newif\ifproofs
\proofstrue

\newcommand{\TZ}[1]{{{#1}}}
\newcommand{\COMMENT}[1]{}
\newcommand{\MD}[1]{{#1}}

\newcommand{\mypar}[1]{\paragraph{\bf #1}}

\newcommand{\ASET}[1]{\{ #1 \}}
\newcommand{\ASETT}[1]{#1}
\newcommand{\subst}[2]{\{\nicefrac{#1}{#2}\}}%
\newcommand{\rulename}[1]{\text{\small[\textsc{#1}]}}

\newcommand{\mkkeyword}[1]{\mathtt{\color{myblue}#1}}

\newcommand{\bigpar}{\mathrel{\Big\arrowvert}}

\newcommand{\annotaterel}[2]{#1_{\mathsf{#2}}}

\newcommand{\cinferrule}[3][]{
  \mprset{fraction={===},
  fractionaboveskip=0.2ex,
  fractionbelowskip=0.4ex}
  \inferrule[#1]{#2}{#3}
}

\definecolor{mymagenta}{rgb}{0.5,0,0.5}
\definecolor{mygreen}{rgb}{0,0.4,0}
\definecolor{myblue}{rgb}{0,0,0.6}
\definecolor{myred}{rgb}{0.4,0,0}
\definecolor{hlcolor}{rgb}{1,0.95,0}

\newcommand{\Type}{\TypeU}
\newcommand{\TypeU}{\SessionTypeS}

\newcommand{\SessionType}{\SessionTypeT}
\newcommand{\SessionTypeT}{T}
\newcommand{\SessionTypeS}{S}
\newcommand{\SessionTypeV}{V}

\newcommand{\var}{\varX}
\newcommand{\varX}{x}
\newcommand{\varY}{y}
\newcommand{\varZ}{z}

\newcommand{\pvar}{\pvarX}
\newcommand{\pvarX}{X}
\newcommand{\pvarY}{Y}
\newcommand{\pvarZ}{Z}

\newcommand{\tvar}{\tvarA}
\newcommand{\tvarA}{\mathbf{t}}

\newcommand{\Process}{\ProcessP}
\newcommand{\ProcessP}{P}
\newcommand{\ProcessQ}{Q}
\newcommand{\ProcessR}{R}

\newcommand{\Context}{\ContextC}
\newcommand{\ContextC}{C}
\newcommand{\AContext}{\mathcal{A}}
\newcommand{\BContext}{\mathcal{B}}
\newcommand{\AContextf}[1]{\AContext [ #1 ]}
\newcommand{\AContextfp}[1]{\AContext' [ #1 ]}

\newcommand{\Value}{\ValueV}
\newcommand{\ValueV}{\ChannelA}

\newcommand{\Expression}{\ExpressionE}
\newcommand{\ExpressionE}{\NameU}

\newcommand{\Queue}{h}
\newcommand{\queuetype}[2]{#1 #2 :}
\newcommand{\queue}[2]{#1 #2 \,\text{\small $\blacktriangleright$}\,}
\newcommand{\QueueType}{\tau}
\newcommand{\remainder}[2]{#1 - #2}

\newcommand{\Channel}{\ChannelA}
\newcommand{\ChannelA}{a}
\newcommand{\ChannelB}{b}
\newcommand{\ChannelC}{c}
\newcommand{\ChannelD}{d}

\newcommand{\Name}{\NameU}
\newcommand{\NameU}{u}

\newcommand{\Tag}{l}

\newcommand{\EmptyEnv}{\emptyset}

\newcommand{\EmptyQueue}{\varnothing}
\newcommand{\EmptyQueueT}{\epsilon}
\newcommand{\msg}[2]{#1\langle#2\rangle}
\newcommand{\imsg}[2]{#1(#2)}
\newcommand{\qconc}{\cdot}

\newcommand{\idle}{\mathbf{0}}
\newcommand{\new}[2]{
  (
  \nu#1#2
  )
}

\newcommand{\send}[3]{#1{!}\msg{#2}{#3}}
\newcommand{\receive}[3]{#1{?}\imsg{#2}{#3}}
\newcommand{\parop}{\mathbin{|}}

\newcommand{\choice}{\oplus}
\newcommand{\invoke}[2]{
  #1\langle#2\rangle
}

\newcommand{\Def}[4]{
  \mkkeyword{def}~#1(#2)=#3~\mkkeyword{in}~#4
}

\newcommand{\DefD}[2]{
  \mkkeyword{def}~#1~\mkkeyword{in}~#2
}

\renewcommand{\hole}{[~]}

\newcommand{\succf}{\mkkeyword{succ}}

\newcommand{\tmsg}[2]{#1\langle#2\rangle}
\newcommand{\tmsgin}[2]{#1(#2)}
\newcommand{\End}{\mkkeyword{end}}

\newcommand{\In}[2]{{?}\tmsgin{#1}{#2}}
\newcommand{\Out}[2]{{!}\tmsg{#1}{#2}}
\newcommand{\trec}{\mu}
\newcommand{\Select}{\bigoplus}
\newcommand{\Branch}{\&}

\newcommand{\UEnv}{\Upgamma}
\newcommand{\pbind}[2]{#1 : \langle#2\rangle}

\newcommand{\LEnv}{\Updelta}
\newcommand{\bind}[2]{#1 : #2}

\newcommand{\dom}{\mathsf{dom}}
\newcommand{\domq}{\dom_{\mathsf{q}}}

\newcommand{\fpv}{\mathsf{fpv}}

\newcommand{\fn}{\mathsf{fc}}

\newcommand{\sn}{\mathsf{sc}}

\newcommand{\sbn}{\gamma}
\newcommand{\sbnF}{\varphi}
\newcommand{\sbnD}{\delta}

\newcommand{\co}[1]{\overline{#1}}

\newcommand{\cprocess}[1][]{
  \ifblank{#1}{
    \mathbfsf{P}
  }{
    \mathbfsf{P}_{#1}
  }
}

\newcommand{\eqdef}{\stackrel{\text{\tiny\upshape def}}{=}}

\newcommand{\red}{\rightarrow}
\newcommand{\reds}{\annotaterel{\red}{s}}
\newcommand{\reda}{\annotaterel{\red}{a}}
\newcommand{\wred}{\rightarrow^\ast}
\newcommand{\wreds}{\annotaterel{\wred}{s}}
\newcommand{\wreda}{\annotaterel{\wred}{a}}

\newcommand{\Red}{\Rightarrow}

\newcommand{\subt}{\leqslant}
\newcommand{\ssubt}{\annotaterel{\leqslant}{s}}

\newcommand{\asubt}{\annotaterel{\leqslant}{a}}

\newcommand{\ssubtTrans}{\ssubt^{+}}
\newcommand{\asubtTrans}{\asubt^{+}}

\newcommand{\nsubt}{\ntriangleleft}
\newcommand{\nssubt}{\annotaterel{\nsubt}{s}}
\newcommand{\nasbut}{\annotaterel{\nsubt}{a}}
\newcommand{\approxs}{\annotaterel{\approx}{s}}
\newcommand{\approxa}{\annotaterel{\approx}{a}}

\newcommand{\wtp}[3]{
  \ifblank{#1}{}{
    #1 \vdash {}
  }
  #2 \triangleright #3  
}

\newcommand{\wtps}[3]{
  \ifblank{#1}{}{
    #1 \annotaterel{\vdash}{s} {}
  }
  #2 \triangleright #3  
}

\newcommand{\wtpa}[3]{
  \ifblank{#1}{}{
    #1 \annotaterel{\vdash}{a} {}
  }
  #2 \triangleright #3  
}

\newcommand{\dual}{\mathrel{\bowtie}}
\newcommand{\dualf}[1]{\overline{#1}}

\newcommand{\error}{\mathtt{error}}

\newcommand{\ENCan}[1]{\langle #1 \rangle} 

\newcommand{\dlsqb}{[\![}
\newcommand{\drsqb}{]\!]}

\newcommand{\e}{{\sf e}}
\newcommand{\cond}[3]{\mkkeyword{if}~ #1 ~\mkkeyword{then} ~#2 ~\mkkeyword{else}~#3}
\newcommand{\val}{{\sf v}}
\newcommand{\true}{\mkkeyword{true}}
\newcommand{\false}{\mkkeyword{false}}
\newcommand{\fsucc}[1]{\mkkeyword{succ}(#1)}
\newcommand{\fneg}[1]{\mkkeyword{neg}(#1)}

\newcommand{\eval}[2]{#1 \downarrow #2}
\newcommand{\valn}{{\sf n}}
\newcommand{\valr}{{\sf i}}
\newcommand{\Econtext}{\mathcal{E}}
\newcommand{\neval}[1]{#1 \not\downarrow}
\newcommand{\B}{B}
\newcommand{\tbool}{\mkkeyword{bool}}
\newcommand{\tnat}{\mkkeyword{nat}}
\newcommand{\tint}{\mkkeyword{int}}
\newcommand{\sep}{\mid}
\newcommand{\subs}{\leq\vcentcolon}
\newcommand{\der}[3]{ #1 \vdash   #2  :#3}
\newcommand{\myrule}[3]{\inferrule[]{#1}{#2}}
\newcommand{\U}{U}

\newcommand{\T}{\SessionTypeT}
\newcommand{\s}{{\sf s}}
\newcommand{\req}[2]{\co{#1}(#2).}
\newcommand{\acc}[2]{#1(#2).}
\newcommand{\res}[1]{(\nu #1)}
\newcommand{\sct}[1]{<\! #1\!>}
\newcommand{\tree}[1]{\mathcal{T}(#1)}
\newcommand{\carriedT}{\text{\,\textsf{E}}}%
\newcommand{\contT}{\text{\,\textsf{C}}}%
\newcommand\asubtB{\mathrel{\sqsubseteq}}

\newtheorem{proposition}[thm]{Proposition}

\newtheorem{example}[thm]{Example}

\begin{document}

\title[On the Preciseness of Subtyping in Session Types]{On the Preciseness of Subtyping in Session Types\rsuper*}

\thanks{{\lsuper*}This work was partly supported
  by the COST Action IC1201 \emph{BETTY}. %
}

\author[T.~Chen]{Tzu-chun Chen\rsuper a}
\address{{\lsuper a}Dept.~of Computer Science, TU Darmstadt,
Hochschulstr. 10,
64289 Darmstadt, Germany} 
\email{tzu-chun.chen@dsp.tu-darmstadt.de}
\thanks{{\lsuper a}Tzu-chun Chen was supported by the ERC grant %
  FP7-617805 \emph{LiVeSoft}.} 

\author[M.~Dezani-Ciancaglini]{Mariangiola Dezani-Ciancaglini\rsuper a}
\address{{\lsuper b}Dip.~di Informatica, Universit\`a di Torino,
Corso Svizzera 185, 10149 Torino, Italy}
\email{dezani@di.unito.it}
\thanks{{\lsuper d}Mariangiola Dezani was partly supported by EU projects %
  H2020-644235 \emph{Rephrase} and H2020-644298 \emph{HyVar}, %
  ICT COST Actions IC1402 , IC1405  and Ateneo/CSP project \emph{RunVar}. %
}
\author[A.~Scalas]{Alceste Scalas\rsuper c}
\address{{\lsuper{c,d}}Dept.~of Computing, Imperial College London, 
180 Queen's Gate,
London SW7 2AZ, UK}
\email{\{alceste.scalas, n.yoshida\}@imperial.ac.uk}
\thanks{{\lsuper c}Alceste Scalas was partly supported by EPSRC
  EP/K011715/1, and was also affiliated with: %
  Dip.~di Matematica e Informatica, Universit\`a di Cagliari, %
  Italy. }

\author[N.~Yoshida]{Nobuko Yoshida\rsuper d}
\address{\vspace{-18 pt}}
\thanks{{\lsuper d}Nobuko Yoshida was partly supported by EPSRC
  EP/K011715/1, EP/K034413/1, EP/L00058X/1 and EP/N027833/1 and 
and EU FP7-612985
  \emph{UpScale}.}




\begin{abstract}
  Subtyping in concurrency has been extensively studied since early
  1990s as one of the most interesting issues in type theory.  The
  correctness of subtyping relations has been usually provided as the
  soundness for type safety.  The converse direction, the
  completeness, has been largely ignored in spite of its usefulness to
  define the largest subtyping relation ensuring type safety.  This
  paper formalises preciseness (i.e. both soundness and completeness)
  of subtyping for mobile processes and studies it for the synchronous
  and the asynchronous session calculi.  We first prove that the
  well-known session subtyping, the branching-selection subtyping, is
  sound and complete for the synchronous calculus. Next we show that
  in the asynchronous calculus, this subtyping is incomplete for
  type-safety: that is, there exist session types $\SessionTypeT$ and
  $\SessionTypeS$ such that $\SessionTypeT$ can safely be considered
  as a subtype of $\SessionTypeS$, but
  $\SessionTypeT \subt \SessionTypeS$ is not derivable by the
  subtyping.  We then propose an asynchronous subtyping system which
  is sound and complete for the asynchronous calculus.  The method
  gives a general guidance to design rigorous channel-based subtypings
  respecting desired safety properties.  Both the synchronous and the
  asynchronous calculus are first considered with linear channels
  only, and then they are extended with session initialisations and
  communications of expressions (including shared channels).
\end{abstract}

\maketitle\vfill

\section{Introduction} \label{sec:introduction}


\mypar{Subtyping in concurrency}
Since Milner first introduced the idea of assigning types to channels in 
the $\pi$-calculus~\cite{MilnerR:polpt}, the subtypings 
which define an ordering over usages of channels have been
recognised as one of the most useful concepts in the studies of 
the $\pi$-calculus. 

The earliest work is a simple 
subtyping between input and output capabilities (called IO-subtyping) 
\cite{PierceSangiorgi95},   
which has been 
extended to and implemented in different areas of concurrency 
~\cite{HR02,PierceTurner:PictDesign,RielyJ:trupartiosoma,Sewell:2010}
and has been continuously studied 
as one of the core subjects in concurrency
~\cite{HMS12,
HMS13}
.  
Later, a generic type system with subtyping was
introduced by Igarashi \etal~\cite{igarashi.kobayashi04:gentypPi}, where   
the subtyping plays a fundamental r\^ole 
in generating a variety of interesting type
systems as its instances.
 
More recently, another subtyping based on  
session types~\cite{GH05} 
has been applied 
to many aspects of software design and implementations
such as web services, programming languages and 
distributed computing 
\cite{Carbone:2012:SCP:2220365.2220367,DemangeonH11,GayVRGC10,HNHYH13,scribble10,event,HU07TYPE-SAFE,scribsite,Vasconcelos09}. The standpoint of session types is that 
communication-centred applications exhibit highly structured
interactions involving, for example, sequencing,
branching, selection and recursion, 
and such a series of interactions can be abstracted as a
{\em session type} through a simple syntax. 
The session subtyping specified along session structures 
is then used for validating a large set of programs, 
giving flexibility to programmers. 

As an example of session subtyping 
\cite{Carbone:2012:SCP:2220365.2220367,DemangeonH11,event,asyncsession}, 
consider a simple protocol between a 
Buyer and a Seller from Buyer's viewpoint: Buyer sends a 
book's title (a string), and Seller sends a quote (an integer). If
Buyer is satisfied by the quote, he then sends his address (a string) and
Seller sends back the delivery date (a date); otherwise he quits the
conversation. This can be described by the session type:
\begin{equation}
\label{AAA}
!\ENCan{\mathtt{string}}.
?(\mathtt{int}).
\{!\mathsf{ok}\ENCan{\mathtt{string}}.?(\mathtt{date}).\mathsf{end}
\ \oplus \ 
!\mathsf{quit}.\mathsf{end}\}
\end{equation}

The prefix $!\ENCan{\mathtt{string}}$ denotes an
  output of a value of type $\mathtt{string}$, whereas
  $?(\mathtt{int})$ denotes an input of a value of type
  $\mathtt{int}$. Instead $\mathsf{ok}$ and $\mathsf{quit}$ are labels distinguishing different branches. The operator $\oplus$ is an internal choice, meaning the process may choose to either send the label $\mathsf{ok}$ with a string and receive a date, or send the label $\mathsf{quit}$.
  The type
  $\mathsf{end}$ represents the termination of the session. 
From Seller's viewpoint the same session is described by the {\em dual} type
\begin{equation}
\label{AAA2}
?(\mathtt{string}).
!\ENCan{\mathtt{int}}.
\{?\mathsf{ok}(\mathtt{string}).!\ENCan{\mathtt{date}}.\mathsf{end}
\ \& \ 
?\mathsf{quit}.\mathsf{end}\}
\end{equation}
in which $\&$ means that the process offers two behaviours: one where it receives $\mathsf{ok}$ with a string and sends a date, and one where it receives $\mathsf{quit}$.

As $\mathtt{nat} \subt \mathtt{real}$ in the standard subtyping, 
a type representing a more confined behaviour
is smaller. 
A selection subtype is a type which 
selects among fewer options (as outputs). The following is 
an example of a subtype of (\ref{AAA}):
\begin{equation}
\label{BBB}
!\ENCan{\mathtt{string}}.
?(\mathtt{int}).
!\mathsf{ok}\ENCan{\mathtt{string}}.?(\mathtt{date}).\mathsf{end}
\end{equation}
Conversely, a branching subtype 
is a type which offers more options (as inputs). 
The following is an example of 
a subtype of (\ref{AAA2}): 
\begin{eqnarray}
\label{BBB2}
?({\mathtt{string}}).
!\ENCan{\mathtt{int}}.
\{?\mathsf{ok}({\mathtt{string}}).!\ENCan{\mathtt{date}}.\mathsf{end}
  \ \&  \ ?\mathsf{quit}.\mathsf{end}
  \ \& \ ?\mathsf{later }.\mathsf{end} \} \!\!\!
\end{eqnarray}
\noindent Intuitively, a type $T$ is a subtype of a type $S$ if $T$ is ready to receive no fewer labels than $S$,
and $T$ potentially sends no more labels than $S$
(in other words, $T$ represents a
more permissive behaviour than $S$) 
\cite{Carbone:2012:SCP:2220365.2220367,DemangeonH11}. 
If we run two processes typed by (\ref{BBB}) and (\ref{BBB2}), 
they are {\em type safe}, 
i.e.~there is no mismatch of labels or types during communication. 
Hence the subtyping is {\em sound} with respect to the type safety. 
An important question, however, 
is still remaining: is this subtyping {\em complete}? I.e.~is this session subtyping the {\em largest}
relation which does not violate type safety? The proof of soundness 
is usually immediate as a corollary of the subject reduction
theorem. But how can we state and prove completeness? 

A type system with a subsumption rule is parametric in the subtyping relation. A larger
subtyping relation will yield a type system accepting
more programs. A subtyping relation is {\em sound} if no typeable program is incorrect. It is {\em complete} if
there is no strictly larger sound subtyping relation. %
Following 
Ligatti \etal \cite{BHLN12}, %
we say that a subtyping relation is
{\em precise} if it is both sound and complete. 
The preciseness  
is a simple operational property
that specifies a relationship between static and dynamic semantics.

\mypar{Preciseness}
To formally define preciseness,
we assume a multi-step reduction between processes $P \wred P'$ \MD{(where $P'$ is possibly the $\error$ process)} as well as  
typing judgements of the form $\wtp{}{P}{\ASET{a:T}}$, assuring that the process $P$ 
has a single free channel $a$ whose type is $T$. 
We also use reduction contexts $C$ in the standard way. The judgement 
$\wtp{}{\Context[\ChannelA:\SessionType]}{\emptyset}$ means that 
filling the hole of $C$ with any process $P$ typed by $a:T$ 
produces a well-typed closed process (formally 
$\wtp{}{\Context[\ChannelA:\SessionType]}{\emptyset}
\iff
\wtp{
  \pvar:\SessionType
}{
  \Context[\invoke\pvar{\ChannelA}]
}{
  \emptyset 
}$, where $X$ is a process variable which does not appear in $C$, see
\S~\ref{sec:synchronous_language}).

Our preciseness definition is an adaptation of the preciseness definition 
for the call-by-value $\lambda$-calculus with sums and product types~\cite{BHLN12}.

\begin{defi}[Preciseness]\rm 
\label{def:preciseness}
A subtyping $\subt$ is \emph{precise} 
  when, for all session types $\SessionTypeT,\SessionTypeS$:
\[
\SessionTypeT \subt \SessionTypeS
\iff
\left(
\begin{array}{@{}l@{}}
\text{there do not exist $\Context$ and $\ProcessP$ such that:}\\
{\wtp{}{\Context[\ChannelA:\SessionTypeS]}{\EmptyEnv}}
\text{ and }
{\wtp{}{\ProcessP}{\ASET{\ChannelA:\SessionTypeT}}}
\text{ and }
\Context[\ProcessP] \wred \error
\end{array}
\right)
\]
When the \emph{only-if} direction ($\Rightarrow$) of this formula
holds, we say that the subtyping is \emph{sound};
when the \emph{if} direction ($\Leftarrow$) holds,
we say that the subtyping is \emph{complete}.
\end{defi}

%
Consider the set of contexts 
$\Context$ such that 
$\wtp{}{\Context[\ChannelA:\SessionTypeS]}{\EmptyEnv}$, %
i.e., contexts with one
hole, where the channel $\ChannelA$ %
is typed as $\SessionTypeS$. %
The \emph{soundness} property of Definition~\ref{def:preciseness} %
says that if we take \emph{any} such $\Context$ %
and fill it with \emph{any} process with $\ChannelA$
typed as $\SessionTypeT \subt \SessionTypeS$, %
the result is safe. %
The \emph{completeness} property, instead, %
says that for all $\SessionTypeT \not\subt \SessionTypeS$, %
we can find some $\Context$ in the set above, %
and fill it with some process with $\ChannelA$ typed as $\SessionTypeT$, %
so that their combination reduces to $\error$.
Soundness is clearly Liskov's Substitution Principle~\cite{LW94}, %
whereas completeness ensures that %
a subtyping relation cannot be safely extended.
Notice that we take a ``must view'' of correctness, asking that a correct process never reduces to an $\error$. 

Here we are interested in {\em syntactically} 
defined subtyping and an {\em operational} notion of 
preciseness. 
Our approach is opposed to semantic subtyping %
as proposed by Frisch \etal~\cite{FCB08}, which is given 
denotationally; %
in addition, the calculus introduced by Frisch \etal~\cite{FCB08} has a 
type case constructor from which completeness follows for free.  
See \S~\ref{sec:related} for a detailed discussion. 

\mypar{Preciseness and impreciseness for the $\pi$-calculus} 
IO-subtyping classifies channels according to their reading and writing capabilities~\cite{PierceSangiorgi95}. %
It is not precise, %
because no operational error can
be detected when a read only channel is used to write, or vice versa, %
by a context without type annotations. %
For a similar reason, the
branching and selection subtyping~\cite{DemangeonH11} is also
imprecise for the $\pi$-calculus. The branching and selection
subtyping is instead precise for the $\pi$-calculus with only linear
channels~\cite{LinearPi}, whose expressivity is limited.  

Igarashi \etal~\cite{igarashi.kobayashi04:gentypPi} %
only state necessary conditions
for subtyping, the aim being that of having the maximum generality. The subtyping relations in the instances of the generic type system depend on the properties (arity-mismatch check, race detection, static garbage-channel collection, deadlock detection) one wants to guarantee. 

These results led us to consider preciseness for two representative session calculi: the synchronous~\cite{HVK,V09} and 
the asynchronous~\cite{mostrous_phd,mostrous09sessionbased,MY15,mostrous_yoshida_honda_esop09} session calculi.


\mypar{Two preciseness results}
Session types have
sufficiently rich structure to assure completeness, 
hence if $\SessionTypeT \not\subt \SessionTypeS$, then 
$T$ and $S$ can be distinguished by suitable contexts and processes.

The first result of this article is preciseness 
of the branching-selection subtyping (dubbed also {\em synchronous subtyping}) described above for the synchronous session calculus.  
Our motivation to study the first result 
is to gently introduce a proof method for preciseness 
and justify the correctness of 
the synchronous subtyping, which is widely used in session-based 
calculi, programming languages and implementations~\cite{Carbone:2012:SCP:2220365.2220367,DemangeonH11,HU07TYPE-SAFE,scribsite}. 

The case of the asynchronous session calculus is more
challenging.  The original session typed calculi~\cite{HVK} are based on
synchronous communication primitives, assumed to be compiled into
asynchronous interactions using queues %
--- i.e., synchronous communications are modelled by asynchronous ones. Later researchers~\cite{CHY08} found that, assuming
{\em ordered} asynchronous communications for binary interactions, one could 
directly express asynchronous non-blocking
interactions. 
One can then assure not only 
the original synchronous safety, but also 
the asynchronous safety, i.e.~deadlock-freedom (every input process will always
receive a message) and orphan message-freedom
(every message in a queue will always be received by an
input process). 

Our first observation is that the branching-selection subtyping %
$\subt$
is not large enough for the
asynchronous calculus, i.e.~there exist session types 
$\SessionTypeT$ and $\SessionTypeS$ such that $\SessionTypeT$ can
safely be considered as a subtype of $\SessionTypeS$, but $\SessionTypeT \subt \SessionTypeS$ 
is not derivable by the subtyping. The reason is natural: 
in the presence of queues, the processes typed by 
the following two non-dual types can run in parallel  
{\em without} reducing to $\error$:
\begin{eqnarray}
\label{CCC}
T_a &= & !\ENCan{\mathtt{int}}.!\ENCan{\mathtt{char}}.
?({\mathtt{string}}).?({\mathtt{nat}}).\mathsf{end} \nonumber \\
\label{CCC2}
T_b &= & !\ENCan{\mathtt{string}}.!\ENCan{\mathtt{nat}}.
?({\mathtt{int}}).?({\mathtt{char}}).\mathsf{end} \nonumber
\end{eqnarray}
since a process typed by $T_a$ can put 
two messages typed by $\mathtt{int}$ and $\mathtt{char}$ in one queue
and a process typed by $T_b$ can put 
two messages typed by $\mathtt{string}$ and $\mathtt{nat}$ in another
queue, and they can receive the two messages from each queue,  
without getting stuck. %
Moreover, if we replace $T_a$ or $T_b$ with any of their respective %
subtypes under $\subt$, we can still safely compose the
corresponding processes in parallel, without reducing to $\error$: %
therefore, %
if we extend $\subt$ to also relate $T_a$  and the dual of $T_b$,
we would not compromise type safety.

The {\em asynchronous subtyping} %
proposed by Mostrous \etal~\cite{mostrous_phd,mostrous09sessionbased,MY15,mostrous_yoshida_honda_esop09} 
permutes the order of messages, for example:
$$T_a \subt \ ?({\mathtt{string}}).?({\mathtt{nat}}).
!\ENCan{\mathtt{int}}.!\ENCan{\mathtt{char}}.\mathsf{end}$$
so that a process typed by $T_a$ can have a type which is dual of $T_b$ by
the subsumption rule. 
This asynchronous permutation is often used as a means of 
messaging optimisation, e.g. as ``messaging overlapping'', in the parallel 
programming community~\cite[\S 6]{NYH12}. Our result demonstrates the preciseness of this subtyping, which was introduced for practical motivations.

We have found that the subtyping of Mostrous' PhD thesis~\cite{mostrous_phd,MY15} is unsound 
if we require the absence of orphan messages. %
If we allow orphan messages and we 
only have deadlock errors, then such a subtyping is sound but not complete.
All this is discussed in \S~\ref{sec:related}.
The other asynchronous subtypings %
introduced by Mostrous \etal~\cite{mostrous09sessionbased,mostrous_yoshida_honda_esop09}, whose targets are the higher-order $\pi$-calculus and the multiparty session types, respectively, are 
sound for deadlock and orphan message errors. 
Hence we simplify them and we adapt them to the binary session $\pi$-calculus.

\mypar{Contributions} 
As far as we are aware, 
this is the first time that completeness of subtypings,
which is solely based on (untyped) operational semantics,  
is formalised and proved in the context of mobile processes. 
We also demonstrate its applicability to two session type disciplines, 
the synchronous and the asynchronous ones. 
The most technical challenge is the proof of completeness
for the asynchronous subtyping, which requires some ingenuity in the definition of its negation relation.  
Key in the proofs is the construction of processes which characterise types. These processes allow us to show also the denotational preciseness of both synchronous and asynchronous subtypings. 

This article is an expanded version of a previous paper presented at PPDP 2014~\cite{CDY14}, including detailed definitions and full proofs, which were omitted. In addition, we %
provide new results about %
the uniqueness of precise subtyping relations %
(Corollary~\ref{lem:ssubt-largest}, Theorem~\ref{lem:asubt-largest}). %
Moreover,  %
we include a new section 
(\S~\ref{ext}) dealing with session initialisation and with communication of expressions (including shared channels), which were not treated: this %
demonstrates that our approach smoothly generalises to the original calculus~\cite{HVK}, showing that the invariance of shared channel types is precise. 

\mypar{Outline} 
\S~\ref{sec:synchronous_language} defines the 
synchronous session calculus and its typing system, 
and proves soundness of 
the branching-selection subtyping $\ssubt$. 
\S~\ref{sec:s:completeness} 
proposes a general 
scheme for showing completeness, %
and proves that $\ssubt$ is the unique complete subtyping %
for the synchronous session
calculus.  \S~\ref{sec:asynchronous_language} defines the asynchronous
session calculus and introduces a new asynchronous subtyping relation
$\asubt$, which is shown to be sound.
\S~\ref{sec:a:completeness} 
proves %
that $\asubt$ is the unique subtyping that both extends $\ssubt$,
and is complete %
for the asynchronous 
calculus. 
This last completeness proof is non-trivial,
since the permutations introduced by 
the asynchronous subtyping rules make
session types unstructured. %
\S~\ref{ext} extends both calculi with shared channels for session initialisation and communications of
expressions as in paper~\cite{HVK}. 
The proof of operational preciseness gives us denotational preciseness of both synchronous and asynchronous subtypings,  as shown in \S~\ref{sec:denotation}. Related work and conclusion are the contents of 
\S~\ref{sec:related} and
\S~\ref{sec:conclusion}, respectively.

\section{Synchronous Session Calculus}
\label{sec:synchronous_language}

\begin{table}
\small%
$$
\begin{array}[t]{@{}c@{\qquad}c@{}}
\begin{array}[t]{@{}rcl@{\quad}l@{}@{}rcl@{\quad}l@{}}
  \Process \hspace{-0.3cm}& ::= \hspace{-0.4cm}& & \textbf{Process} \\
  &   & \idle & \text{(nil)} &
  & | & \invoke\pvar{\tilde\Expression} & \text{(variable)} \\
  & | & \sum_{i\in I} \receive\NameU{\Tag_i}{\var_i}.\Process_i & \text{(input)} &
  & | & \send\NameU\Tag{\Expression'}.\Process & \text{(output)} \\
  & | & \Process \parop \Process & \text{(parallel)}  &
  & | & 
  \Process \choice \Process & \text{(choice)}\\
  & | & \DefD{D}{\Process} & \text{(definition)} &
  & | & \new\ChannelA\ChannelB \Process & \text{(restriction)} \\
  & | & \error & \text{(error)}
  \end{array}
\quad\;\;
\begin{array}[t]{@{}rcl@{\quad}l@{}}
  D & ::= & &\hspace{-0.5cm}\textbf{Declaration} \\
&&  \hspace{-0.4cm}
\pvar (\tilde \var) = \Process 
  \\\\
  \NameU & ::= & & \hspace{-0.5cm}\textbf{Identifiers} \\
  &   & \Channel & \hspace{-0.5cm}\text{(linear channel)} \\
  & | & \var & \hspace{-0.5cm}\text{(channel variable)}
  \end{array}
\end{array}
$$
\caption{\label{tab:sync:syntax} Syntax of synchronous processes.\strut}
\end{table}
This section starts by introducing syntax and semantics of a simplification
of the most widely studied synchronous session calculus~\cite{HVK,yoshida.vasconcelos:language-primitives}. 
Since our main 
focus is on subtypings between session types, we only allow exchanges of linear channels. 
The obtained calculus is similar to that  presented by Vasconcelos in 2009~\cite{V09}.
We then define the typing system and prove soundness of subtyping as defined in Definition~\ref{def:preciseness}. 
\S~\ref{ext} will consider session initialisations and communication of  expressions (including shared channels).

\subsection{Syntax}
\label{subsec:syntax}
A \emph{session} is a series of 
interactions between two
parties, possibly with branching and recursion, and serves as a unit
of abstraction for describing communication protocols. 
We use the following base sets: \emph{channel variables}, ranged over by
$x,y,z\dots$; \emph{linear channels}, ranged over by $a,b$;
\emph{identifiers} (channel variables and linear channels), ranged over by $u,u'\dots$;
\emph{labels}, ranged over by $l,l',\dots$;
\emph{process variables}, ranged over by $X,Y,\dots$; 
and \emph{processes}, ranged over by $P,Q\dots$. The syntax is given in Table~\ref{tab:sync:syntax}.

Session communications are performed between 
an output process $\send\NameU\Tag{\Expression'}.\Process$ and an input process $\sum_{i\in I} \receive\NameU{\Tag_i}{\var_i}.\Process_i $ (the $\Tag_i$ are pairwise distinct), where the former sends a channel  
choosing one of the branches offered by the latter. In $\sum_{i\in I} \receive\NameU{\Tag_i}{\var_i}.\Process_i $ and 
 $\send\NameU\Tag{\Expression'}.\Process$ 
the identifier
$u$ is the {\em subject} of input and output, respectively. 
The choice $P \choice Q$ internally chooses either $P$ or $Q$. In many session calculi~\cite{Carbone:2012:SCP:2220365.2220367,HVK,mostrous_yoshida_honda_esop09} the conditional plays the r\^ole  of the choice.
The process $\DefD{D}{\Process}$ is a recursive agent and 
$\invoke\pvar{\tilde\Expression}$ is a recursive variable. %
We postulate \emph{guarded recursion}, %
i.e.~recursive variables can occur in recursive agent
declarations %
only after an input/output prefix; %
for instance, %
$\DefD{\pvarX (\tilde \var) = \invoke\pvarY{\tilde\Expression}}{\ProcessP}$ %
and $\DefD{\pvarX (\tilde \var)=\invoke\pvarY{\tilde\Expression}
  \choice \Process'}{\ProcessP}$ %
are \emph{not} syntactically valid, %
whereas %
$\DefD{\pvarX (\tilde \var) =
  \send\NameU\Tag{\Expression'}.\invoke\pvarY{\tilde\Expression''}%
}{\ProcessP}$ is valid. %
The process $\new\ChannelA\ChannelB \Process$ is a restriction which binds 
two channels, $\ChannelA$ and $\ChannelB$ in $\Process$, making them {\em co-channels}, i.e. allowing them to communicate (see rule \rulename{r-com-sync} in Table~\ref{tab:sync:red}).
This double-restriction is commonly used in the recent literature of session types~\cite{GH05,
V09}.
We often omit $\idle$ from the tail of processes.

The \emph{bindings} for variables
are in inputs and declarations, %
those for channels are in restrictions, %
and those for process variables are in declarations.  The derived notions of bound
and free identifiers, alpha equivalence, 
and
substitution are standard. We use 
Barendregt  convention~\cite[\S 2.1.13]{B84} %
that no bound name can occur free or in two different bindings.

By $\fpv(P)$/$\fn(P)$ we denote the set of 
\emph{free process variables}/\emph{free channels} in $P$.   
By $\sn(P)$ we denote the set of  \emph{free subject channels} in $P$, defined 
by:\label{sn}
\[
\begin{array}{rcl}
  \sn(\Name \Out{\Tag}{\Name'} . \Process)%
  &=&%
  \fn{(\Name)} \;\cup\; \sn(\Process)%
  \\%
  \sn(\sum_{i \in I} \Name \In{\Tag_i}{\var_i} . \Process_i)%
  &=&%
  \fn{(\Name)} \;\cup\;\bigcup_{i\in I} \sn(\Process_i)\\
  \sn(\DefD{D}{\ProcessP})
  &=& \sn(\Process)
\end{array}
\qquad\qquad%
\text{where:}
\quad%
\begin{array}{r@{\hskip 1mm}c@{\hskip 1mm}l}
  \fn(\var) &=& \emptyset%
  \\%
  \fn(\ChannelA) &=& \{a\}%
\end{array}
\]
and as expected 
in the other cases. %
Note that we need to use $\fn{(\Name)}$, since $\Name$ can be either a channel or a variable.

\subsection{Operational semantics}
\label{subsec:semantics}
\begin{table}
\small%
\begin{math}
\displaystyle
\begin{array}[t]{@{}c@{}}
  \inferrule[\rulename{s-par 1}]{}{
    \idle \parop \Process \equiv \Process
  }
  \quad
  \inferrule[\rulename{s-par 2}]{}{
    \ProcessP \parop \ProcessQ \equiv \ProcessQ \parop \ProcessP
  }
  \quad
  \inferrule[\rulename{s-par 3}]{}{
    \ProcessP \parop (\ProcessQ \parop \ProcessR)
    \equiv
    (\ProcessP \parop \ProcessQ) \parop \ProcessR
  }
 \quad 
 \inferrule[\rulename{s-ch 1}]{}{
  \ProcessP \oplus \ProcessQ \equiv \ProcessQ \oplus \ProcessP
  }\quad
  \inferrule[\rulename{s-ch 2}]{}{
  (\ProcessP \oplus \ProcessQ) \oplus R \equiv \ProcessQ \oplus (\ProcessP \oplus R)
  }
  \\\\
   \inferrule[\rulename{s-res 1}]{}
  {
    \new\ChannelA\ChannelB \idle
    \equiv
   \idle
  }
  \qquad
  \inferrule[\rulename{s-res 2}]{}
  {
    \new\ChannelA\ChannelB \ProcessP \parop \ProcessQ
    \equiv
    \new\ChannelA\ChannelB (\ProcessP \parop \ProcessQ)
  }
  \qquad
  \inferrule[\rulename{s-res 3}]{}
  {
    \new\ChannelA\ChannelB
    \new\ChannelC\ChannelD
    \Process
    \equiv
    \new\ChannelC\ChannelD
    \new\ChannelA\ChannelB
    \Process
  }
   \qquad
  \inferrule[\rulename{s-def 1}]{}{
  \DefD{D}{\idle} \equiv \idle  
  } \\\\
  \inferrule[\rulename{s-def 2}]{}
  {
    \DefD{D}{\new \ChannelA \ChannelB \Process}
    \equiv  
    \new \ChannelA \ChannelB
    (\DefD{D}{\Process}) 
  } 
    \qquad  \inferrule[\rulename{s-def 3}]{}
  {
    ( \DefD{D}{\Process} ) \mid Q 
    \equiv
    \DefD{D}{(\Process \mid Q)}
  }
  \\\\
  \inferrule[\rulename{s-def 4}]{}
  {
    \DefD{D}{\DefD{D'}{\Process}}
    \equiv
    \DefD{D'}{\DefD{D}{\Process}}
  }
\end{array}
\end{math}
\caption{\label{tab:sync:congruence} 
  Structural congruence for synchronous processes.\strut%
}
\end{table}
\begin{table}
$$
\begin{array}[t]{@{}c@{}}
  \inferrule[\rulename{r-com-sync}]{
    k \in I
  }{
    \new \ChannelA \ChannelB (
      \send\ChannelA {\Tag_k} \ChannelC.\Process
      \parop
      \sum_{i\in I} \receive\ChannelB{\Tag_i}{\var_i}.\ProcessQ_i
    )
    \red
    \new \ChannelA \ChannelB (
      \Process
      \parop
      \ProcessQ_k \subst{\ChannelC}{\var_k}
    )
  }
  \\\\
    \inferrule[\rulename{r-def}]{}{
    \Def\pvar{\tilde\var}\ProcessP{
      (\invoke\pvar{\tilde \ChannelA}
      \parop
      \ProcessQ)
    }
    \red
    \Def\pvar{\tilde\var}\ProcessP{
      (\ProcessP
      \subst{\tilde\ChannelA}{\tilde \var} 
      \parop
      \ProcessQ)
    }
  }
  \\ \\
   \inferrule[\rulename{r-choice}]{}{
    \ProcessP \choice \ProcessQ \red \ProcessP
  }
  \qquad\qquad
    \inferrule[\rulename{r-context}]{
    \Process \red \Process'
  }
  { 
    \Context[\Process] \red \Context[\Process']
  }
  \qquad\qquad
  \inferrule[\rulename{r-struct}]{
    \ProcessP \equiv \ProcessP'
    \\
    \ProcessP' \red \ProcessQ'
    \\
    \ProcessQ' \equiv \ProcessQ
  }{
    \ProcessP \red \ProcessQ
  }
\end{array}
$$
\caption{\label{tab:sync:red} 
Reduction of synchronous processes.
}
\end{table}

Table~\ref{tab:sync:red} gives the reduction relation between
synchronous processes which do not contain free variables. 
It uses the following evaluation context:\label{pc}
$$
  \Context\hole ::= \hole
  \bigpar
  \Context\hole \parop \Process
  \bigpar
  \new\ChannelA\ChannelB \Context\hole
  \bigpar
  \DefD{D}{\Context\hole}
$$
and the structural rules of Table~\ref{tab:sync:congruence}.

In Table~\ref{tab:sync:red}, \rulename{r-com-sync}  
is the main communication rule between input and output  
at two co-channels $a$ and $b$, where the label $l_k$ is 
selected and channel $c$ is instantiated into the $k$-th input branch. 
Other rules are standard. 

\begin{table}
$$
\begin{array}[t]{@{}c@{}}
   \inferrule[\rulename{err-mism-sync}]
  { \forall i \in I : \Tag \not = \Tag_i
  }
  { 
    \new\ChannelA\ChannelB
        (
        \send\ChannelA\Tag\ChannelC.\ProcessP
        \parop
        \sum_{i\in I} \receive\ChannelB{\Tag_i}{\var_i}.\ProcessQ_i
        )
        \red \error
  }
\quad 
  \inferrule[\rulename{err-new-sync}]{
    \ChannelA\in\sn(\ProcessP) \quad
    \ChannelB\not\in\fn(\ProcessP)
  }{
    {\new\ChannelA\ChannelB\ProcessP}
    \red \error
  }
   \quad
  \inferrule[\rulename{err-context}]{}{
     \Context[\error] \red \error
  }
\\\\
  \inferrule[\rulename{err-out-out-sync}]{}{
    \new \ChannelA \ChannelB 
    (\send\ChannelA\Tag\ChannelC.\ProcessP \parop
    \send\ChannelB{\Tag'}{\ChannelC'}.\ProcessQ)
    \red \error
  }
   \qquad
  \inferrule[\rulename{err-in-in-sync}]{}{
    \new \ChannelA \ChannelB 
    (\sum_{i\in I} \receive\ChannelA{\Tag_i}{\var_i}.\ProcessP_i \parop
    \sum_{j\in J} \receive\ChannelB{\Tag'_j}{\var'_j}.\ProcessQ_j)
    \red \error
  }
\end{array}
$$
\caption{\label{tab:sync:red:err} 
Error reduction for synchronous processes.}
\end{table}

We also define error reduction, which is crucial for 
stating the preciseness theorem. Our guideline in this definition (both for the synchronous and, later, for the asynchronous semantics) is the following sentence from a seminal paper by Honda \etal~\cite{HVK}: ``The typeability of a program ensures two possibly communicating processes always own compatible communication patterns.'' In our case it amounts to require the duality of the communications offered by co-channels. %

The error reduction rules are listed in Table~\ref{tab:sync:red:err}. %
Rule 
\rulename{err-mism-sync} is a mismatch between the output and input labels.
Rule \rulename{err-new-sync} represents an error situation where one of two
co-channels ($b$) is missing. Rule \rulename{err-out-out-sync} gives 
an error when two co-channels are both subjects of outputs, destroying the
duality of sessions. Similarly rule \rulename{err-in-in-sync} gives 
an error when two co-channels are both subjects of inputs.  We do not consider errors due to non linear use of channels, %
since they are statically prevented by the typing rules %
and cannot be introduced by changing the definition of subtyping. %
Obviously, arbitrary processes can be stuck without reducing to $\error$: 
this can happen, for instance, %
due to the lack of a companion process, %
or if a process interacts on multiple interleaved sessions. %
A simple example is the process $\ChannelA \In{\Tag}{\var}$, %
which is deadlocked; %
intriguing examples of deadlocks caused by session interleaving %
can be written %
using process variables with more than one parameter. %

We denote by $\reds$ the reduction relation 
for the synchronous processes, generated by the rules in Tables~\ref{tab:sync:red} and~\ref{tab:sync:red:err},
 and by $\wreds$ the reflexive and transitive closure of $\reds$.

Proposition~\ref{lem:sync-errors-persistent}, %
says that if a process can reduce to $\error$ in one step, %
then a different reduction produces a process with the same property.

\begin{proposition}
  \label{lem:sync-errors-persistent}%
  If
  $\Process \reds \error$ and $\Process \reds \Process' \neq \error$, then
  $\Process' \reds \error$.%
\end{proposition}
\begin{Proof}
  By cases on the rule giving $\Process \reds \error$. %
  The statement holds vacuously for rules %
  \rulename{err-mism-sync}, %
  \rulename{err-out-out-sync} %
  and \rulename{err-in-in-sync} %
  (Table~\ref{tab:sync:red:err}): %
  in such cases, $\Process \reds \Process'$ implies $\Process' = \error$.
  
  In the case \rulename{err-new-sync}, %
  we have %
  $\Process = \new\ChannelA\ChannelB\ProcessP_0$ 
  with $\ChannelA\in\sn(\ProcessP_0)$ %
  and $\ChannelB\not\in\fn(\ProcessP_0)$, %
  and $\Process \reds \ProcessP' \neq \error$. %
  Such a transition can only fire by rule \rulename{r-context}, %
  and thus $\ProcessP' = \new\ChannelA\ChannelB\ProcessP_1$ %
  and $\ProcessP_0 \reds \ProcessP_1$. %
  By induction on the derivation of the latter transition, %
  we can verify that %
  $\ChannelA\in\sn(\ProcessP_1)$ and $\ChannelB\not\in\fn(\ProcessP_1)$.
  Hence, again by rule \rulename{err-new-sync}, %
  we conclude %
  $\new\ChannelA\ChannelB\ProcessP_1 = \Process' \reds \error$.
  
  In the case \rulename{err-context}
  we have $\Process = \Context[\error]$. %
  If $\Process \reds \Process' \neq \error$, %
  then the reduction is fired inside the context $\Context\hole$: %
  by induction on the derivation of the transition, %
  we can verify that $\exists \Context'\hole: %
  \Process' = \Context'[\error]$. %
  Hence, again by rule \rulename{err-context}, %
  we conclude %
  $\Process' \reds \error$.
\end{Proof}

\subsection{Typing synchronous processes}
\label{subsec:types} 

The syntax of \emph{synchronous session types}, ranged over by $T$ and $S$, is:
$$
\begin{array}{rcl}
\SessionType, \SessionTypeS & \;::=\; &
\Branch_{i\in I} \In{\Tag_i}{\Type_i}.\SessionTypeT_i 
\;\;\parop\;\;
\Select_{i\in I} \Out{\Tag_i}{\Type_i}.\SessionTypeT_i
\;\;\parop\;\;
\tvar 
\;\;\parop\;\;
\trec\tvar.\SessionType
\;\;\parop\;\;
 \End 
\end{array}
$$
The {\em branching type} 
$\Branch_{i\in I} \In{\Tag_i}{\Type_i}.\SessionTypeT_i$
describes a channel willing to 
branch on an incoming label $\Tag_i$, 
receive a channel of type $\Type_i$, and
then continue its interaction as prescribed by $\SessionTypeT_i$.  
The {\em selection type} 
$\Select_{i\in I} \Out{\Tag_i}{\Type_i}.\SessionTypeT_i$ is its dual: it describes a channel willing to 
send a label $\Tag_i$ with a channel of type $\Type_i$, and
then continue its interaction as prescribed by $\SessionTypeT_i$. In branching and in selection types:
\begin{itemize}\item  the labels are pairwise distinct; 
\item 
the types of the exchanged channels are closed. %
\end{itemize}
We omit $\Branch$ and $\oplus$ and labels
when there is only one branch.
We use $\tvar$ to range over type variables. The type
$\trec\tvar.\SessionType$ is a {\em recursive type}. 
We assume that recursive types are \emph{contractive}, %
i.e.~$\trec\tvar_1.\trec\tvar_2\ldots\trec\tvar_n.\tvar_1$ %
is not a type. %
The type
$\End$ represents the termination of 
a session and it is often omitted. 

We take an equi-recursive view of types~\cite[Chapter 20, \S 2]{PierceBC:typsysfpl}, %
considering two types with the same regular tree as equal. %
\label{def:type-tree}%
Table~\ref{tree} defines {\em coinductively}~\cite[Chapter 21, \S 2.1]{PierceBC:typsysfpl} %
the {\em tree of a type $\SessionType$} (notation $\tree\SessionType$), where each label $\Tag$ generates two edges $\Tag^\carriedT$ and $\Tag^\contT$, %
pointing respectively to %
the \emph{exchanged} and the \emph{continuation} sub-trees.
Figure~\ref{te} shows (part of) the infinite tree %
$\tree{\trec\tvar.\Out{\Tag_1}{\End}.\tvar\oplus\Out{\Tag_2}{\End}.\End}$. %
\label{def:continuation-path}%
We will mainly focus on \emph{continuation paths}, %
i.e.~(possibly infinite) sequences of edges %
$\Tag_1^\contT,\ldots,\Tag_n^\contT$ starting from tree roots.

\begin{table}
\[
\begin{array}{c}
  \tree{\Branch_{1\leq i\leq n} \In{\Tag_i}{\Type_i}.\SessionTypeT_i} \;=\;%
  \xymatrix@R-1pc{%
    &&&\ar@{-}[dll]_{\Tag_1^{\carriedT}}\ar@{-}[ddl]_{\Tag_1^{\contT}}\ar@{-}[drr]^{\Tag_n^{\contT}}\ar@{-}[ddr]^{\Tag_n^{\carriedT}}\&\\
    &\tree{\Type_1}&&&&\tree{\SessionTypeT_n}\\
    &&\tree{\SessionTypeT_1}&\cdots&\tree{\Type_n}
  }%
  \\\\[0.5cm]%
  \tree{\Select_{1\leq i\leq n} \Out{\Tag_i}{\Type_i}.\SessionTypeT_i} \;=\;%
  \xymatrix@R-1pc{%
    &&&\ar@{-}[dll]_{\Tag_1^{\carriedT}}\ar@{-}[ddl]_{\Tag_1^{\contT}}\ar@{-}[drr]^{\Tag_n^{\contT}}\ar@{-}[ddr]^{\Tag_n^{\carriedT}}\oplus\\
    &\tree{\Type_1}&&&&\tree{\SessionTypeT_n}\\
    &&\tree{\SessionTypeT_1}&\cdots&\tree{\Type_n}
  }%
  \\\\[0.5cm]%
\tree{\trec\tvar.\SessionType}=\tree{\SessionType\subst{\trec\tvar.\SessionType}{\tvar}}
\qquad\qquad\qquad
 \tree{\End}=\End
 \end{array}
 \]
 \caption{
   Trees of session types.%
 }%
 \label{tree}
 \end{table}
 
 \begin{figure}
   \[
   \xymatrix@R-1pc{%
     &&\oplus\ar@{-}[dll]_{\Tag_1^{\carriedT}}\ar@{-}[ddl]_{\Tag_1^{\contT}}\ar@{-}[ddr]^{\Tag_2^{\carriedT}}\ar@{-}[drr]^{\Tag_2^{\contT}}&&&\\
     \End&&&&\End\\
     &\oplus\ar@{-}[dl]_{\Tag_1^{\carriedT}}\ar@{-}[dd]_{\Tag_1^{\contT}}\ar@{-}[ddr]^{\Tag_2^{\carriedT}}\ar@{-}[drr]^{\Tag_2^{\contT}}&&\End\\
     \End&&&\End&\\
     &_{\vdots}\oplus_{\vdots}&\End%
   }\]
\caption{%
  The tree of %
  $\trec\tvar.\Out{\Tag_1}{\End}.\tvar\oplus\Out{\Tag_2}{\End}.\End$.%
}%
\label{te}
\end{figure}

In the examples we use infix notation for $\&$ and $\oplus$ and ground types ($\mathtt{int}, \mathtt{bool}, \ldots)$ for messages. The extension to ground types is given in \S~\ref{ext}. %
\label{remark:closed-types}%
Unless otherwise noted, %
our definitions and statements will always refer to closed types.%

As usual {\em session duality}~\cite{HVK} 
plays an important r\^ole  for session types. %
\label{def:dualf}%
The function $\dualf T$, defined below, yields the
dual of the (possibly open) session type $T$. 
$$
\begin{array}{c}
\dualf{\Branch_{i\in I} \In{\Tag_i}{\Type_i}.\SessionTypeT_i}
=\Select_{i\in I} \Out{\Tag_i}{\Type_i}.\dualf{\SessionTypeT_i} 
\qquad
\dualf{\Select_{i\in I} \Out{\Tag_i}{\Type_i}.\SessionTypeT_i}
=\Branch_{i\in I} \In{\Tag_i}{\Type_i}.\dualf{\SessionTypeT_i} 
\\[1mm]
\dualf{\tvar} = \tvar
\qquad
\dualf{\trec\tvar.\SessionType} = \trec \tvar .\dualf{\SessionType}
\qquad 
\dualf{\End} = \End 
\end{array}
$$
We write $\SessionTypeT_1 \dual \SessionType_2$ if 
$\SessionType_2 = \dualf{\SessionType_1}$.
Note that $\tree{\dualf\SessionTypeT}$
can be obtained from $\tree\SessionTypeT$ %
by turning branching nodes into selection nodes (and \emph{vice versa}) in all continuation paths, %
without altering the exchanged sub-trees.

\begin{table}[h!]
$$
\begin{array}{@{}c@{}}
\inferrule[\rulename{sub-end}]{}
{\End\subt\End}
\qquad 
\cinferrule[\rulename{sub-bra}]{
  \forall i\in I: \TypeU_i \subt \TypeU'_i
  \quad \SessionTypeT_i \subt \SessionTypeT'_i
}{
 \Branch_{i\in I \cup J}
  \In{\Tag_i}{\TypeU_i}.\SessionTypeT_i
    \subt
    \Branch_{i\in I}
  \In{\Tag_i}{\TypeU'_i}.\SessionTypeT'_i
}
  \qquad
\cinferrule[\rulename{sub-sel}]{
  \forall i\in I: \TypeU'_i \subt \TypeU_i
  \quad \SessionTypeT_i \subt \SessionTypeT'_i
}{
  \Select_{i\in I}
  \Out{\Tag_i}{\TypeU_i}.\SessionTypeT_i
  \subt
  \Select_{i\in I \cup J}
  \Out{\Tag_i}{\TypeU'_i}.\SessionTypeT'_i
 }
\end{array}
$$
\caption{\label{tab:sync:ssubt} Synchronous subtyping.}
\end{table}

Table~\ref{tab:sync:ssubt} defines the subtyping. 
Note that the double line in rules indicates
that the rules should be interpreted {\em coinductively}.
We follow the ordering of the branching-selection originally adopted by Honda \etal~\cite{Carbone:2012:SCP:2220365.2220367,DemangeonH11,event,asyncsession,
mostrous09sessionbased,mostrous_phd,MY15,mostrous_yoshida_honda_esop09}.
Rule \rulename{sub-bra} states that 
the branching which offers fewer branches is a supertype of 
the one with more branches; 
and rule \rulename{sub-sel} is its dual (see the explanations in \S~\ref{sec:introduction}). 
We write $T \ssubt S$ if $T \subt S$ is derived by the rules in 
Table~\ref{tab:sync:ssubt}. 
Reflexivity of $\ssubt$ is immediate and
transitivity of $\ssubt$ can be shown in the standard way, see Theorem~\ref{thm:trans:ssubt} in Appendix~\ref{proof:soundness}. 

The typing judgements for synchronous processes take the following form:
$$
\wtps{\UEnv}{\Process}{\LEnv}
$$
where $\UEnv$ is the {\em shared environment} 
which associates process variables to sequences of session types and 
$\LEnv$ is the {\em session environment} 
which associates identifiers to session types.
They are defined by:
$$
\UEnv:: = 
\emptyset \mid 
\UEnv, \pbind{\pvar}{\tilde\SessionType}
\quad \quad \quad 
\LEnv::=
\emptyset \mid
\LEnv, \Name: \SessionType
$$
We write 
$\LEnv_1, \LEnv_2$ for 
$\LEnv_1\cup \LEnv_2$ when 
$\dom(\LEnv_1) \cap \dom(\LEnv_2)=\emptyset$. 
We say that  
$\LEnv$ is {\em end-only} if
$\Name:\SessionType \in \LEnv$ implies $\SessionType = \End$.

We define a pre-order between the session environments which reflects subtyping.
More precisely, $\LEnv_1 \ssubt \LEnv_2$ if:
$$
\hspace{-3mm}
\begin{array}{rl}
&\hspace{-2mm} 
\Name \in \dom(\LEnv_1) \cap \dom(\LEnv_2) \text{ implies }
\LEnv_1(\Name) \ssubt \LEnv_2(\Name)\\
&\hspace{-2mm} 
\Name \in \dom(\LEnv_1)  \text{ and } \Name \not\in \dom(\LEnv_2)  \text{ imply }
\LEnv_1(\Name) = \End\\
& \hspace{-2mm}\Name \not\in \dom(\LEnv_1)  \text{ and }\Name \in \dom(\LEnv_2)  \text{ imply }
\LEnv_2(\Name) = \End
\end{array}
$$
We write $\LEnv_1 \approxs \LEnv_2$ if $\LEnv_1 \ssubt \LEnv_2$ and $\LEnv_2 \ssubt \LEnv_1$. It is easy to verify that $\LEnv$ is end-only iff $\LEnv\approxs\EmptyEnv$.

\begin{table}
$$
\begin{array}[t]{@{}c@{}}
 \inferrule[\rulename{t-sub}]{
    \wtp{\UEnv}{\Process}{\LEnv}
    \\
    \LEnv \ssubt \LEnv'
  }{
    \wtp{\UEnv}{\Process}{\LEnv'}
  }
    \qquad
  \inferrule[\rulename{t-idle}]{}{
  \wtp{\UEnv}{\idle}{\emptyset}
  }
    \qquad
  \inferrule[\rulename{t-var}]{}{
    \wtp{
    \UEnv,
    \pbind{\pvar}{\tilde\SessionType}
    }{
      \invoke\pvar{\tilde\Name}
    }{\ASET{\bind{\tilde \Name}{\tilde \SessionType}}}
  }
\\[5.5mm]
  \inferrule[\rulename{t-input}]{
    \forall i\in I:
    \wtp{\UEnv}{
      \Process_i
    }{
      \LEnv,
      \ASETT{
      \bind{\Name}{\SessionTypeT_i},
      \bind{\var_i}{\SessionTypeS_i}}
    }  }{
    \wtp{\UEnv}{
      \sum_{\mathclap{i\in I}}
      \receive{\Name}{\Tag_i}{\var_i}.\Process_i
    }{
      \LEnv,
      \ASETT{
        \bind{\Name}{
          \Branch_{i\in I}
          \In{\Tag_i}{\Type_i}.\SessionTypeT_i
        }}
    }
  }
\qquad
  \inferrule[\rulename{t-output}]{
    \wtp{\UEnv}{
      \Process
    }{
      \LEnv,
      \ASETT{\bind{\NameU}{\SessionTypeT}}
    }
  }{
    \wtp{\UEnv}{
      \send{\NameU}{\Tag}{\Name'}.\Process
    }{
      \LEnv,
      \ASETT{
        \bind{\Name}{\Out{\Tag}{\SessionTypeS}.\SessionTypeT},
        \bind{\Name'}{\SessionTypeS}}
    }
  }
\\[5.5mm]
  \inferrule[\rulename{t-par}]{
    \wtp{\UEnv}{
      \Process_1
    }{
      \LEnv_1
    }
    \\
    \wtp{\UEnv}{
      \Process_2
    }{
      \LEnv_2
    }
  }{
    \wtp{\UEnv}{
      \Process_1 \parop \Process_2
    }{
      \LEnv_1, \LEnv_2
    }
  }
  \qquad
  \inferrule[\rulename{t-choice}]{
    \wtp{\UEnv}{
      \Process_1
    }{
      \LEnv
    }
    \\
    \wtp{\UEnv}{
      \Process_2
    }{
      \LEnv
    }
  }{
    \wtp{\UEnv}{
      \Process_1 \choice \Process_2
    }{
      \LEnv
    }
  }
  \\[5.5mm]
  \inferrule[\rulename{t-def}]{
    \wtp{
      \UEnv,
      \pbind{\pvar}{\tilde\SessionType}
    }{
      \ProcessP
    }{
        \ASET{\bind{\tilde\var}{\tilde\SessionType}}
    }
    \\
    \wtp{
      \UEnv,
      \pbind{\pvar}{\tilde\SessionType}
    }{
      \ProcessQ
    }{
      \LEnv
    }
  }{
    \wtp{\UEnv}{
      \Def{\pvar}{\tilde\var}{\ProcessP}{\ProcessQ}
    }{
      \LEnv
    }
  }
  \qquad
  \inferrule[\rulename{t-new-sync}]{
    \wtp{\UEnv}{
      \Process
    }{
      \LEnv,
      \ASETT{
      \bind{\ChannelA}{\SessionType_1},
      \bind{\ChannelB}{\SessionType_2}}
      \quad
      \SessionTypeT_1 \dual \SessionType_2
    }
  }{
    \wtp{\UEnv}{
      \new{\ChannelA}{\ChannelB}\Process
    }{
      \LEnv
    }
  }
\end{array}
$$
\caption{\label{tab:sync:typing} Typing rules for synchronous processes.
}
\end{table}

\bigskip

Table~\ref{tab:sync:typing} gives the typing rules. 
They are standard 
in session calculi~\cite{GH05}.   
Rule \rulename{t-idle} is the introduction rule for the nil process.
To type an input process, rule 
\rulename{t-input}
requires the type $S_i$ of variable $x_i$ and the type 
$T_i$ of channel $u$ for the continuation $P_i$. 
In the resulting
session environment, the type $u$ has the branching type 
in which $u$ receives $S_i$ and then continues with~$T_i$ for each
label $l_i$.
The rule for typing output processes is similar and simpler. 
In rule \rulename{t-par}, the session environment of $P_1\parop  P_2$ is the
disjoint union of the environments $\Delta_1$ and $\Delta_2$ for the two
processes, reflecting the linear nature of channels. 
Contrarily, in rule \rulename{t-choice}, the two processes share the same
session environment, since at most one of them will be executed.
Rules
\rulename{t-var} and \rulename{t-def} 
deal with process calls and definitions, requiring 
the channel parameters have the types which are assumed in the shared environment. 
Rule \rulename{t-var} gives these types to  the arguments of the process variable. 
In \rulename{t-def}, the parameters of the process associated with the process variable must be typed with these types.  
The assumption on parameter types is also used to type the body of the definition. 
Rule \rulename{t-new-sync} is a standard rule for name binding, where we ensure 
the co-channels have dual types.
Finally, %
rule \rulename{t-sub} is the standard rule for subtyping: %
a process $\ProcessP$ whose channels are typed according to the
session environment $\LEnv$ %
can be used in a 
 type derivation requiring a less demanding environment $\LEnv'$. This is the key rule which allows, %
as described in \S~\ref{sec:introduction}, %
to use a process $\ProcessP$ with a channel $\ChannelA$ of type~(\ref{BBB}) %
in a derivation where $\ChannelA$ has the supertype~(\ref{AAA}). %
As a consequence, %
a larger subtyping relation allows to type more processes.

We write $\wtps{\UEnv}{\Process}{\LEnv}$ 
if $\Process$ is typed using the rules in Table~\ref{tab:sync:typing}. 

\subsection{Soundness of synchronous subtyping}
Our type system enjoys the standard property of subject reduction. Notice that session environments are unchanged since only bound channels can be reduced.
\begin{thm}[Subject reduction for synchronous processes]\label{thm:sbr:sync}
If $\wtps{\UEnv}{\Process}{\LEnv}$ and 
$\Process \wreds \ProcessQ$, then
\mbox{$\wtps{\UEnv}{\ProcessQ}{\LEnv}$.}
\end{thm}

From subject reduction we can easily derive that well-typed processes cannot produce error. 
\begin{cor}\label{pro:comsafe:sync}
  If $\wtps{\UEnv}{\Process}{\LEnv}$, 
  then $\Process \not \wreds \error$.
\end{cor}
\begin{Proof}
By  Theorem \ref{thm:sbr:sync},
$\Process \wreds \error$ implies $\wtps{\UEnv}{\error}{\LEnv}$,
which is impossible
because $\error$ has no type.
\end{Proof}

The proof of soundness theorem follows easily.
\begin{thm}\label{thm:s:sound}
  The synchronous subtyping relation $\ssubt$ is sound  for the synchronous calculus.
\end{thm}

The proofs of Theorems~\ref{thm:sbr:sync} and~\ref{thm:s:sound} are given in Appendix~\ref{proof:soundness}. 

\section{Completeness for Synchronous Subtyping}
\label{sec:s:completeness}
This section proves the first main result, 
completeness of synchronous subtyping, which together with soundness shows the preciseness theorem. 
We shall take the following three steps.

\begin{itemize}
\item {\bf [Step~1]} \ For each type $T$ and identifier $u$, we 
define a {\em characteristic process} $\cprocess{}(u,T)$ typed by $u:T$,  
which offers the series of interactions described by $T$ on identifier $u$.

\item {\bf [Step~2]} \ We characterise the negation of the subtyping relation 
by inductive rules (notation $\nssubt$). 

\item {\bf [Step~3]} \ %
  \label{step-iii}%
  We leverage characteristic processes %
  to prove that if $\SessionTypeT \nssubt \SessionTypeS$, %
  then there exist $\ProcessP, \ProcessQ$ %
  such that\; %
  ${\wtp{}{\ProcessP}{\ASET{\ChannelA:\SessionTypeT}}}$ %
  \;and\; %
  ${\wtp{}{\ProcessQ}{\ASET{\ChannelB:\dualf \SessionTypeS}}}$ %
  \;and\; %
  $( \nu \ChannelA\ChannelB )(\ProcessP \parop \ProcessQ) \wred \error$.
  Hence, %
  by suitably choosing $\ProcessP$, $\ProcessQ$, %
  and 
  $\Context\hole$
  in the definition of preciseness %
  (Definition~\ref{def:preciseness}), %
  we achieve completeness.%
\end{itemize}
The same three steps will be used for the completeness proof in the asynchronous case. 

\mypar{Characteristic synchronous processes} 
The characteristic synchronous processes are defined following the structure of types. For each type we build a process with a single identifier offering the communications prescribed by the type. We also create auxiliary processes for exchanged identifiers. 

\begin{defi}[Characteristic synchronous processes]
\label{s:cprocesses} \rm 
The characteristic process offering communication $T$ on identifier $u$
for the synchronous calculus, denoted by $\cprocess(\Name,\SessionType)$,
is defined by induction on (possibly open) session types:\\
$$
\begin{array}{rcl}
\cprocess(\Name,\SessionType)
 &\eqdef&
\begin{cases}
  \sum_{i\in I}
  \cprocess[?](\Name,\Tag_i,\Type_i,\SessionType_i)
  &\text{if }  \SessionType = \Branch_{i\in I} \In{\Tag_i}{\Type_i}.\SessionType_i
  \\
  \bigoplus_{i\in I}
  \cprocess[!](\Name,\Tag_i,\Type_i,\SessionType_i)
  &\text{if } \SessionType = \Select_{i\in I}
  \Out{\Tag_i}{\Type_i}.\SessionType_i
  \\
  \invoke{\pvar_\tvar}{\Name} &\text{if }  \SessionType = \tvar  \\
  \Def{\pvar_\tvar}{\var}{
    \cprocess(\var,\SessionTypeS)
  }{
    \invoke{\pvar_\tvar}{\Name}
  }
  &\text{if } 
  \SessionType = \trec\tvar.\SessionTypeS\\
   \idle &\text{if }  \SessionType = \End 
\end{cases}
\\[1mm]
\cprocess[?](\Name,\Tag,\Type,\SessionType)
 &\eqdef&
   \receive{\Name}{\Tag}{\var}.(
    \cprocess(\Name,\SessionType)
    \parop
    \cprocess(\var,\SessionTypeS)
  )
\\[1mm]
\cprocess[!](\Name,\Tag,\Type,\SessionType)
 &\eqdef&
  \new\ChannelA\ChannelB(
    \send{\Name}{\Tag}{\ChannelA}.\cprocess(\Name,\SessionType)
    \parop
    \cprocess(\ChannelB,\co\SessionTypeS)
  )
\end{array}
$$
\end{defi}

\begin{table}
\begin{math}
\displaystyle
\begin{array}{@{}c@{}}
 \inferrule[\rulename{n-end r}]{ 
  \SessionTypeT \ne \End
  }{
    \End \nsubt \SessionTypeT
   }
  \qquad
   \inferrule[\rulename{n-end l}]
  {  \SessionTypeT \ne \End
  }{
    \SessionTypeT \nsubt \End
  }
  \\\\
  \inferrule[\rulename{n-brasel}]
  {}{
    \textstyle
    \Branch_{i\in I}
    \In{\Tag_i}{\Type_i}.\SessionTypeT_i
    \nsubt
    \Select_{j\in J}
    \Out{\Tag'_j}{\TypeU'_j}.\SessionTypeT'_j
  }
 \qquad
  \inferrule[\rulename{n-selbra-sync}]{}{
    \textstyle
    \Select_{j\in J}
    \Out{\Tag'_j}{\TypeU'_j}.\SessionTypeT'_j
    \nsubt
    \Branch_{i\in I}
    \In{\Tag_i}{\Type_i}.\SessionTypeT_i
  }
  \\\\
  \inferrule[\rulename{n-label-bra}]{
    \exists j\in J~
    \forall i\in I:
        \Tag_i \ne \Tag'_j
  }{
    \Branch_{i\in I} \In{\Tag_i}{\Type_i}.\SessionTypeT_i
    \nsubt
    \Branch_{j\in J} \In{\Tag'_j}{\Type'_j}.\SessionTypeT'_j
  }
\qquad
  \inferrule[\rulename{n-label-sel}]{
  \exists i\in I~
    \forall j\in J:
    \Tag_i \ne \Tag'_j
  }{
    \textstyle
    \Select_{i\in I} \Out{\Tag_i}{\Type_i}.\SessionTypeT_i
    \nsubt
    \Select_{j\in J} \Out{\Tag'_j}{\Type'_j}.\SessionTypeT'_j
  }
  \\\\
  \inferrule[\rulename{n-exch-bra}]
{
    \exists i\in I~
    \exists j\in J:
    \Tag_i = \Tag'_j~~~
    \Type_i \nsubt \Type'_j
  }{
    \Branch_{i\in I} \In{\Tag_i}{\Type_i}.\SessionTypeT_i
    \nsubt
    \Branch_{j\in J} \In{\Tag'_j}{\Type'_j}.\SessionTypeT'_j
  }
  \qquad
  \inferrule[\rulename{n-exch-sel}]{
    \exists i\in I~
    \exists j\in J:
    \Tag_i = \Tag'_j~~~
\Type'_j \nsubt \Type_i
  }{
    \textstyle
    \Select_{i\in I} \Out{\Tag_i}{\Type_i}.\SessionTypeT_i
    \nsubt
    \Select_{j\in J} \Out{\Tag'_j}{\Type'_j}.\SessionTypeT'_j
  }
 \\\\
  \inferrule[\rulename{n-cont-bra}]
{
    \exists i\in I~
    \exists j\in J:
    \Tag_i = \Tag'_j~~~
    \SessionTypeT_i \nsubt \SessionTypeT'_j
  }{
    \Branch_{i\in I} \In{\Tag_i}{\Type_i}.\SessionTypeT_i
    \nsubt
    \Branch_{j\in J} \In{\Tag'_j}{\Type'_j}.\SessionTypeT'_j
  }
 \qquad
  \inferrule[\rulename{n-cont-sel}]{
    \exists i\in I~
    \exists j\in J:
    \Tag_i = \Tag'_j ~~~
    \SessionTypeT_i \nsubt \SessionTypeT'_j
  }{
    \textstyle
    \Select_{i\in I} \Out{\Tag_i}{\Type_i}.\SessionTypeT_i
    \nsubt
    \Select_{j\in J} \Out{\Tag'_j}{\Type'_j}.\SessionTypeT'_j
  }
\end{array}
\end{math}
\caption{\label{tab:negsubtype} 
Negation of synchronous subtyping.}
\end{table}

A branching type is mapped to the inputs 
$\cprocess[?](\Name,\Tag_i,\Type_i,\SessionType_i)$ $(i\in I)$, 
which uses the input channel $x$ in 
$\cprocess(x,\SessionTypeS_i)$. 
A selection type is mapped to 
the choice between the outputs 
$\cprocess[!](\Name,\Tag_i,\Type_i,\SessionType_i)$ $(i\in I)$,
where the fresh channel $a$ carried by $u$ will be received 
by the dual input, which will interact with the process
$\cprocess(\ChannelB,\co\SessionTypeS_i)$.
A recursive type is mapped in a definition associated to the characteristic process of the type body. The process body of this definition is just a call to 
the process variable associated to the recursive type variable. Type $\End$ is mapped to $\idle$. %
  Note that our characteristic processes %
  interact sequentially on a \emph{single} session %
  --- and in case of recursion, %
  they only have one parameter: %
  this is sufficient %
  to capture the errors we are interested in, %
  since they do not depend on multiple sessions being interleaved. %

For example if $\SessionType=\trec\tvar.\Out{\Tag_1}{\End}.\tvar\oplus\Out{\Tag_2}{\Out{\Tag_3}\End.\End}.\End$, then 
$$\cprocess(\ChannelA,\SessionType)\quad=\quad\Def{\pvar_\tvar}{\var}{
    P
  }{
    \invoke{\pvar_\tvar}{\ChannelA}
  }$$
  where 
$$\begin{array}{lll}P&=&\cprocess[!](\ChannelA,\Tag_1,\End,\tvar)\oplus\cprocess[!](\ChannelA,\Tag_2,\Out{\Tag_3}\End.\End,\End)\\
&=& \new\ChannelB{\ChannelB'}(
    \send{\ChannelA}{\Tag_1}{\ChannelB}.\cprocess(\ChannelA,\tvar)
    \parop
    \cprocess(\ChannelB',\End))\oplus
    \new\ChannelC{\ChannelC'}(
    \send{\ChannelA}{\Tag_2}{\ChannelC}.\cprocess(\ChannelA,\End)
    \parop
    \cprocess(\ChannelC',\In{\Tag_3}\End.\End))
  \\
  &=& \new\ChannelB {\ChannelB'}(\send{\ChannelA}{\Tag_1}{\ChannelB}. \invoke{\pvar_\tvar}{\ChannelA}\parop\idle) \oplus
   \new\ChannelC{\ChannelC'}(\send{\ChannelA}{\Tag_2}{\ChannelC}.\idle \parop \receive{\ChannelC'}{\Tag_3}{\var}.( \cprocess(\ChannelC',\End) \parop \cprocess(\var,\End)))\\
  &=& \new\ChannelB {\ChannelB'}(\send{\ChannelA}{\Tag_1}{\ChannelB}. \invoke{\pvar_\tvar}{\ChannelA}\parop\idle) \oplus
  \new\ChannelC{\ChannelC'}(\send{\ChannelA}{\Tag_2}{\ChannelC}.\idle \parop \receive{\ChannelC'}{\Tag_3}{\var}.(\idle \parop\idle))\\
&\equiv& \new\ChannelB {\ChannelB'}(\send{\ChannelA}{\Tag_1}{\ChannelB}. \invoke{\pvar_\tvar}{\ChannelA} ) \oplus \new\ChannelC{\ChannelC'}(\send{\ChannelA}{\Tag_2}{\ChannelC}\parop \receive{\ChannelC'}{\Tag_3}{\var})
\end{array}$$
We can easily check that characteristic processes are well typed as expected. 
\begin{lem}\label{cpts}$\wtps{\;}{\cprocess(\Name,\SessionType)}{\ASET{\Name:\SessionType}}$.\end{lem}

\mypar{Rules for negation of synchronous subtyping}
\label{subsec:s:negation}
Table~\ref{tab:negsubtype} defines the rules which characterise when a
type is not a subtype of another type. %
We formulate these rules inductively, to simplify the completeness proof. %
Rules \rulename{n-end r} and \rulename{n-end l} say that 
$\End$ cannot be a super or subtype of a type different from $\End$. 
Rule \rulename{n-brasel} says that a branching type cannot be subtype of a
selection type. Rule \rulename{n-selbra-sync} is its dual. Rules
\rulename{n-label-bra} and \rulename{n-label-sel} represent the cases 
in which the labels do not conform to the subtyping rules. Rules 
\rulename{n-exch-bra} and \rulename{n-exch-sel} represent the cases in which
carried types do not match the subtyping rules. Lastly,
rules 
\rulename{n-cont-bra} and \rulename{n-cont-sel} represent the cases in which
 continuations do not match the subtyping rules. Notice that if rule \rulename{sub-bra} holds, then the rules \rulename{n-$\star$-bra} do not hold, and if rule \rulename{sub-sel} holds, then the rules \rulename{n-$\star$-sel} do not hold, where $\star\in\{\text{\small{\textsc{label, exch, cont}}}\}$.
We write $T \nssubt S$ if $T \nsubt S$ is derived by the rules in 
Table~\ref{tab:negsubtype}.

In Lemma~\ref{lem:s:negation}, %
we show that $\nssubt$ is the negation of the synchronous subtyping. %
This result will be used (in the \emph{``only if''} direction) %
in the proof of Theorem~\ref{thm:preciseness:sync}.
\begin{lem}\label{lem:s:negation} If
  $\T\ssubt \SessionTypeS$ is not derivable %
  if and only if %
  $\T \nssubt \SessionTypeS$ is derivable. 
\end{lem}
\begin{Proof} 
If $\SessionTypeT \nssubt \SessionTypeS$, then we can show
$\SessionTypeT \not\ssubt \SessionTypeS$ by induction on the
derivation of $\SessionTypeT \nssubt \SessionTypeS$. %
We develop just two cases (the others are similar):
\begin{itemize}
\item%
 base case \rulename{n-brasel}.\quad %
 Then, %
 $\SessionTypeT = \Branch_{i\in
   I}\In{\Tag_i}{\Type_i}.\SessionTypeT_i$ %
 and %
 $\SessionTypeS = \Select_{j\in
   J}\Out{\Tag'_j}{\TypeU'_j}.\SessionTypeT'_j$. %
 We can verify that $\SessionTypeT$ and $\SessionTypeS$ do not
 match %
 the conclusion of \rulename{sub-end}, nor \rulename{sub-bra}, nor
 \rulename{sub-sel} %
 --- hence, we conclude $\SessionTypeT \not\ssubt \SessionTypeS$;
\item%
 inductive case \rulename{n-cont-sel}. %
 Then, %
 $\SessionTypeT = \Select_{i\in I} \Out{\Tag_i}{\Type_i}.\SessionTypeT_i$ %
 and %
 $\SessionTypeS = \Select_{j\in J} \Out{\Tag'_j}{\Type'_j}.\SessionTypeT'_j$; %
 moreover, %
 $\exists i\in I,j\in J: \Tag_i = \Tag'_j$ and %
 $\SessionTypeT_i \nsubt \SessionTypeT'_j$ %
 --- and thus, by the induction hypothesis, %
 $\SessionTypeT_i \not\ssubt \SessionTypeT'_j$. %
 We now notice that $\SessionTypeT \ssubt \SessionTypeS$ %
 could only possibly hold by rule \rulename{sub-sel} %
 --- but, since $\SessionTypeT_i \not\ssubt \SessionTypeT'_j$, %
 at least one of the coinductive premises of such a rule is 
 not satisfied. %
 Hence, we conclude $\SessionTypeT \not\ssubt \SessionTypeS$.
\end{itemize}
\emph{Vice versa}, %
if $\SessionTypeT \not\ssubt \SessionTypeS$ %
we construct a derivation of $\T \nssubt \SessionTypeS$ %
by looking at a ``failing derivation'' of 
$\SessionTypeT \ssubt \SessionTypeS$. %
If we try to apply the subtyping rules to show
$\SessionTypeT \ssubt \SessionTypeS$, %
there exists a derivation branch that fails after $n$ steps, %
\ie that reaches two types $\SessionTypeT', \SessionTypeS'$ %
whose trees do \emph{not} match the conclusion of
\rulename{sub-end}, nor \rulename{sub-bra}, nor \rulename{sub-sel}. %
Note that no alternative derivation exists, %
because the rules in Table~\ref{tab:sync:ssubt} do not overlap, %
hence at most one of them can be applied at each step. %
We prove $\SessionTypeT \nssubt \SessionTypeS$ by induction on $n$, %
turning the failing coinductive derivation branch %
into a derivation of depth $n+1$ %
which concludes %
$\SessionTypeT \nssubt \SessionTypeS$:
\begin{itemize}
\item%
  base case $n=0$.\quad %
  The derivation fails immediately, %
  i.e.~$\SessionTypeT'=\SessionTypeT$ and
  $\SessionTypeS'=\SessionTypeS$. %
  By cases on the possible shapes of $\SessionTypeT$ and $\SessionTypeS$, %
  we construct %
  a derivation which concludes %
  $\SessionTypeT \nssubt \SessionTypeS$ %
  in $1 = n+1$ steps, %
  by one of the axioms %
  \rulename{n-end r}, \rulename{n-end l}, \rulename{n-brasel},
  \rulename{n-selbra-sync}, \rulename{n-label-bra},
  \rulename{n-label-sel};
\item%
  inductive case $n=m+1$.\quad %
  The shapes of $\SessionTypeT,\SessionTypeS$ match the conclusion of
  rule \rulename{sub-bra} (resp.~\rulename{sub-sel}), %
  but there is some coinductive premise $\SessionTypeT' \ssubt \SessionTypeS'$ %
  whose sub-derivation has a branch that fails after $m$ steps. %
  By the induction hypothesis, %
  there exists a derivation of depth $m+1$
  that concludes $\SessionTypeT' \nssubt \SessionTypeS'$; %
  using this as a premise, %
  by one of the rules \rulename{n-exch-bra} or \rulename{n-cont-bra} %
  (resp.~\rulename{n-exch-sel} or \rulename{n-cont-sel}), %
  we construct a derivation of depth $(m+1)+1 = n+1$ %
  which concludes $\SessionTypeT \nssubt \SessionTypeS$.\qedhere
\end{itemize}
\end{Proof}

\noindent The main theorem for synchronous subtyping can now be proved. 

\begin{thm}[Completeness for synchronous subtyping]\label{thm:preciseness:sync}
The synchronous subtyping relation $\ssubt$ is complete for the synchronous calculus. 
\end{thm}
\begin{Proof}
We need to produce $\ProcessP,\ProcessQ$ as described in
{\bf Step 3} on page~\pageref{step-iii}. %
We let $\ProcessP = \cprocess(\ChannelA,\SessionTypeT)$ %
and $\ProcessQ = \cprocess(\ChannelB,\co\SessionTypeS)$, %
and we show that %
$$
  \new\ChannelA\ChannelB(
    \cprocess(\ChannelA,\SessionTypeT)
    \parop
    \cprocess(\ChannelB,\co\SessionTypeS)
  )
  \wreds
  \error
$$
where $ \cprocess(\ChannelA,\SessionTypeT)$, $ \cprocess(\ChannelB,\co\SessionTypeS)$ are characteristic synchronous processes.
The proof is by induction on the derivation of %
$\T \nssubt \SessionTypeS$. 

\smallskip

{\em Case} \rulename{n-end r}: 
$\SessionTypeT = \End$ and $\SessionTypeS \not= \End$.
$$
\begin{array}{l}
\new\ChannelA\ChannelB
 (
 \cprocess(\ChannelA,\SessionTypeT)
 \parop
 \cprocess(\ChannelB,\co{\SessionTypeS}))
 =
\new \ChannelA \ChannelB
(
 \idle \parop \cprocess(\ChannelB, \co\SessionTypeS)
)
 \reds \error
\end{array}
$$
by rule $\rulename{err-new-sync}$:
in fact, we have %
$\ChannelA \not \in \fn(\idle)$, %
and
$\SessionTypeS \not= \End$ implies %
that $\cprocess(\ChannelB, \co\SessionTypeS)$ 
is a (possibly recursive) %
internal or external choice, %
and thus %
$\ChannelB \in \sn{(\cprocess(\ChannelB, \co\SessionTypeS))}$ %
(by Definition~\ref{s:cprocesses}).%

{\em Case} \rulename{n-end l}: 
$\SessionTypeT \not = \End$ and $\SessionTypeS = \End$.
The proof is as in the previous case. 

\smallskip

{\em Case} \rulename{n-brasel}: 
$\SessionTypeT = 
\Branch_{i\in I}\In{\Tag_i}{\SessionTypeT'_i}.\SessionTypeT_i$ and 
$\SessionTypeS = 
\Select_{j\in J}\Out{\Tag'_j}{\SessionTypeS'_j}.\SessionTypeS_j$.
$$
\begin{array}{l}
\new\ChannelA\ChannelB
 (
 \cprocess(\ChannelA,\SessionTypeT)
 \parop
 \cprocess(\ChannelB,\co{\SessionTypeS}))
 =

\new \ChannelA \ChannelB
(
 \sum_{i \in I} \cprocess[?](\ChannelA, \Tag_i, \SessionTypeT'_i, \SessionTypeT_i)
 \parop
 \sum_{j \in J} \cprocess[?](\ChannelB, \Tag'_j, \SessionTypeS'_j, \co{\SessionTypeS_j})
)
 \reds \error
\end{array}
$$
by rule $\rulename{err-in-in-sync}$.

\smallskip

{\em Case} \rulename{n-selbra-sync}:
$\SessionTypeT = 
\Select_{i\in I}\Out{\Tag_i}{\SessionTypeT'_i}.\SessionTypeT_i$ and
$\SessionTypeS = 
\Branch_{j\in J}\In{\Tag'_j}{\SessionTypeS'_j}.\SessionTypeS_j$.
$$
\begin{array}{l}
\new\ChannelA\ChannelB
 (
 \cprocess(\ChannelA,\SessionTypeT)
 \parop
 \cprocess(\ChannelB,\co\SessionTypeS))
 =
\new \ChannelA \ChannelB
(
 \bigoplus_{i \in I} \cprocess[!](\ChannelA, \Tag_i, \SessionTypeT'_i, \SessionTypeT_i)
 \parop
 \bigoplus_{j \in J} \cprocess[!](\ChannelB, \Tag'_j, \SessionTypeS'_j, \co{\SessionTypeS_j})
) 
\reds
\error
\end{array}
$$
by rule $\rulename{err-out-out-sync}$. 

\smallskip

{\em Case} \rulename{n-label-bra}:
$\SessionTypeT= \Branch_{i\in I} \In{\Tag_i}{\SessionTypeT'_i}.\SessionTypeT_i$, 
$\SessionTypeS = \Branch_{j\in J} \In{\Tag'_j}{\SessionTypeS'_j}.\SessionTypeS_j$, 
and $ \exists k\in J$ such that $\forall i\in I:\Tag'_k \ne \Tag_i$.
$$
\begin{array}{l}
\new\ChannelA\ChannelB
 (
 \cprocess(\ChannelA,\SessionTypeT)
 \parop
 \cprocess(\ChannelB,\co\SessionTypeS))
 =

\new \ChannelA \ChannelB
(
 \sum_{i \in I} 
 \cprocess[?](\ChannelA, \Tag_i, \SessionTypeT'_i, \SessionTypeT_i)
 \parop
 \bigoplus_{j \in J} 
 \cprocess[!](\ChannelB, \Tag'_j, \SessionTypeS'_j, \co{\SessionTypeS_j})
)
\reds \\
\new \ChannelA \ChannelB
(
 \sum_{i \in I}
 \cprocess[?](\ChannelA, \Tag_i, \SessionTypeT'_i, \SessionTypeT_i)
 \parop
 \cprocess[!](\ChannelB, \Tag'_k, \SessionTypeS'_k, \co{\SessionTypeS_k})
)
\reds
\error
\end{array}
$$
by rule $\rulename{err-mism-sync}$.

\smallskip

{\em Case} \rulename{n-label-sel}:
$\SessionTypeT = \Select_{i\in I} \Out{\Tag_i}{\SessionTypeT'_i}.\SessionTypeT_i$,
$\SessionTypeS = \Select_{j\in J} \Out{\Tag'_j}{\SessionTypeS'_j}.\SessionTypeS_j$ and
$\exists k \in I$ such that $\forall j \in J:\Tag_k \ne \Tag'_j$.
$$
\begin{array}{l}
\new\ChannelA\ChannelB
 (
 \cprocess(\ChannelA,\SessionTypeT)
 \parop
 \cprocess(\ChannelB,\co{\SessionTypeS}))
 =

\new \ChannelA \ChannelB
(
 \bigoplus_{i \in I} 
 \cprocess[!](\ChannelA, \Tag_i, \SessionTypeT'_i, \SessionTypeT_i)
 \parop
 \sum_{j \in J} \cprocess[?](\ChannelB, \Tag'_j, \SessionTypeS'_j, \co{\SessionTypeS_j})
)
\reds
\\
\new \ChannelA \ChannelB
(
 \cprocess[!](\ChannelA, \Tag_k, \SessionTypeT'_k, \SessionTypeT_k)
 \parop
 \sum_{j \in J} 
 \cprocess[?](\ChannelB, \Tag'_j, \SessionTypeS'_j, \co{\SessionTypeS_j})
)
\reds
\error
\end{array}
$$
by rule $\rulename{err-mism-sync}$.

\smallskip

{\em Case} \rulename{n-exch-bra}: 
$\SessionTypeT =  
\Branch_{i\in I} \In{\Tag_i}{\SessionTypeT'_i}.\SessionTypeT_i$,
$\SessionTypeS = 
\Branch_{j\in J} \In{\Tag'_j}{\SessionTypeS'_j}.\SessionTypeS_j$, 
and $\exists k \in I~\exists k' \in J$ such that $\Tag_{k} = \Tag'_{k'}$
and $
    \SessionTypeT'_k \nssubt \SessionTypeS'_{k'}
     $.
$$
\begin{array}{l}
\new \ChannelA \ChannelB
    ( 
    \cprocess(\ChannelA,\SessionTypeT)
    \parop
    \cprocess(\ChannelB, \co \SessionTypeS) 
    )
 =
\new \ChannelA \ChannelB    
( \sum_{i \in  I}
  \cprocess[?](\ChannelA,\Tag_i,\SessionTypeT'_i,\SessionTypeT_i)
  \parop   \bigoplus_{j \in J }
  \cprocess[!](\ChannelB,\Tag'_j,\SessionTypeS'_j,\co{\SessionTypeS_j})
) 
\reds\\  \new \ChannelA \ChannelB    
( \sum_{i \in  I}
  \cprocess[?](\ChannelA,\Tag_i,\SessionTypeT'_i,\SessionTypeT_i)
       \parop
       \new {\ChannelC} {\ChannelD} (
    \send{\ChannelB}{\Tag_{k}}{{\ChannelC}}.
    \cprocess(\ChannelB,\co{\SessionTypeS_{k'}})
    \parop
    \cprocess({\ChannelD} ,\co{\SessionTypeS'_{k'}})) 
) 
 \reds\\
\new \ChannelA \ChannelB    
(  
  \new \ChannelC \ChannelD
  ( 
    \cprocess(\ChannelA, \SessionTypeT_{k})
    \parop
    \cprocess(\ChannelC, \SessionTypeT'_{k})
    \parop 
    \cprocess(\ChannelB, \co{\SessionTypeS_{k'}})
    \parop 
    \cprocess(\ChannelD, \co{\SessionTypeS'_{k'}})
    ) )
\equiv\Context
[ \new \ChannelC \ChannelD
    (\cprocess(\ChannelC, \SessionTypeT'_{k})
    \parop
    \cprocess(\ChannelD, \co{\SessionTypeS'_{k'}})
   )
]
\end{array}
$$
where
$$
\ContextC\hole =
\new \ChannelA \ChannelB
(  
    \cprocess(\ChannelA,\SessionTypeT_{k})
    \parop
    \cprocess(\ChannelB,\co{\SessionTypeS_{k'}})
    )\parop
    \hole.
$$
By induction, 
$$ \new \ChannelC \ChannelD
    (\cprocess(\ChannelC, \SessionTypeT'_{k})
    \parop
    \cprocess(\ChannelD, \co{\SessionTypeS'_{k'}})
   )
  \wreds \error$$
then by rule \rulename{err-context}, we conclude
$$
\Context
[ \new \ChannelC \ChannelD
    (\cprocess(\ChannelC, \SessionTypeT'_{k})
    \parop
    \cprocess(\ChannelD, \co{\SessionTypeS'_{k'}})
   )]
\wreds \error$$

{\em Case} \rulename{n-exch-sel}: 
$\SessionTypeT = \Select_{i\in I} \Out{\Tag_i}{\SessionTypeT'_i}.\SessionTypeT_i$,
$\SessionTypeS =  \Select_{j\in J} \Out{\Tag'_j}{\SessionTypeS'_j}.\SessionTypeS_j$, 
and $\exists k \in I~\exists k' \in J$ %
such that $\Tag_{k} = \Tag'_{k'}$ and
$\SessionTypeS'_{k'} \nssubt \SessionTypeT'_{k}$.
$$
\begin{array}{l}
\new \ChannelA \ChannelB
    (\cprocess(\ChannelA,\SessionTypeT)
    \parop
    \cprocess(\ChannelB, \co \SessionTypeS) )
 \reds \new \ChannelA \ChannelB
     \new {\ChannelC} {\ChannelD} ((
    \send{\ChannelA}{\Tag_{k}}{{\ChannelC}}.
    \cprocess(\ChannelA,\SessionTypeT_{k})
    \parop
    \cprocess({\ChannelD} ,\co{\SessionTypeT'_{k}}))\parop 
  \sum_{j \in J}
  \cprocess[?](\ChannelB,\Tag'_j,\SessionTypeS'_j,\co{\SessionTypeS_j})
 ) 
 \reds\\
\new \ChannelA \ChannelB
    \new {\ChannelC} {\ChannelD} ( 
    \cprocess( \ChannelA,\SessionTypeT_{k} )
    \parop
    \cprocess( {\ChannelD} ,\co{\SessionTypeT'_{k}} )
    \parop
    \cprocess( \ChannelB,\co{\SessionTypeS_{k'}}  ) \parop
    \cprocess( \ChannelC, \SessionTypeS'_{k'}  ) 
    )
\equiv
\Context[
\new \ChannelC \ChannelD
    (\cprocess(\ChannelC, \SessionTypeS'_{k'})
    \parop
    \cprocess(\ChannelD, \co{\SessionTypeT'_{k}})
    )
]
\end{array}
$$
where
$$
\ContextC\hole=
\new \ChannelA \ChannelB
(  
    \cprocess(\ChannelA,\SessionTypeT_{k})
    \parop
    \cprocess(\ChannelB,\co{\SessionTypeS_{k'}})
   ) \parop
    \hole
$$
By induction, 
$$ \new \ChannelC \ChannelD
    (\cprocess(\ChannelC, \SessionTypeS'_{k'})
    \parop
    \cprocess(\ChannelD, \co{\SessionTypeT'_{k}})
   )
  \wreds \error$$
then by rule \rulename{err-context}, we conclude
$$
\Context
[ \new \ChannelC \ChannelD
    (\cprocess(\ChannelC, \SessionTypeS'_{k'})
    \parop
    \cprocess(\ChannelD, \co{\SessionTypeT'_{k}})
   )]
\wreds \error
$$

{\em Case} \rulename{n-cont-bra}: 
$\SessionTypeT =  
\Branch_{i\in I} \In{\Tag_i}{\SessionTypeT'_i}.\SessionTypeT_i$,
$\SessionTypeS = 
\Branch_{j\in J} \In{\Tag'_j}{\SessionTypeS'_j}.\SessionTypeS_j$, 
and $\exists k \in I~\exists k' \in J$ such that %
$\Tag_{k} = \Tag'_{k'}$ and $\SessionTypeT_k \nssubt \SessionTypeS_{k'}$.
As in the previous case
$$
\begin{array}{l}
\new \ChannelA \ChannelB
    (\cprocess(\ChannelA,\SessionTypeT)
    \parop
    \cprocess(\ChannelB, \co \SessionTypeS) )
 \reds^*
\new \ChannelA \ChannelB    
  \new \ChannelC \ChannelD
  ( 
    \cprocess( \ChannelA, \SessionTypeT_{k} )
    \parop
    \cprocess(\ChannelC, \SessionTypeT'_{k}  )
    \parop 
    \cprocess(\ChannelB, \co{\SessionTypeS_{k'}}  )
    \parop 
    \cprocess(\ChannelD, \co{\SessionTypeS'_{k'}} )
    ) 
\equiv\\
\Context[
\new \ChannelA \ChannelB    
(  \cprocess(\ChannelA, \SessionTypeT_{k}) \parop
   \cprocess(\ChannelB, \co{\SessionTypeS_{k'}})
)]
\end{array}
$$
where
$$
\ContextC\hole  =
\new \ChannelC \ChannelD
(  
    \cprocess(\ChannelC, \SessionTypeT'_{k})
    \parop
   \cprocess(\ChannelD, \co{\SessionTypeS'_{k'}})
  )  \parop
    \hole
$$
By induction,
 $$\new \ChannelA \ChannelB 
(\cprocess(\ChannelA,\SessionTypeT_k) \parop 
\cprocess(\ChannelB, \co{\SessionTypeS_{k'}}))  
\wreds \error$$
then by rule \rulename{err-context}, we conclude 
$$\Context[
\new \ChannelA \ChannelB    
(  \cprocess(\ChannelA, \SessionTypeT_{k}) \parop
   \cprocess(\ChannelB, \co{\SessionTypeS_{k'}})
)] \wreds \error$$

\smallskip

{\em Case} \rulename{n-cont-sel}: 
$\SessionTypeT = \Select_{i\in I} \Out{\Tag_i}{\SessionTypeT'_i}.\SessionTypeT_i$,
$\SessionTypeS =  \Select_{j\in J} \Out{\Tag'_j}{\SessionTypeS'_j}.\SessionTypeS_j$, 
and $\exists k \in I ~\exists k' \in J$ such that %
$\Tag_{k} = \Tag'_{k'}$ and $\SessionTypeT_k \nssubt \SessionTypeS_{k'}$.
As in the previous case
$$
\begin{array}{l}
\new \ChannelA \ChannelB
    (\cprocess(\ChannelA,\SessionTypeT)
    \parop
    \cprocess(\ChannelB, \co \SessionTypeS) )
 \reds
\new \ChannelA \ChannelB
 \new \ChannelC \ChannelD
  (   \cprocess(\ChannelA,\SessionTypeT_{k})
      \parop
      \cprocess(\ChannelB,\co{\SessionTypeS_{k'}}) 
      \parop
      \cprocess(\ChannelC, \SessionTypeS'_{k'})
      \parop
      \cprocess(\ChannelD, \co{\SessionTypeT'_{k}})  
  ) 
\equiv\\
\ContextC [
\new \ChannelA \ChannelB
    (\cprocess(\ChannelA,\SessionTypeT_{k})
    \parop
    \cprocess(\ChannelB,\co{\SessionTypeS_{k'}}) 
    )
]
\end{array}
$$
where
$$
\ContextC  =
\new \ChannelC \ChannelD
(  
    \cprocess(\ChannelC, \SessionTypeS'_{k'})
    \parop
    \cprocess(\ChannelD, \co{\SessionTypeT'_{k}})
    )\parop
    [ \; \; ]
$$
By induction,
$$ \new \ChannelA \ChannelB
    (\cprocess(\ChannelA,\SessionTypeT_{k})
    \parop
    \cprocess(\ChannelB,\co{\SessionTypeS_{k'}}) 
    )
  \wreds \error$$
then by rule \rulename{err-context}, we conclude
$$
\ContextC [
\new \ChannelA \ChannelB
    (\cprocess(\ChannelA,\SessionTypeT_{k})
    \parop
    \cprocess(\ChannelB,\co{\SessionTypeS_{k'}}) 
    )
]
\wreds \error
$$
Summing up, we proved that $\SessionType\not\ssubt\SessionTypeS$ implies %
$
  \new\ChannelA\ChannelB(
    \cprocess(\ChannelA,\SessionTypeT)
    \parop
    \cprocess(\ChannelB,\co\SessionTypeS)
  )
  \wreds
  \error
$: %
hence, the subtyping relation $\ssubt$ %
is complete for the synchronous calculus, %
according to Definition~\ref{def:preciseness}.%
\end{Proof}
%
%
  Corollary~\ref{lem:ssubt-largest} shows that %
  $\ssubt$ is the unique precise subtyping for the synchronous calculus. %
  Notably, this result %
  is based on the definitions of %
  typing system and preciseness %
  we adopted for our treatment; %
  a similar result could also be shown for the Gay-Hole
  type system and subtyping~\cite{GH05}, as discussed in~\S\ref{sec:related}
  (paragraph ``Choices of typing system and subtyping''). %
\begin{cor}
  \label{lem:ssubt-largest}%
  $\ssubt$ is the unique precise subtyping for the
  synchronous calculus.
\end{cor}
\begin{Proof}
  \def\ssubti{\annotaterel{\sqsubseteq}{s}}%
  Take a reflexive and transitive relation %
  $\mathord{\ssubti} \not\subseteq \mathord{\ssubt}$ %
  --- i.e., %
  $\exists \SessionTypeS, \SessionTypeT$ %
  such that $\SessionTypeS \ssubti \SessionTypeT$ %
  but $\SessionTypeS \not\ssubt \SessionTypeT$. %
  We prove that $\ssubti$ is an unsound subtyping. %
  By the proof of Theorem~\ref{thm:preciseness:sync}, %
  $\SessionTypeT \not\ssubt \SessionTypeS$ implies %
  $\new\ChannelA\ChannelB(
  \cprocess(\ChannelA,\SessionTypeT)
  \parop
  \cprocess(\ChannelB,\co\SessionTypeS)
  ) \wreds \error$. %
  Therefore, %
  by Definition~\ref{def:preciseness}, %
  $\ssubti$ is not a sound subtyping. %
  We conclude %
  that if $\ssubti$ is sound, %
  then $\mathord{\ssubti} \subseteq \mathord{\ssubt}$; %
  hence, $\ssubt$ is the largest sound subtyping %
  for the synchronous calculus, %
  and therefore the unique precise one.
\end{Proof}

\section{Asynchronous Session Calculus}
\label{sec:asynchronous_language}
``Asynchrony'' in communication means that 
message outputs are {\em non-blocking}. %
Given a pair of co-channels $\ChannelA$ and $\ChannelB$,
we model asynchrony %
using \emph{two} FIFO queues: %
one queue delivers messages from $\ChannelA$ to $\ChannelB$, %
and the other from $\ChannelB$ to $\ChannelA$. %
This double-queue model %
\emph{preserves message order}, %
resembling %
communication over a TCP/IP-like network: %
this is the most common formulation for asynchronous sessions, %
and follows recent formalisms in 
literature~\cite{CDY07,
GV10,event,
mostrous_phd,mostrous09sessionbased,MY15}. 

\subsection{Syntax and operational semantics}
\label{subsec:a:syntax_semantics}

Table~\ref{tab:async:syntax} shows the asynchronous session calculus obtained by
extending the synchronous calculus of Table~\ref{tab:sync:syntax}
with 
queues. A queue $\queue \ChannelA \ChannelB \Queue$
is used by channel $\ChannelA$ to enqueue messages in $\Queue$ and  
by channel $\ChannelB$ to dequeue messages from $\Queue$. 
We extend the definition of the set of free channels to queues
by $\fn(\EmptyQueue)=\emptyset$, $\fn(\queue \ChannelA \ChannelB \Queue)=\{ a,b\}\cup \fn(h)$,  
$\fn(\msg\Tag\Value)=\{ a\}$, and
$\fn(\Queue_1 \qconc\Queue_1) = \fn(\Queue_1) \cup \fn(\Queue_2)$.

We use the structural congruence defined by adding
the rules in Table~\ref{tab:async:congruence} to the rules of Table~\ref{tab:sync:congruence}.
Rule \rulename{s-null} represents garbage collection of empty queues.

\begin{table}[b]
$$
\begin{array}[t]{@{}c@{\qquad}c@{}}
\begin{array}[t]{@{}rcl@{\quad}l@{}}
  \Process \hspace{-3mm}& ::= \hspace{-3mm}& & \textbf{Process} \\
  &   & \vdots & \text{from Table~\ref{tab:sync:syntax}} \\
  & | & 
  \queue\ChannelA\ChannelB \Queue & \text{(queue)}
\end{array}
\qquad \qquad 
\begin{array}[t]{@{}rcl@{\quad}l@{}@{}rcl@{\quad}l@{}}
  \Queue & ::= & & \textbf{Queue} \\
  &   & \EmptyQueue & \text{(empty)} &
  & | & \msg\Tag\Value & \text{(message)} \\
  & | & \Queue \qconc \Queue & \text{(composition)}
\end{array}
\end{array}
$$
\caption{\label{tab:async:syntax} Syntax of asynchronous processes.\strut}
\end{table}
\begin{table}
$$
\begin{array}[t]{@{}c@{}}
  \inferrule[\rulename{s-null}]{}{
    \new\ChannelA\ChannelB
    (\queue \ChannelA \ChannelB \EmptyQueue \parop
     \queue \ChannelB \ChannelA \EmptyQueue)
    \equiv
    \idle
   }\qquad
   \inferrule[\rulename{s-queue-equiv}]{ \Queue   \equiv \Queue'}{
   \queue \ChannelA \ChannelB \Queue   \equiv
     \queue \ChannelA \ChannelB \Queue'
   }
  \\\\
  \inferrule[\rulename{s-queue 1}]{}{
    \EmptyQueue \qconc \Queue
    \equiv
    \Queue
  }
  \qquad
  \inferrule[\rulename{s-queue 2}]{}{
    \Queue \qconc \EmptyQueue
    \equiv
    \Queue
  }
  \qquad
  \inferrule[\rulename{s-queue 3}]{}{
    \Queue_1 \qconc (\Queue_2 \qconc \Queue_3)
    \equiv
    (\Queue_1 \qconc \Queue_2) \qconc \Queue_3
  }
\end{array}
$$
  \caption{\label{tab:async:congruence} Structural congruence for
    asynchronous processes.\strut}
\end{table}
\begin{table}
$$
\begin{array}[t]{@{}c@{}}
  \inferrule[\rulename{r-send-async}]{}{
      \queue\ChannelA\ChannelB \Queue
    \parop
    \send\ChannelA\Tag\ChannelC.\Process
        \red
      \queue\ChannelA\ChannelB \Queue \qconc \msg{\Tag}{\ChannelC}
    \parop
    \Process
      }
  \qquad
  \inferrule[\rulename{r-receive-async}]{
    k\in I
  }{
    \textstyle
    \queue\ChannelA\ChannelB \msg{\Tag_k}{\ChannelC} \qconc \Queue
    \parop
    \sum_{i\in I} \receive\ChannelB{\Tag_i}{\var_i}.\Process_i
    \red
    \queue\ChannelA\ChannelB \Queue
    \parop
    \Process_k\subst\ChannelC{\var_k}
    }
\end{array}
$$
\caption{\label{tab:async:red} Reduction of asynchronous processes.\strut}
\end{table}

The reduction rules for asynchronous processes 
are obtained from the reduction rules of synchronous processes given in Table
\ref{tab:sync:red} by 
replacing rule \rulename{r-com-sync} with 
\rulename{r-send-async} and \rulename{r-receive-async}, shown in Table~\ref{tab:async:red}.
Rule \rulename{r-send-async} enqueues messages and rule
\rulename{r-receive-async} dequeues messages. 

We write $P\reda Q$ if $P \red Q$ is derived
by the rules of Table~\ref{tab:sync:red} but rule 
\rulename{r-com-sync} and by the rules of Table~\ref{tab:async:red}. 

\subsection{Errors in asynchronous processes}
\label{subsec:a:errors}
Like in the synchronous case, 
also in the asynchronous case the errors arise when 
the duality of the communications offered by co-channels
is lost. The presence 
of queues makes the
definition of the error reductions 
for asynchronous communications not trivial. In addition to the other errors, we need to identify 
the following classical error situations 
(the terminology is adopted from 
works %
on Communicating Finite State Machines~\cite{CeceF05,
GMY84,Villard11}):

\textbf{(1) deadlocks:}
there are inputs waiting to dequeue messages from queues which will be forever empty.  

\textbf{(2) orphan message errors:}
there are messages in queues which  
will never be received by corresponding inputs, i.e. orphan messages will remain forever in queues. 

\medskip

Both errors are important, 
since we want to ensure every input can receive a message and every message in a queue can be read.
These errors correspond to the following processes:

\textbf{(1) deadlocks:} 
a pair of co-channels are waiting for inputs and their queues are both
empty, or one channel is waiting for an input over an empty queue and
the co-channel only occurs as a queue name. In the first case the process has the shape$$\new \ChannelA \ChannelB 
    (\sum_{i\in I} \receive\ChannelA{\Tag_i}{\var_i}.\ProcessP_i \parop
    \sum_{j\in J} \receive\ChannelB{\Tag'_j}{\var'_j}.\ProcessQ_j\parop\queue\ChannelB\ChannelA\EmptyQueue\parop\queue\ChannelA\ChannelB\EmptyQueue)$$
and in the second case the process 
 has the shape
$$\new \ChannelA \ChannelB 
    (\sum_{i\in I} \receive\ChannelA{\Tag_i}{\var_i}.\ProcessP_i \parop
 \queue\ChannelB\ChannelA\EmptyQueue\parop\queue\ChannelA\ChannelB\Queue)$$

\textbf{(2) orphan message errors:}
a queue is not empty, but the corresponding channel will never appear as subject of an input.
 I.e.~the process has the shape 
${\new\ChannelA\ChannelB(\ProcessP\parop\queue\ChannelB\ChannelA\Queue)}$
where $\Queue$ is not empty, 
but $\ProcessP$ will neither reduce to a process which performs an input on channel $\ChannelA$, nor pass the channel $\ChannelA$ to an outer process. 

\bigskip

To define statically the second error situation, we 
need to compute an over approximation (denoted by $\sbnF(\ProcessP)$) of the set of free channels 
which {\em might} eventually appear as subjects of \emph{inputs} by {\em reducing} a process $\ProcessP$. %
The definition of $\sbnF(\ProcessP)$ requires some care. %
We cannot simply modify the mapping $\sn(\ProcessP)$ 
(page \pageref{sn}), %
since it only takes into account the subjects which occur free in $\ProcessP$, %
without considering its unfoldings in case of recursion. %
Moreover, we cannot take the whole set of channels occurring free in $\ProcessP$, because it is too large: for example,
a recursive process which only sends messages will always contain both the channels of the subjects and of the objects of the outputs,
but it will never read a message on its queue. 
We also need to carefully consider the reductions of $\ProcessP$, %
as shown for example by the process:
\begin{equation}
\label{eq:sbnF}
\ProcessP =\ChannelB \In{\Tag_0}{\var} . \var \In{\Tag_1}{\varY} 
\parop 
\ChannelC \Out{\Tag_0}{\ChannelA}
\parop
\queue{\ChannelC}{\ChannelB} \EmptyQueue
\end{equation}
Notice that $\ProcessP$ does not contain inputs with subject $\ChannelA$, but 
$$
\ProcessP   \reda  
\ChannelB \In{\Tag_0}{\var} . \var \In{\Tag_1}{\varY}
\parop
\queue{\ChannelC}{\ChannelB} \msg{\Tag_0}{\ChannelA}
 \reda 
\ChannelA \In{\Tag_1}{\varY} 
\parop
\queue{\ChannelC}{\ChannelB} \EmptyQueue
$$
and this last process has an input with subject $\ChannelA$. 
Hence to define $\sbnF(\ProcessP)$, 
we need to take care of channels carried by outputs as well as channels occurring in messages inside queues. 
Another delicate case in the definition of $\sbnF(\ProcessP)$ comes from recursive definitions. For example the process $$\ProcessQ\;\;=\;\;\Def{\pvar}{\var} 
{\send{\ChannelA}{\Tag}{\var}} 
{\invoke{\pvar}{\ChannelC}\parop 
\receive{\ChannelB}{\Tag}{\varY} . \receive{\varY}{\Tag}{\varZ}
\parop \queue \ChannelA \ChannelB \EmptyQueue 
\parop \queue \ChannelB \ChannelA \EmptyQueue} $$ does not contain inputs with subject $\ChannelC$, but
$$
\begin{array}{lll}
\ProcessQ& 
\reda&
\Def{\pvar}{\var} 
{\send{\ChannelA}{\Tag}{\var}} 
{\send{\ChannelA}{\Tag}{\ChannelC} \parop 
\receive{\ChannelB}{\Tag}{\varY} . \receive{\varY}{\Tag}{\varZ}
\parop \queue \ChannelA \ChannelB \EmptyQueue 
\parop \queue \ChannelB \ChannelA \EmptyQueue}\\ 
& 
\reda&
\Def{\pvar}{\var} 
{\send{\ChannelA}{\Tag}{\var}} 
{
\receive{\ChannelB}{\Tag}{\varY} . \receive{\varY}{\Tag}{\varZ}
\parop \queue \ChannelA \ChannelB \msg{\Tag}{\ChannelC} 
\parop \queue \ChannelB \ChannelA \EmptyQueue}\\ 
& 
\reda&
\Def{\pvar}{\var} 
{\send{\ChannelA}{\Tag}{\var} } 
{\receive{\ChannelC}{\Tag}{\varZ}
\parop \queue \ChannelA \ChannelB \EmptyQueue 
\parop \queue \ChannelB \ChannelA \EmptyQueue}
\qquad\text{(note the input with subject $\ChannelC$)}
\end{array}
$$

In~Definition~\ref{def:sbn} below, %
  we introduce $\sbnF{(\ProcessP)}$ using %
two auxiliary mappings: $\sbn$ and $\sbnD$.
Let $\tilde D$ range over sets of declarations and $D\cdot\tilde D$ denote the addition of $D$ to the set $\tilde D$. Let $\chi$ range over sets of process variables and $\invoke\pvar{\tilde\Expression}\cdot\chi$ denote the addition of $\invoke\pvar{\tilde\Expression}$ to the set $\chi$.

\begin{table}
$$
\begin{array}{rclrcl}
\sbnD(\invoke\pvar{\tilde\Expression},\tilde D,\chi)      & = &      \begin{cases}
\sbnD(\ProcessP\subst{\tilde\Expression}{\tilde\var},\tilde D,\invoke\pvar{\tilde\Expression}\cdot\chi)       &      \text{if } \invoke\pvar{\tilde\Expression}\not\in\chi \text{ and }\\ &\pvar(\tilde\var)=\ProcessP\in\tilde D\\
    \emptyset      &        \text{otherwise}
\end{cases}&\qquad
\sbnD(\idle,\tilde D,\chi)       &=&      \emptyset
\end{array}$$
{\small $$\begin{array}{r@{\hskip 1.5mm}c@{\hskip 1.5mm}lr@{\hskip 1.5mm}c@{\hskip 1.5mm}l}
\sbnD(\sum_{i\in I} \receive\Expression{\Tag_i}{\var_i}.\Process_i,\tilde D,\chi)      & = &      
\ASET{\Expression} \cup \bigcup_{i\in I}\sbnD( \Process_i,\tilde D,\chi)\setminus\ASET{\var_i} &
\sbnD(\send\Expression\Tag{\Expression'}.\ProcessP,\tilde D)& = &\ASET{\Expression'} \cup 
\sbnD(\ProcessP,\tilde D,\chi) \\
\sbnD(\ProcessP_1 \parop\ProcessP_2,\tilde D,\chi)     & = &     
\sbnD(\ProcessP_1,\tilde D,\chi) \cup \sbnD(\ProcessP_2,\tilde D,\chi) 
&  \sbnD(\queue \ChannelA \ChannelB \Queue,\tilde D,\chi)      &=&      \sbn(\Queue)
\\
\sbnD(\ProcessP_1  \choice \ProcessP_2,\tilde D,\chi)     & = &      
\sbnD(\ProcessP_1,\tilde D,\chi) \cup \sbnD(\ProcessP_2,\tilde D,\chi) 
&
\sbnD((\nu \ChannelA \ChannelB) \ProcessP,\tilde D,\chi)       & = &      \sbnD(\ProcessP,\tilde D,\chi)\setminus\ASET{\ChannelA, \ChannelB}
\\
\sbnD(\DefD{D}{\ProcessP},\tilde D,\chi)      & = &      \sbnD(\ProcessP, D\cdot \tilde D,\chi)
&
\end{array}
$$}
\caption{The mapping $\sbnD$. 
} \label{sbnD}
\end{table}

\begin{defi} \label{def:sbn}\rm 
(Mappings $\sbn,\sbnD,$ and $\sbnF$)
The mapping $\sbn$ is defined by induction on queues:
$$\begin{array}{c}
\sbn(\EmptyQueue)=  \emptyset \qquad
\sbn(\msg\Tag\ChannelA)=  \ASET{\ChannelA} \qquad
 \sbn(\Queue_1 \qconc \Queue_2 )=  \sbn(\Queue_1) \cup\sbn(\Queue_2 )
 \end{array}$$
The mapping $\sbnD$ is defined (by induction on processes) in Table~\ref{sbnD}, using $\sbn$. 
The mapping $\sbnF$ is defined as $\sbnF(\ProcessP)=\sbnD(\ProcessP,\emptyset,\emptyset)$. 
\end{defi}
\noindent%
The mapping $\sbn$ applied to a message queue $\Queue$ gives the set
of channels which occur in $\Queue$. %
The mapping $\sbnD$ has three arguments: a process $\ProcessP$, a set of
declarations and a set of process invocations. %
The result is the set of free channels which \emph{might} occur as
input subjects %
along the reductions of $\ProcessP$: %
the declarations can be used in these
reductions, and the set of process invocations memorises
the process calls already considered. %
The mapping $\sbnD$ uses the mapping $\sbn$ for dealing with
queues. The mapping $\sbnF$ is the mapping $\sbnD$ applied to a
process, an empty set of declarations, and an empty set of process
invocations. %
Notably, if $\ProcessP$ only contains recursive process
definitions $\pvar{(\varX)} = \ProcessP'$ %
(i.e., with one parameter $\varX$), %
and all recursive calls to $\pvar$ in $\ProcessP'$ %
have the form $\invoke{\pvar}{\varX}$ %
(i.e., each call to $\pvar$ reuses $\varX$), %
then $\sbnF(\ProcessP) = \sbnD(\ProcessP, \emptyset, \emptyset)$ %
is well founded %
and each recursive call is unfolded only once %
(see Example~\ref{ex:sbnF:recursion}). %
We will see that these properties are %
enjoyed by characteristic \emph{asynchronous} processes, %
that will be introduced 
in Definition~\ref{async:charprocess} %
for studying the preciseness of subtyping in the asynchronous calculus.

For example, for the process $\ProcessP$ shown in 
\eqref{eq:sbnF}, we obtain $\sbnF(\ProcessP)=\{a,b\}$. 
\begin{example} 
\label{ex:sbnF:recursion} 
{\em Assume
$$
\Process \quad= \quad
\DefD{\pvar(\var) =\send{\ChannelA}{\Tag}{\var}.\invoke{\pvar}\var} 
{\invoke{\pvar}{\ChannelC}\parop 
\receive{\ChannelB}{\Tag}{\varY} . \receive{\varY}{\Tag}{\varZ}
\parop \queue \ChannelA \ChannelB \EmptyQueue 
\parop \queue \ChannelB \ChannelA \EmptyQueue}
$$
Then we get:
$$
\begin{array}{rcl}
\sbnF(\Process)
&=& 
\sbnD(\Process, \emptyset, \emptyset) \\
&=& 
\sbnD(
\invoke{\pvar}{\ChannelC}\parop 
\receive{\ChannelB}{\Tag}{\varY} . \receive{\varY}{\Tag}{\varZ}
\parop \queue \ChannelA \ChannelB \EmptyQueue 
\parop \queue \ChannelB \ChannelA \EmptyQueue, 
D, \emptyset)
\\
& = &
\sbnD(\invoke{\pvar}{\ChannelC}, D, \emptyset) \cup
\sbnD(\receive{\ChannelB}{\Tag}{\varY}.\receive{\varY}{\Tag}{\varZ}, D, \emptyset) \cup
 \sbnD(\queue \ChannelA \ChannelB \EmptyQueue, D, \emptyset) \cup 
\sbnD(\queue \ChannelB \ChannelA \EmptyQueue, D, \emptyset)
\\
& = &
\sbnD(\send{\ChannelA}{\Tag}{\ChannelC}.\invoke{\pvar}{\ChannelC}, D,\invoke{\pvar}{\ChannelC})\cup 
\ASET{\ChannelB} \cup (\sbnD(\receive{\varY}{\Tag}{\varZ}, D, \emptyset) \setminus \ASET{\varY})
\\
& = & 
\ASET{\ChannelC}\cup 
\ASET{\ChannelB} 
=\ASET{\ChannelC,\ChannelB}
\end{array}
$$
where $D$ is $\pvar(\var) =\send{\ChannelA}{\Tag}{\var}.\invoke{\pvar}\var$. 
The evaluation of  $\sbnF(\Process)$ tells that
$\ChannelC$ can become the subject of an input. 
We illustrate this fact also by the following reduction:
$$
\begin{array}{lll}
\Process& 
\reda&
\Def{\pvar}{\var} 
{\send{\ChannelA}{\Tag}{\var}.\invoke{\pvar}\var} 
{\send{\ChannelA}{\Tag}{\ChannelC}.\invoke{\pvar}{\ChannelC} \parop 
\receive{\ChannelB}{\Tag}{\varY} . \receive{\varY}{\Tag}{\varZ}
\parop \queue \ChannelA \ChannelB \EmptyQueue 
\parop \queue \ChannelB \ChannelA \EmptyQueue} 
\\
& 
\reda&
\Def{\pvar}{\var} 
{\send{\ChannelA}{\Tag}{\var}.\invoke{\pvar}\var} \invoke{\pvar}{\ChannelC} \parop
{
\receive{\ChannelB}{\Tag}{\varY} . \receive{\varY}{\Tag}{\varZ}
\parop \queue \ChannelA \ChannelB \msg{\Tag}{\ChannelC} 
\parop \queue \ChannelB \ChannelA \EmptyQueue} 
\\
& 
\reda&
\Def{\pvar}{\var} 
{\send{\ChannelA}{\Tag}{\var}.\invoke{\pvar}\var } \invoke{\pvar}{\ChannelC} \parop
{\receive{\ChannelC}{\Tag}{\varZ}
\parop \queue \ChannelA \ChannelB \EmptyQueue 
\parop \queue \ChannelB \ChannelA \EmptyQueue} 
\end{array}
$$}
\end{example}

\begin{table}
$$
\begin{array}[t]{@{}c@{}}
  \inferrule[\rulename{err-mism-async}]
  { \forall i \in I : \Tag \not = \Tag_i
  }
  { 
        \queue\ChannelA\ChannelB \msg{\Tag}{\ChannelC} \qconc \Queue
        \parop
        \sum_{i\in I} \receive\ChannelB{\Tag_i}{\var_i}.\Process_i
        \red \error
  }
  \\\\
\inferrule[\rulename{err-in-in-async}]{}
 {\new \ChannelA \ChannelB
    (\sum_{i\in I} \receive\ChannelA{\Tag_i}{\var_i}.\ProcessP_i \parop
    \sum_{j\in J} \receive\ChannelB{\Tag'_j}{\var'_j}.\ProcessQ_j\parop\queue\ChannelB\ChannelA\EmptyQueue\parop\queue\ChannelA\ChannelB\EmptyQueue)
    \red \error  }
\\\\
\inferrule[\rulename{err-in-async}]{
}
{
    {\new \ChannelA \ChannelB 
    (\sum_{i\in I} \receive\ChannelA{\Tag_i}{\var_i}.\ProcessP_i \parop
 \queue\ChannelB\ChannelA\EmptyQueue\parop\queue\ChannelA\ChannelB\Queue)}
    \red \error
  }
 \qquad
\inferrule[\rulename{err-orph-mess-async}]{
    \ChannelA\not\in\sbnF(\ProcessP) \quad\fpv(\ProcessP)=\emptyset\quad \Queue\not=\EmptyQueue
  }{
    {\new\ChannelA\ChannelB(\ProcessP\parop\queue\ChannelB\ChannelA\Queue)}
    \red \error
  }
\end{array}
$$
\caption{Error reduction for asynchronous processes.%
}\label{tab:async:red:err} 
\end{table}

The error reduction rules for asynchronous processes are \rulename{err-context} of Table~\ref{tab:sync:red:err} and the rules of Table~\ref{tab:async:red:err}. 
Rule \rulename{err-mism-async} 
deals with a label mismatch between a message on the top of the queue and an input. 
Rule \rulename{err-in-in-async} gives an error when two processes with restricted channels are in deadlock waiting to read from empty queues. 
Rule \rulename{err-in-async} deals with the case of one process waiting to read from an empty queue which will never contain a message, since there are no occurrences of the unique channel that can enqueue messages. 
Rule \rulename{err-orph-mess-async}
corresponds to the orphan message error.  Clearly $\new\ChannelA\ChannelB(\ProcessP\parop\queue\ChannelB\ChannelA\Queue)$ could reduce in other ways, take for example $\ProcessP=\send{\ChannelB}{\Tag}{\ChannelC}$.
In this rule the condition $\fpv(\ProcessP)=\emptyset$ assures that we consider all needed declarations in computing $\sbnF(\ProcessP)$.

\begin{defi}\label{def:async-errors}
  We write $\ProcessP \reda \error$ if  $\ProcessP \red \error$ can be derived using the rules of Table~\ref{tab:async:red:err}
  plus rule \rulename{err-context} of Table~\ref{tab:sync:red:err}. The notation $\wreda$ is used with the expected meaning.
\end{defi}

Let $\cprocess(\ChannelA,\SessionTypeT)$ range over characteristic asynchronous processes as in Definition~\ref{async:charprocess}.
We will use the definitions above to show that 
a typeable process 
$\cprocess(\ChannelA,\SessionTypeT) \parop
\cprocess(\ChannelB,\SessionTypeS)
\parop \queue \ChannelB \ChannelA\EmptyQueue
\parop \queue \ChannelA \ChannelB\EmptyQueue$ does not reduce to
$\error$. %
\emph{Vice versa}, %
we will also show that 
if the process $\cprocess(\ChannelA,\SessionTypeT) \parop
\cprocess(\ChannelB,\SessionTypeS)
\parop \queue \ChannelB \ChannelA\EmptyQueue
\parop \queue \ChannelA \ChannelB\EmptyQueue$ cannot be typed, then there are $\SessionTypeT'\ssubt\SessionTypeT$ and $\SessionTypeS'\ssubt\SessionTypeS$ such that 
$\cprocess(\ChannelA,\SessionTypeT') \parop
\cprocess(\ChannelB,\SessionTypeS')
\parop \queue \ChannelB \ChannelA\EmptyQueue
\parop \queue \ChannelA \ChannelB\EmptyQueue$ reduces to $\error$.
%
As in the synchronous case, arbitrary processes can be stuck without reducing to $\error$: this happens, for instance, %
  if a process inputs/outputs on a channel without a corresponding queue. %
Note that the type system we will introduce in \S~\ref{tap} is an
adaptation of those by Mostrous
\etal~\cite{mostrous09sessionbased,mostrous_yoshida_honda_esop09} to
our calculus, and they do not aim at avoiding deadlock or orphan
message errors.

Similarly to Proposition~\ref{lem:sync-errors-persistent}, %
if an asynchronous process can reduce to $\error$ in one step, %
then a different reduction produces a process with the same property.
\begin{proposition}
  \label{lem:async-errors-persistent}%
  If %
  $\Process \reda \error$ and $\Process \reda \Process' \neq \error$, then
  $\Process' \reda \error$.%
\end{proposition}
\begin{Proof}
  By cases on the rule giving $\Process \reda \error$ %
  (Definition~\ref{def:async-errors}). %
  The statement holds vacuously for %
  \rulename{err-mism-async}, \rulename{err-in-in-async} and 
  \rulename{err-in-async} %
  (Table~\ref{tab:async:red:err}): %
  in such cases, $\Process \reda \Process'$ implies $\Process' = \error$.

  In the case \rulename{err-orph-mess-async}, %
  we have %
  $\Process = %
  {\new\ChannelA\ChannelB(\ProcessQ\parop\queue\ChannelB\ChannelA\Queue)}
  \reda \error$; %
  moreover, %
  $\ChannelA\not\in\sbnF(\ProcessQ)$, %
  $\fpv(\ProcessQ)=\emptyset$ and $\Queue\not=\EmptyQueue$. %
  If $\Process \reda \Process' \neq \error$, %
  then (by \rulename{r-context}) it must be %
  $\Process' = %
  {\new\ChannelA\ChannelB(\ProcessQ'\parop\queue\ChannelB\ChannelA\Queue')}$ %
  with %
  $\ProcessQ\parop\queue\ChannelB\ChannelA\Queue \reda %
  \ProcessQ'\parop\queue\ChannelB\ChannelA\Queue'$: %
  by induction on the derivation of this transition, %
  we can verify that %
  $\ChannelA\not\in\sbnF(\ProcessQ')$, %
  $\fpv(\ProcessQ')=\emptyset$ and $\Queue'\not=\EmptyQueue$. %
  Hence, again by rule \rulename{err-orph-mess-async}, %
  we conclude %
  $\Process' \reda \error$.
  
  The case \rulename{err-context} (Table~\ref{tab:sync:red:err}) %
  is proved similarly to Proposition~\ref{lem:sync-errors-persistent}.%
\end{Proof}

\subsection{Asynchronous subtyping}
\label{subsec:async}
The asynchronous subtyping is not only essential for 
the completeness result, 
but it is also important in practice. 
As observed by Yoshida, Honda \etal~\cite{NYH12,
YVPH08},
implementing this subtyping is a key tool 
for maximising message-overlapping 
in the high-performance computing environments. 
To explain the
usefulness of the asynchronous subtyping, consider: 
$$\begin{array}{c}
P_1 = a?\Tag(y_1).a\Out{\Tag}{5}.Q_1
\quad 
P_2 = b\Out{\Tag}{Large\_datum}.b?\Tag(y_2).Q_2
\end{array}
$$
First $P_2$ sends a large datum on channel $b$; then after receiving it, 
$P_1$ sends $5$ to $P_2$.  
We note that $P_1$'s output is blocked until this large datum is
received. 
Since the value replacing $y_1$ does not influence the subsequent output 
at $a$, process $P_1$ can be optimised by sending the small datum ``5'' first, so
that once the large datum is put in the queue at $ba$, process $P_2$ can immediately
receive the small datum.  Thus a better version of    $P_1$ is $P'_1$ defined by:
$$P_1' = a\Out{\Tag}{5}.a?\Tag(y_1).Q_1
$$
Asynchronous 
subtyping specifies
safe permutations of actions, 
by which we can refine a local protocol 
to maximise asynchrony  
without violating session safety. 

To define asynchronous subtyping, %
the notion of asynchronous context of types is handy~\cite{mostrous_phd,MY15}. An asynchronous context is a sequence of branchings containing holes that we index in order to distinguish them. %

\begin{defi}[Asynchronous context] \label{def:async:context}\rm
$\;$
$$\AContext \;\;:: =\;\;
[ \; ]^{n} \;\parop\;
\Branch_{i\in I} \In{\Tag_i}{\Type_i}.\AContext_i
$$
We write $\AContext[\;]^{n\in N}$ to denote a context 
with  holes  indexed by elements of $N$
and $\AContext[\SessionTypeT_n]^{n\in N}$ to denote the same context when the hole 
$[ \; ]^n$ has been filled with $\SessionTypeT_n$.
\end{defi}

We naturally extend the definition of type tree %
(page~\pageref{def:type-tree}) %
to contexts, letting $\tree{[\;]^n} = [\;]^n$. %
For an example, see Figure~\ref{fig:context-tree-example}. %
Note that each continuation path %
of a context tree %
is \emph{finite}, %
only connects branching nodes, %
and ends on a hole.

\begin{figure}
  \[
  \xymatrix@R-1pc{%
    &&\&\ar@{-}[dll]_{\Tag_0^{\carriedT}}\ar@{-}[ddl]_{\Tag_0^{\contT}}\ar@{-}[ddr]^{\Tag_3^{\carriedT}}\ar@{-}[drr]^{\Tag_3^{\contT}}&&&\\
      \tree{\SessionTypeS_0}&&&&[\;]^3\\
    &\&\ar@{-}[dl]_{\Tag_1^{\carriedT}}\ar@{-}[dd]_{\Tag_1^{\contT}}\ar@{-}[ddr]^{\Tag_2^{\carriedT}}\ar@{-}[drr]^{\Tag_2^{\contT}}&&\tree{\SessionTypeS_3}\\
     \tree{\SessionTypeS_1}&&&[\;]^2&\\
    &[\;]^1&\tree{\SessionTypeS_2}%
  }\]
  \caption{The tree of\; %
    $%
    \AContext \,=\, %
    \In{\Tag_0}{\SessionTypeS_0} . %
    \left(%
      \In{\Tag_1}{\SessionTypeS_1} . [\;]^1%
      \,\&\,
      \In{\Tag_2}{\SessionTypeS_2} . [\;]^2%
    \right)%
    \;\&\;%
    \In{\Tag_3}{\SessionTypeS_3} . [\;]^3%
    $.%
  }%
  \label{fig:context-tree-example}%
\end{figure}

\begin{example}
\label{ex:tr-out}
{\em 
Let $N = \ASET{1,2}$ and
$$\hspace{-7pt}
\begin{array}{l}
\SessionType_1   =  
\Out{\mathsf m}{\SessionTypeS_{m}} . \SessionTypeT_{m}  \oplus 
\Out{\mathsf p}{\SessionTypeS_{p}} . \SessionTypeT_{p},
\qquad
\SessionType_2   =  
\Out{\mathsf m}{\SessionTypeS'_{m}} . \SessionTypeT'_{m}  \oplus 
\Out{\mathsf p}{\SessionTypeS'_{p}} . \SessionTypeT'_{p}  \oplus 
\Out{\mathsf q}{\SessionTypeS_{q}}. \SessionTypeT_{q}.
\end{array}
$$
Assume
$\AContext = 
\In{\mathsf r}{\SessionTypeS_{r}} . [ \; ]^{1} ~\&~
\In{\mathsf s}{\SessionTypeS_{s}} . [ \; ]^{2},
$
then
$$\begin{array}{c}
\AContext[\SessionType_1]^1[\SessionType_2]^2 =
\In{\mathsf r}{\SessionTypeS_{r}} . (
\Out{\mathsf m}{\SessionTypeS_{m}} . \SessionTypeT_{m}\oplus
\Out{\mathsf p}{\SessionTypeS_{p}} . \SessionTypeT_{p})
~\&~
\In{\mathsf s}{\SessionTypeS_{s}}.(
\Out{\mathsf m}{\SessionTypeS'_{m}} . \SessionTypeT'_{m}\oplus
\Out{\mathsf p}{\SessionTypeS'_{p}} . \SessionTypeT'_{p} \oplus
\Out{\mathsf q}{\SessionTypeS_{q}}. \SessionTypeT_{q}).
\end{array}
$$}
\end{example}



\label{def:branch-in-t}%
To define asynchronous subtyping %
we also need to introduce the predicate %
$\mathord{\&}\in \SessionTypeT$, %
which holds if %
each continuation path 
of $\tree{\SessionTypeT}$ %
contains at least one branching. %
The regularity
of the trees corresponding to session types assures that the branchings occur at finite levels.
More precisely $\& \in \SessionTypeT$ holds if we can derive it from the following axioms and inductive rules:%
\[
\inferrule[]{}{\& \in \Branch_{i\in I} \In{\Tag_i}{\Type_i}.\SessionTypeT_i}%
\qquad%
\inferrule[]{%
  \forall i \in I\quad\& \in \SessionTypeT_i%
}{%
  \& \in\Select_{i\in I} \Out{\Tag_i}{\Type_i}.\SessionTypeT_i%
}%
\qquad%
\inferrule[]{%
  \& \in \SessionType 
}{%
  \& \in \trec\tvar.\SessionType}%
\]
  where we consider also open session types; %
  note that such a predicate holds up-to unfolding, %
  \ie %
  $%
  \& \in \trec\tvar.\SessionType \text{ iff }%
  \& \in \SessionType \subst{\trec\tvar.\SessionType}\tvar%
  $ %
  (Proposition~\ref{lem:branch-in-t:unfolding} in Appendix~\ref{app:async_lang}). %
We write $\& \in \AContext$ if $\AContext$ is a branching, i.e. it is not a single hole.

\begin{example}
  \label{ex:branch-in-t}%
  {\em We have: %
  \[
  \& \in%
  \Out{\Tag}{\SessionTypeS}.\In{\Tag'}{\SessionTypeS'}.\End%
  \quad%
  \& \in%
  \trec\tvar . \Out{\Tag}{\SessionTypeS}.\In{\Tag'}{\SessionTypeS'}.\tvar%
  \quad%
  \neg\left(\& \in%
  \trec\tvar . \left(%
    \Out{\Tag}{\SessionTypeS}.\In{\Tag'}{\SessionTypeS'}.\tvar%
    \oplus%
    \Out{\Tag''}{\SessionTypeS''}.\tvar%
  \right)\right)%
  \]
  In the first and second case, the tree of each type %
  has just one continuation path, with a selection followed by a branching, %
  and we have corresponding (finite) derivations under the rules above. %
  In the third case, %
  the recursion guarded by the $\Tag''$-labelled output %
  generates an \emph{infinite} continuation path without branchings: %
  this yields an infinite (and thus, invalid) derivation
  under the inductive rules for 
  $\& \in \SessionTypeT$; %
  since no finite derivations
  exist, the predicate does not hold.}
\end{example}

We can now define the asynchronous subtyping relation
$T \asubt S$, which holds if $T \subt S$ is derived by the rule:
$$\cinferrule[\rulename{sub-perm-async}]{\forall i\in I ~\forall n\in N: \quad
\TypeU_i^{n}\subt\TypeU_i\quad
\SessionTypeT_i\subt\AContext[\SessionTypeT_i^{n}]^{n \in N}\quad
\& \in \AContext
\quad \&\in\SessionTypeT_i}{\;\\
  \textstyle
   \Select_{i\in I} \Out{\Tag_i}{\Type_i}.\SessionTypeT_{i}
  \subt
  \AContext[\Select_{i\in I \cup J_n} \Out{\Tag_i}{\Type_i^{n}}.\SessionTypeT_i^{n}
  ]^{n\in N}}
$$
together with the rules in 
Table~\ref{tab:sync:ssubt}.
%

Rule \rulename{sub-perm-async} allows 
the asynchronous safe permutation explained above. It postpones a selection after an unbounded but finite number of branchings, and the selections inside these branchings can be bigger according to rule \rulename{sub-sel} of Table~\ref{tab:sync:ssubt}. The conditions $\& \in \AContext$
and $ \&\in\SessionTypeT_i$ for all $i \in I$ are crucial for the
soundness of this rule. Without these conditions we get the subtyping
of Mostrous PhD thesis~\cite{mostrous_phd,MY15} %
(for further discussion on this topic, see \S~\ref{sec:related}, paragraph ``Subtyping of Mostrous PhD thesis'').

Reflexivity of  $\asubt$ is immediate, while the
proof of transitivity requires some ingenuity, see Appendix~\ref{app:async_lang}.
\begin{thm} \label{asubtrans}
The relation $\asubt$ is transitive.
\end{thm}

\begin{figure}\label{as}
{\small\centerline{$
\begin{array}{c}
\cinferrule{\SessionTypeS^r_m\subt \SessionTypeS_m \quad \SessionTypeS^s_m \subt \SessionTypeS_m \quad \SessionTypeS^r_p\subt \SessionTypeS_p \quad \SessionTypeS^s_p\subt\SessionTypeS_p\quad T_m\subt \In{\mathsf r}{\SessionTypeS_{r}} . \SessionTypeT_r ~\&~ \In{\mathsf s}{\SessionTypeS_{s}} . \SessionTypeT_s\quad
T_p\subt \In{\mathsf r}{\SessionTypeS_{r}} . \SessionTypeT'_r ~\&~ \In{\mathsf s}{\SessionTypeS_{s}} . \SessionTypeT'_s}
{
\Out{\mathsf m}{\SessionTypeS_{m}} . T_m \oplus \Out{\mathsf p}{\SessionTypeS_{p}} . T_p \,\subt\,
\In{\mathsf r}{\SessionTypeS_{r}} .(\Out{\mathsf m}{\SessionTypeS_{m}^r} . \SessionTypeT_r \oplus \Out{\mathsf p}{\SessionTypeS_{p}^r} . \SessionTypeT'_r \oplus \Out{\mathsf q}{\SessionTypeS_{q}}. \SessionTypeT_{q})  ~\&~ \In{\mathsf s}{\SessionTypeS_{s}} .(\Out{\mathsf m}{\SessionTypeS_{m}^s} . \SessionTypeT_s \oplus \Out{\mathsf p}{\SessionTypeS_{p}^s} . \SessionTypeT'_s) 
}  
\end{array}
$}}
\caption{Application of \rulename{sub-perm-async}, %
  where \mbox{$T_m=\In{\mathsf r}{\SessionTypeS_{r}} . \SessionTypeT_r ~\&~ \In{\mathsf s}{\SessionTypeS_{s}} . \SessionTypeT_s~\&~ \In{\mathsf u}{\SessionTypeS_{u}} . \SessionTypeT_u$} 
and $T_p=\In{\mathsf r}{\SessionTypeS'_{r}} . \SessionTypeT'_r ~\&~ \In{\mathsf s}{\SessionTypeS_{s}} . \SessionTypeT'_s$ and we assume $S'_r\subt S_r$.%
}\label{sa}
\end{figure}

\begin{example}\rm (Asynchronous subtyping){\em 
\begin{enumerate}\item  We show
$T_1\asubt S_1$, where $T_1=\mu \tvar.\Out{\Tag}{T'}.\In{\Tag'}{S'}.\tvar$ and $S_1=\mu \tvar.\In{\Tag'}{S'}.\Out{\Tag}{T'}.\tvar$.
If we assume $T_1\subt S_1$, we obtain 
\begin{center}
$\Out{\Tag}{T'}.\In{\Tag'}{S'}.T_1 \subt \In{\Tag'}{S'}.\Out{\Tag}{T'}.S_1$ 
\end{center}
by rule \rulename{sub-perm-async}, which is $T_1\subt S_1$ by folding. In this way we coinductively proved $T_1\asubt S_1$.
\item  We show $T_2\asubt S_2$, where  $T_2=\Out{\Tag}{T'}.T_1$ and $S_2=\In{\Tag'}{S'}.S_1$ and $T_1,S_1$ are as in previous example.
We assume $T_2\subt S_2$. We get\\[4pt]
\centerline{\hspace{-30pt}$\begin{array}{llll}
\Out{\Tag}{T'}.\Out{\Tag}{T'}.\In{\Tag'}{S'}.T_1&\subt&\In{\Tag'}{S'}.  \Out{\Tag}{T'}.\Out{\Tag}{T'}.T_1
& \text{by rule \rulename{sub-perm-async}}\\[4pt]
&\subt&\In{\Tag'}{S'}.  \Out{\Tag}{T'}.\In{\Tag'}{S'}.S_1
& \text{by the assumption $T_2\subt S_2$}\\[4pt]
&\subt&\In{\Tag'}{S'}.  \In{\Tag'}{S'}.\Out{\Tag}{T'}.S_1
& \text{by rule \rulename{sub-perm-async}}
\end{array}$} 
which is $T_2\subt S_2$ by folding. In this way we proved $T_2\asubt S_2$ using transitivity.
\item Choosing $\AContext$ as in Example~\ref{ex:tr-out}, Figure~\ref{sa} gives an application of rule 
\rulename{sub-perm-async}. The rightmost premises can be written 
\begin{center}
$T_m\subt \AContext [T_r]^1[T_s]^2$ and $T_p\subt \AContext [T'_r]^1[T'_s]^2$
\end{center}
and they 
hold by rule \rulename{sub-bra}. The left-hand-side of the conclusion is a selection between the outputs $\Out{\mathsf m}{\SessionTypeS_{m}} . T_m$ and $\Out{\mathsf p}{\SessionTypeS_{p}} . T_p$. The right-hand-side of the conclusion 
can be written  
\begin{center}
\small
$\AContext [\Out{\mathsf m}{\SessionTypeS_{m}^r} . \SessionTypeT_r \oplus \Out{\mathsf p}{\SessionTypeS_{p}^r} . \SessionTypeT'_r \oplus \Out{\mathsf q}{\SessionTypeS_{q}}. \SessionTypeT_{q}]^1[\Out{\mathsf m}{\SessionTypeS_{m}^s} . \SessionTypeT_s \oplus \Out{\mathsf p}{\SessionTypeS_{p}^s} . \SessionTypeT'_s]^2$\end{center} 
Notice that selections are moved inside branching (possibly making smaller the types of the sent channels) and extra selections (in this case $\Out{\mathsf q}{\SessionTypeS_{q}}. \SessionTypeT_{q}$) can be added. 
\end{enumerate} }
\end{example}

\noindent By definition $\asubt$ is an extension of $\ssubt$, and the example at the beginning of this subsection shows that $\ssubt$ is not complete for the asynchronous calculus.
\subsection{Typing asynchronous processes}\label{tap}
Since processes now include queues, we need queue types defined by:
$$\begin{array}{rcl}
\QueueType &:: = &\EmptyQueueT \parop \tmsg{\Tag}{\Type}
\parop \QueueType \qconc \QueueType
\end{array}
$$
where we assume 
associativity of $\qconc$ and $\QueueType \qconc \EmptyQueueT  = \EmptyQueueT \qconc
\QueueType = \QueueType$. We also extend session environments as follows:
$$
\begin{array}{rcl}
\LEnv &::= &\ldots \mid 
\LEnv, \ASETT{\queuetype \ChannelA \ChannelB \QueueType}
\end{array}
$$
The added new element 
$\queuetype \ChannelA \ChannelB \QueueType$ 
is the type of messages in the queue 
$\queue \ChannelA \ChannelB \Queue$.

We denote by $\domq(\LEnv)$ the set of local queues which occur in $\LEnv$. 
Two session environments
$\LEnv_1$ and $\LEnv_2$ {\em agree}
if\\
\centerline{$\dom(\LEnv_1) \cap \dom(\LEnv_2)=
\domq(\LEnv_1) \cap \domq(\LEnv_2) = \emptyset$.}
If $\LEnv_1$ and $\LEnv_2$ agree, 
their composition $\LEnv_1,\LEnv_2$ is given by
$\LEnv_1, \LEnv_2   = \LEnv_1\cup \LEnv_2$ as in the synchronous
case. 
We also define $\LEnv_1 \asubt \LEnv_2$ by: 
$$\begin{array}{l}
\Name \in \dom(\LEnv_1) \cap \dom(\LEnv_2) 
\text{ implies }
\LEnv_1(\Name) \asubt \LEnv_2(\Name)  \text{ and}\\
\Name \in \dom(\LEnv_1) \text{ and }\Name \not\in \dom(\LEnv_2)  \text{ imply }
\LEnv_1(\Name) = \End \text{ and}\\
\Name \not\in \dom(\LEnv_1) \text{ and } \Name \in \dom(\LEnv_2) \text{ imply }
\LEnv_2(\Name) = \End  \text{ and }
\\
\domq(\LEnv_1) = \domq(\LEnv_2)  \text{ and }
\ChannelA \ChannelB\in\domq(\LEnv_1)  \text{ implies }
\LEnv_1(\ChannelA \ChannelB) = \LEnv_2(\ChannelA \ChannelB)\end{array}
$$
We write $\LEnv_1 \approxa \LEnv_2$ 
if $\LEnv_1 \asubt \LEnv_2$ and $\LEnv_2 \asubt \LEnv_1$. 

We need to take into account the interplay between the session type of a channel and the queue type of the queue dequeued by this channel. Following the literature~\cite{GV10,mostrous_phd,mostrous09sessionbased,MY15}
we define the {\em session remainder} of a session type $\SessionType$ and a queue type $\QueueType$ (notation $\remainder{\SessionType}{\QueueType}$). 
The session type $\remainder{\SessionType}{\QueueType}$ is obtained from $\SessionType$ by erasing all branchings in $\SessionType$ that have corresponding selections in $\QueueType$. Clearly the session remainder is defined only if 
$\SessionType$ and $\QueueType$ agree on labels and on types of exchanged channels. More formally we define:
$$
\begin{array}[t]{@{}c@{}}
\inferrule[\rulename{rm-empty}]
{}
 {
\remainder{\SessionType}{\EmptyQueueT} = \SessionType
}
 \qquad
\inferrule[\rulename{rm-bra}]
 {
   \remainder{\SessionType_k}{\QueueType} = \SessionType' 
 \quad {\Type_k\asubt\Type}
   \quad k \in I
 }
 {
  \remainder{\Branch_{i\in I} \In{\Tag_i}{\Type_i}.\SessionTypeT_i}
   {\tmsg{\Tag_k}{\Type} \qconc  \QueueType} = \SessionType'
 }  
 \qquad
\inferrule[\rulename{rm-sel}]
 { 
   \forall i \in I :
   \remainder{\SessionType_i}{ \QueueType} = \SessionType'_i  
 }
 {
\textstyle
  \remainder{\Select_{i\in I} \Out{\Tag_i}{\Type_i}.\SessionTypeT_i}
   { \QueueType} =
   \Select_{i\in I} \Out{\Tag_i}{\Type_i}.\SessionTypeT'_i
 }  
\end{array}
$$

\begin{table}
$$
\begin{array}[t]{@{}c@{}}
\inferrule[\rulename{t-new-async}]{
     \wtp{\UEnv}{\Process}
      {
      \LEnv,
      \ASETT{
      \bind{\ChannelA}{\SessionType_1},
      \bind{\ChannelB}{\SessionType_2},
      \queuetype \ChannelB \ChannelA \QueueType_1,
      \queuetype \ChannelA \ChannelB \QueueType_2}
     }
      \quad
      \remainder{\SessionType_1}{\QueueType_1}
      \dual
      \remainder{\SessionType_2}{\QueueType_2}
    }{
      \wtp{\UEnv}{\new{\ChannelA}{\ChannelB}\Process}{\LEnv}
  }
  \\\\
  \inferrule[\rulename{t-empty-q}]{}{
    \wtp{\UEnv}
     {
      \queue \ChannelB\ChannelA \EmptyQueue
      }
     {
      \queuetype \ChannelB \ChannelA \EmptyQueueT
     }
     }
 \qquad
  \inferrule[\rulename{t-message-q}]{
    \wtp{\UEnv}{
    \queue \ChannelB \ChannelA  \Queue
    }
    {
    \LEnv, \ASETT{\queuetype \ChannelB \ChannelA  \QueueType}
    }
  }{  
    \wtp{\UEnv}{
     \queue \ChannelB \ChannelA \Queue \qconc \msg{\Tag}{\ChannelC}
    }{ \LEnv,
       \ASETT{\bind{\ChannelC}{\SessionTypeS},
       \queuetype \ChannelB \ChannelA 
       \QueueType  \qconc \tmsg{\Tag}{\SessionTypeS}}
    }
  }
\end{array}
$$
\caption{\label{tab:async:typing} 
Typing rules for asynchronous processes and queues.\strut}
\end{table}
The typing rules
for asynchronous processes are obtained from the rules of Table~\ref{tab:sync:typing} 
by replacing rule \rulename{t-new-sync} with rule \rulename{t-new-async} and $\ssubt$ with
$\asubt$ in rule \rulename{t-sub} and
adding the rules for typing the queues. 
Table~\ref{tab:async:typing} gives all the new rules.  
In rule \rulename{t-new-async} we take into account not only the types of the channels, but also those of the queues, and we require duality between their remainders. 
Rule \rulename{t-empty-q}
types the empty queue. 
Rule \rulename{t-message-q}
says how the type of a queue changes when one message is added. 

\subsection{Soundness of asynchronous subtyping}
\label{subsec:a:sound}
 
Reduction of session environments is standard in session calculi, to take into account how communications modify the types of free channels and queues~\cite{Carbone:2012:SCP:2220365.2220367,HVK,yoshida.vasconcelos:language-primitives}. 
In the synchronous case 
only restricted channels can exchange messages. 
We could 
reduce only restricted channels also in the asynchronous case, but
this would make the reduction rules heavier. 
\begin{table}
$$
\begin{array}{c}
 \inferrule[\rulename{tr-in}]{
    {\SessionTypeS_k\asubt \SessionTypeS}\qquad k\in I
  }{
    \ChannelB : \Branch_{i\in I} \In{\Tag_i}{\Type_i}.\SessionTypeT_i,
    \queuetype \ChannelA \ChannelB \tmsg{\Tag_k}{\SessionTypeS} \qconc\QueueType  
    \Red
    \ChannelB :\SessionTypeT_k,
    \queuetype \ChannelA \ChannelB \QueueType
  }
 \qquad
  \inferrule[\rulename{tr-res}]{
    \LEnv_1 \Red \LEnv'_1
  }{
    \LEnv_1, \LEnv_2 \Rightarrow \LEnv'_1, \LEnv_2
  }
  \\\\
 \inferrule[\rulename{tr-out}]{
   \textstyle 
   \forall n \in N ~ \exists i_n\in I_n:~\Tag_{i_n}^{n}=\Tag\quad\Type_{i_n}^{n} \asubt \Type
  }{
\bind{\ChannelA}{\textstyle\AContextf{\Select_{i\in I_n} \Out{\Tag_i^{n}}{\Type_i^{n}}.\SessionTypeT_i^{n}}^{n \in N}},
    \queuetype \ChannelA \ChannelB \QueueType
    \Red
    \bind{\ChannelA}{\AContextf{\SessionTypeT^{n}_{i_n}}^{n \in N}},
    \queuetype \ChannelA \ChannelB \QueueType\qconc \tmsg{\Tag}{\SessionTypeS}
}
\end{array}
$$
\caption{\label{tab:redtype} Reduction of asynchronous session environments.}\label{subsec:a:types}
\end{table}
Table~\ref{tab:redtype} defines the reduction between session environments.  
Rule \rulename{tr-in} simply corresponds to the dequeue of a message. 
Rule \rulename{tr-out} takes into account 
the asynchronous subtyping:
we need to choose one type out of a selection under a context, since a typeable process might 
contain a selection followed by several branches, thanks to rule \rulename{sub-perm-async}. 
The following example illustrates rule 
\rulename{tr-out}.  
\begin{example}{\em Let $\SessionTypeT_1,  \SessionTypeT_2 ,\AContext$ 
be defined as in Example~\ref{ex:tr-out} and assume that there is $\SessionTypeS$ such that 
$\SessionTypeS_{m},\SessionTypeS_{m}'\asubt\SessionTypeS$ and that there is no $\SessionTypeS'$ such that 
$\SessionTypeS_{p},\SessionTypeS_{p}'\asubt\SessionTypeS'$. 
By rule \rulename{tr-out},
only branch $\mathsf m$ can be triggered to output,
since branch $\mathsf q$ is only at hole 2, 
and the above assumption forbids to choose branch $\mathsf p$.
$$
\begin{array}{l}
\ChannelA  :  
\AContext[\SessionType_1]^1[\SessionType_2]^2,
\queuetype \ChannelA \ChannelB \QueueType
  \Red 
\ChannelA  :  
\In{\mathsf r}{\SessionTypeS_{r}} .\SessionTypeT_{m}
~ \& ~
\In{\mathsf s}{\SessionTypeS_{s}} .\SessionTypeT'_{m},
\queuetype \ChannelA \ChannelB \QueueType\qconc \tmsg{\mathsf m}{\SessionTypeS}
\end{array}
$$}
\end{example}

\medskip

In order to get subject reduction we cannot start from an arbitrary session environment. For example the process
$$\receive\ChannelA{\Tag}{\var}.\send\ChannelA{\Tag'}{\var+1}\parop\queue\ChannelB\ChannelA{\msg\Tag{\rm true}}$$
can be typed with the session environment
$$\{\ChannelA:\In\Tag{{\mathtt{int}}}.\Out{\Tag'}{{\mathtt{int}}}.\End,\queuetype\ChannelB\ChannelA{\tmsg{\Tag}{{\mathtt{bool}}}}\}$$
but it reduces to $\send\ChannelA{\Tag'}{{\rm true}+1}$ which cannot be typed. As usual~\cite{Carbone:2012:SCP:2220365.2220367,HVK,yoshida.vasconcelos:language-primitives} we restrict to balanced session environments according to the following definition.
\begin{defi}[Balanced session environments] \label{def:balanced:async} \rm
A session environment $\LEnv$ is {\em balanced}
if:
\begin{enumerate}
\item\label{def:balanced:1} $\bind{\ChannelA}{\SessionTypeT}, 
\queuetype \ChannelB \ChannelA \QueueType \in \LEnv$ imply that  $\remainder\SessionTypeT\QueueType$ is defined; and
\item\label{def:balanced:2} $\bind{\ChannelA}{\SessionTypeT}, 
\queuetype \ChannelB \ChannelA \QueueType, \bind{\ChannelB}{\SessionTypeT'}, 
\queuetype \ChannelA \ChannelB \QueueType' \in \LEnv$ imply that  
$\remainder{\SessionTypeT}{\QueueType} \dual 
\remainder{\SessionTypeT'}{\QueueType'}$.
\end{enumerate}
\end{defi}
\noindent
Notice that $\remainder{\In\Tag{{\mathtt{int}}}.\Out{\Tag'}{{\mathtt{int}}}.\End}{\tmsg{\Tag}{{\mathtt{bool}}}}$ is undefined, if we extend in the obvious way the definition of session remainder. 

\medskip

It is easy to verify that reduction preserves balancing of session environments.
\begin{lem}
\label{lem:redsessiontype:async}  
If $ 
\LEnv\Red\LEnv'$
and $\LEnv$ is  balanced, then $\LEnv'$ is balanced.
\end{lem}
We can now state subject reduction:
\begin{thm}[Subject reduction for asynchronous processes] 
\label{thm:sbr:async}
If $\wtpa{\UEnv}{\Process}{\LEnv}$
and $\LEnv$ is balanced and $\Process \wreda \ProcessQ$, then there is
$\LEnv'$ such that
$\LEnv \Red^\ast \LEnv'$ 
and
$\wtpa{\UEnv}{\ProcessQ}{\LEnv'}$.
\end{thm}
Also the assurance that well-typed processes cannot go wrong requires balanced session environments. 
\begin{cor}
\label{pro:comsafe:async}
  If $\wtpa{\UEnv}{\Process}{\LEnv}$ and $\LEnv$ is balanced, 
  then $\Process \not \wreda \error$.
\end{cor}
\begin{Proof}
By  Theorem \ref{thm:sbr:async},
$\Process \wreda \error$ implies $\wtpa{\UEnv}{\error}{\LEnv'}$ for some $\LEnv'$,
which is impossible
because $\error$ has no type.
\end{Proof}

Lastly we get:
\begin{thm}\label{thm:a:sound}
  The asynchronous subtyping relation $\asubt$ is sound  for the asynchronous calculus.
\end{thm}

The proofs of Theorems~\ref{thm:sbr:async} and~\ref{thm:a:sound} are given in Appendix~\ref{app:async_lang}.

\section{Completeness for Asynchronous Subtyping}
\label{sec:a:completeness}

We start this section by remarking that the synchronous subtyping is incomplete for the asynchronous calculus. In fact the synchronous subtyping is strictly included in the asynchronous subtyping. For example 
$\SessionTypeT \nssubt \SessionTypeS$ by rule \rulename{n-selbra-sync} 
\TZ{but} $\SessionTypeT \asubt \SessionTypeS$, where $T= \Out{\Tag}{\SessionTypeT'}.\In{\Tag'}{\Type'}$
and $S= \In{\Tag'}{\Type'}.\Out{\Tag}{\SessionTypeT'}$.
Then soundness of the asynchronous subtyping implies incompleteness of the synchronous subtyping. 

We show completeness for asynchronous subtyping following
the three steps described in \S~\ref{sec:s:completeness}.  In the
third step we need to add two queues for exchanging messages,
see the proof of Theorem~\ref{thm:preciseness:async}.
The proofs are more tricky than in the synchronous case since 
the asynchronous subtyping makes the shapes of types less structured. 
The first difficulty is to define 
the negation $\nasbut$ inductively and prove that 
it implies the non-derivability of $\asubt$.  
The second difficulty is to catch the error states arising after an unbounded number of message enqueues, since rule \rulename{sub-perm-async} can exchange a selection with an unbounded number of branchings.

\mypar{Characteristic asynchronous processes}
\label{subsec:chara:aprocess}
The definition of characteristic processes for the asynchronous case differs from that of the synchronous one only for outputs, since the creation of a new pair of restricted channels requires also the 
creation of the corresponding queues. 
\begin{defi}[Characteristic asynchronous processes]
\label{async:charprocess} 
\rm 
The characteristic process offering communication $T$ on identifier $u$
for the asynchronous calculus, denoted by $\cprocess(\Name,\SessionType)$,
is defined as in Definition~\ref{s:cprocesses}, but for the case of $\cprocess[!](\Name,\Tag,\Type,\SessionType)$, which is now:
$$\cprocess[!](\Name,\Tag,\Type,\SessionType)
 \eqdef 
  \new\ChannelA\ChannelB(
    \send{\Name}{\Tag}{\ChannelA}.\cprocess(\Name,\SessionType)
    \parop
    \cprocess(\ChannelB,\co\SessionTypeS)
    \parop \queue \ChannelB \ChannelA\EmptyQueue
    \parop \queue \ChannelA \ChannelB\EmptyQueue
  )$$
\end{defi}
\noindent
For example if $\SessionType=\Out{\Tag_1}{\End}\oplus\Out{\Tag_2}{\Out{\Tag_3}\End.\End}.\End$, then 
$$\begin{array}{lll}
\cprocess(\ChannelA,\SessionType)&=& \cprocess[!](\ChannelA,\Tag_1,\End,\End)\oplus\cprocess[!](\ChannelA,\Tag_2,\Out{\Tag_3}\End.\End,\End)\\
&=& \new\ChannelB{\ChannelB'}(
    \send{\ChannelA}{\Tag_1}{\ChannelB}.\cprocess(\ChannelA,\End)
    \parop
    \cprocess(\ChannelB',\End)
    \parop \queue \ChannelB {\ChannelB'}\EmptyQueue
    \parop \queue {\ChannelB'} \ChannelB\EmptyQueue
    )\oplus
\\
&& 
\new\ChannelC{\ChannelC'}(
    \send{\ChannelA}{\Tag_2}{\ChannelC}.\cprocess(\ChannelA,\End)
    \parop
    \cprocess(\ChannelC',\In{\Tag_3}\End.\End)
   \parop\queue \ChannelC {\ChannelC'}\EmptyQueue
    \parop \queue {\ChannelC'} \ChannelC\EmptyQueue
  )\\
  &=& \new\ChannelB{\ChannelB'}(
    \send{\ChannelA}{\Tag_1}{\ChannelB}.\idle
    \parop
    \idle
    \parop \queue \ChannelB {\ChannelB'}\EmptyQueue
    \parop \queue {\ChannelB'} \ChannelB\EmptyQueue
    )\oplus\\
&& \new\ChannelC{\ChannelC'}(
    \send{\ChannelA}{\Tag_2}{\ChannelC}.\idle
    \parop \receive{\ChannelC'}{\Tag_3}{\var}.(
    \cprocess(\ChannelC',\End)\parop \cprocess(\var,\End))
   \parop\queue \ChannelC {\ChannelC'}\EmptyQueue
    \parop \queue {\ChannelC'} \ChannelC\EmptyQueue
  )\\
   &= &\new\ChannelB{\ChannelB'}(
    \send{\ChannelA}{\Tag_1}{\ChannelB}.\idle
    \parop
    \idle
    \parop \queue \ChannelB {\ChannelB'}\EmptyQueue
    \parop \queue {\ChannelB'} \ChannelB\EmptyQueue
     )  \oplus   \\
&& 
\new{\ChannelC}{\ChannelC'  }(
    \send{\ChannelA}{\Tag_2}{\ChannelC}.\idle
    \parop \receive{\ChannelC'}{\Tag_3}{\var}.(
    \idle\parop \idle  )
   \parop \queue \ChannelC {\ChannelC'}\EmptyQueue
    \parop \queue {\ChannelC'} \ChannelC\EmptyQueue
    )\\
&\equiv& \new\ChannelB{\ChannelB'} (\send{\ChannelA}{\Tag_1}{\ChannelB} \parop \queue \ChannelB {\ChannelB'}\EmptyQueue
    \parop \queue {\ChannelB'} \ChannelB\EmptyQueue)\oplus 
\new\ChannelC{\ChannelC'}(\send{\ChannelA}{\Tag_2}{\ChannelC}\parop \receive{\ChannelC'}{\Tag_3}{\var}\parop \queue \ChannelC {\ChannelC'}\EmptyQueue
    \parop \queue {\ChannelC'} \ChannelC\EmptyQueue)
\end{array}$$

\bigskip

Similarly to Lemma~\ref{cpts} we get:
\begin{lem}\label{cpta}$\wtpa{\;}{\cprocess(\Name,\SessionType)}{\ASET{\Name:\SessionType}}$.\end{lem}

\mypar{Rules for negation of the asynchronous subtyping}
\label{subsec:a:negation}
\begin{table}[b]
$$
\begin{array}{c}
{\inferrule[\rulename{n-label-async}]{
    \exists i_0\in I ~\exists n_0\in N~
    \forall j\in J_{n_0}:
    \Tag_{j}^{n_0}\not=\Tag_{i_0}}
    {\textstyle \Select_{i\in I} \Out{\Tag_i}{\Type_i}.\SessionTypeT_i
     \nsubt
     \AContextf{\Select_{j\in J_n} \Out{\Tag_j^{n}}{\Type_j^{n}}.\SessionTypeT_j^{n}}^{n \in N}}}
      \\\\
     \inferrule[\rulename{n-exch-async}]{
     \exists i_0\in I  ~\exists n_0\in N
     ~\exists j_{0}\in J_{n_0}:
	 \Tag_{j_{0}}^{n_0}=\Tag_{i_0}
     \quad
     \Type_{j_{0}}^{n_0} \nsubt \Type_{i_0}}
     {\textstyle \Select_{i\in I} \Out{\Tag_i}{\Type_i}.\SessionTypeT_i
     \nsubt
     \AContextf{\Select_{j\in J_n} \Out{\Tag_j^{n}}{\Type_j^{n}}.\SessionTypeT_j^{n}}^{n \in N}}
     \\\\
     \inferrule[\rulename{n-cont-async}]{
      \forall i\in I  ~\forall n\in N
     ~\exists j_{i,n}\in J_{n}:
	 \Tag_{j_{i,n}}^{n}=\Tag_{i}\quad
	\exists i_0\in I~
     \SessionTypeT_{i_0} \nsubt \AContextf{\SessionTypeT_{j_{i_0,n}}^{n}}^{n \in N}
     }
    {\textstyle {\Select}_{i\in I} \Out{\Tag_i}{\Type_i}.\SessionTypeT_i
     \nsubt
     \AContextf{\Select_{j\in J_n} \Out{\Tag_j^{n}}{\Type_j^{n}}.\SessionTypeT_j^{n}}^{n \in N}}
\\\\
    {\inferrule[\rulename{n-bra-async}]
{\&\not \in \SessionTypeT}
    {\SessionTypeT\nsubt\Branch_{i\in I} \In{\Tag_i}{\Type_i}.\SessionTypeT_i}}
     \qquad
    {\inferrule[\rulename{n-sel-async}]
{\oplus\not \in \SessionTypeT}
    {\textstyle \Select_{i\in I} \Out{\Tag_i}{\Type_i}.\SessionTypeT_i\nsubt\SessionTypeT}}
\end{array}
$$
\caption{\label{tab:negasubtype} 
Negation of asynchronous subtyping.}%
\end{table}
%
%
The negation rules of asynchronous subtyping are obtained from the rules of Table~\ref{tab:negsubtype} {\em excluding} rule
\rulename{n-selbra-sync} plus the rules of
Table~\ref{tab:negasubtype}. %
Rule
\rulename{n-label-async} deals with the case 
that the selection cannot find a matching label inside the $n_0$-th hole. %
Rule \rulename{n-exch-async} considers
a mismatch between carried types inside the $n_0$-th hole. Rule  
\rulename{n-cont-async} considers 
a mismatch between continuation 
types, again inside the $n_0$-th hole. The asynchronous context in these rules allows to consider selection surrounded by branchings.  %
These three rules become 
the rules \rulename{n-label-sel}, \rulename{n-exch-sel} and 
\rulename{n-cont-sel} of Table~\ref{tab:negsubtype}, 
respectively, when the context $\AContext$ is just one hole. 
Rule \rulename{n-bra-async} assures that a type without branchings cannot be a subtype of a branching type;
the predicate $\mathord{\Branch}\not \in \SessionTypeT$  is the negation of the predicate $\mathord{\Branch} \in \SessionTypeT$ (see page~\pageref{def:branch-in-t}).  %
More precisely $\& \not\in \SessionTypeT$ holds %
iff we can derive it from
the following axioms and inductive rules:%
\label{def:branch-not-in-t}%
\[\inferrule[]{}{\& \not\in \End}\qquad \inferrule[]{}{\& \not\in \tvar}\qquad \inferrule[]{\exists i \in I\quad\& \not\in \SessionTypeT_i}{\& \not\in\Select_{i\in I} \Out{\Tag_i}{\Type_i}.\SessionTypeT_i}\qquad \inferrule[]{\&\not \in \SessionType }{ \& \not\in \trec\tvar.\SessionType}\]
where we consider also open session types; %
such a predicate holds up-to unfolding %
and is the complement of $\& \in \SessionType$ %
(Propositions~\ref{lem:branch-not-in-t:unfolding} %
and~\ref{lem:branch-not-in-t:complement} in Appendix~\ref{ac}).
Dually, %
rule \rulename{n-sel-async} assures that a type without selections cannot be a subtype of a selection type. %
\label{def:sel-in-t}%
The predicate $\mathord{\oplus}\in \SessionTypeT$ holds if each
continuation path of $\tree\SessionTypeT$ contains at least one
selection and the predicate $\mathord{\oplus}\not \in \SessionTypeT$
is the negation of the predicate $\mathord{\oplus} \in
\SessionTypeT$. The definitions of these predicates are  analogous to
those of $\& \in \SessionTypeT$ and $\& \not\in \SessionTypeT$. %
We write $T \nasbut S$ if $T \nsubt S$ is generated by the rules in 
Table~\ref{tab:negasubtype} and Table~\ref{tab:negsubtype} excluding rule 
\rulename{n-selbra-sync}. %

For example, by rule \rulename{n-cont-async},
$$
\begin{array}{l}
\Out{\mathsf m}{\SessionTypeS_{m}}.(\In{\mathsf r}{\SessionTypeS_{r}} . {\SessionTypeT_{r}}  \&
\In{\mathsf s}{\SessionTypeS_{s}} . \End)\nasbut \In{\mathsf r}{\SessionTypeS_{r}} . \Out{\mathsf m}{\SessionTypeS_{m}}.\SessionTypeT_{r} \& \In{\mathsf s}{\SessionTypeS_{s}} . \Out{\mathsf m}{\SessionTypeS_{m}}. \Out{\mathsf p}{\SessionTypeS_{p}}. \SessionTypeT_{p}
\end{array}
$$
since
$\In{\mathsf r}{\SessionTypeS_{r}} . {\SessionTypeT_{r}}  \& 
\In{\mathsf s}{\SessionTypeS_{s}} . \End\nasbut$
$\In{\mathsf r}{\SessionTypeS_{r}} . {\SessionTypeT_{r}}  \& 
\In{\mathsf s}{\SessionTypeS_{s}} . \Out{\mathsf p}{\SessionTypeS_{p}}. \SessionTypeT_{p}.
$

\medskip

In Lemma~\ref{pro:s:negation:async}, %
we show that $\nasbut$ is the negation of the asynchronous subtyping. %
This result will be used (in the \emph{``only if''} direction) %
in the proof of Theorem~\ref{thm:preciseness:async}.

\begin{lem}\label{pro:s:negation:async} 
  $\T\asubt \SessionTypeS$ is not derivable %
  if and only if %
  $\T \nasbut \SessionTypeS$ is derivable.
\end{lem}

\begin{Proof} 
If $\SessionTypeT \nasbut \SessionTypeS$, then we can show
$\SessionTypeT \not\asubt \SessionTypeS$ by induction on the
derivation of $\SessionTypeT \nasbut \SessionTypeS$. %
The proof is similar to that for %
``\,$\SessionTypeT \nssubt \SessionTypeS$ implies
$\SessionTypeT \not\ssubt \SessionTypeS$\,''
(Lemma~\ref{lem:s:negation}) %
--- removing the base case \rulename{n-selbra-sync}, %
and adding the cases for the rules in Table~\ref{tab:negasubtype}. %
We can further remove the cases %
\rulename{n-label-sel}, \rulename{n-exch-sel} and
\rulename{n-cont-sel}, %
since (as discussed above) they are subsumed respectively by %
\rulename{n-label-async},
\rulename{n-exch-async} and \rulename{n-cont-async}, %
when $\AContext$ is just one hole. %
We show the detailed proofs for some of the new cases (the omitted proofs 
are
similar):%
\begin{itemize}
\item%
 base case \rulename{n-bra-async}.\quad %
 We have %
 $\SessionTypeT \nsubt \SessionTypeS%
 = \Branch_{i\in I} \In{\Tag_i}{\Type_i}.\SessionTypeT_i$,
 with $\& \not\in \SessionTypeT$. %
 Since $\SessionTypeS$ is a branching, %
 $\SessionTypeS \asubt \SessionTypeT$ cannot hold %
 by \rulename{sub-end} nor \rulename{sub-sel}; %
 moreover, it cannot hold by \rulename{sub-bra}, %
 since $\SessionTypeT$ cannot be a branching %
 (otherwise, we would have the contradiction $\& \in \SessionTypeT$); %
 also \rulename{sub-perm-async} is ruled out: %
 otherwise, we would have %
 $\SessionTypeT = \Select_{i\in J}\Out{\Tag_i}{\Type_i}.\SessionTypeT_{i}$ %
 and %
 $\forall i\in J:\&\in\SessionTypeT_i$, %
 and thus the contradiction $\& \in \SessionTypeT$. %
 Hence, we conclude $\SessionTypeS \not\asubt \SessionTypeT$;
\item%
 base case \rulename{n-sel-async}.\quad %
 We have %
 $\Select_{i\in I} \Out{\Tag_i}{\Type_i}.\SessionTypeT_i%
 = \SessionTypeT \nsubt \SessionTypeS$,
 with $\oplus \not\in \SessionTypeS$. %
 Since $\SessionTypeT$ is a selection, %
 $\SessionTypeS \asubt \SessionTypeT$ cannot hold %
 by \rulename{sub-end} nor \rulename{sub-bra}; %
 moreover, it cannot hold by \rulename{sub-sel}, %
 since $\SessionTypeS$ cannot be a selection %
 (otherwise, we would have the contradiction $\oplus \in \SessionTypeS$); %
 also \rulename{sub-perm-async} is ruled out: %
 otherwise, we would have %
 $\SessionTypeS = %
 \AContext[\Select_{i\in I \cup J_n} \Out{\Tag_i}{\Type_i^{n}}.\SessionTypeT_i^{n}
 ]^{n\in N}$, %
 and thus the contradiction $\oplus \in \SessionTypeS$. %
 Hence, we conclude $\SessionTypeS \not\asubt \SessionTypeT$;
\item%
 inductive case \rulename{n-cont-async}.\quad %
 Then, we have: %
 \begin{itemize}
 \item%
   $\SessionTypeT = \Select_{i\in I} \Out{\Tag_i}{\Type_i}.\SessionTypeT_i$; %
 \item%
   $\SessionTypeS =%
   \AContextf{\Select_{j\in J_n} \Out{\Tag_j^{n}}{\Type_j^{n}}.\SessionTypeT_j^{n}}^{n \in N}$; %
 \item%
   $\exists i_0\in I, n_0\in N, j_{0}\in J_{n_0}:%
   \Tag_{j_{0}}^{n_0}=\Tag_{i_0}$
   and %
   $\SessionTypeT_{i_0} \nsubt
   \AContextf{\SessionTypeT_{j_{0}}^{n}}^{n \in N}$. %
 \end{itemize}
 From the last item, %
 by the induction hypothesis we have %
 $\SessionTypeT_{i_0} \not\asubt%
 \AContextf{\SessionTypeT_{j_{0}}^{n}}^{n \in N}$. %
 We observe that, %
 since $\SessionTypeT$ is a selection, %
 $\SessionTypeT \asubt \SessionTypeS$ %
 could only possibly hold by rule \rulename{sub-perm-async}, %
 or \rulename{sub-sel} when $\AContext$ is just one hole. %
 Since %
 $\SessionTypeT_{i_0} \not\asubt \AContextf{\SessionTypeT_{j_{0}}^{n}}^{n \in N}$,
 in both cases %
at least one of the coinductive premises of the candidate rule is 
not satisfied.  %
 Hence, we conclude $\SessionTypeT \not\asubt \SessionTypeS$. %
\end{itemize}
\emph{Vice versa}, %
the proof for %
$\SessionTypeT \not\asubt \SessionTypeS$ implies %
$\SessionTypeT \nasbut \SessionTypeS$ %
is similar to that for %
$\SessionTypeT \not\ssubt \SessionTypeS$ implies %
$\SessionTypeT \nssubt \SessionTypeS$ %
(Lemma~\ref{lem:s:negation}): %
we take a tentative derivation %
for $\SessionTypeT \asubt \SessionTypeS$ %
with a branch that fails after $n$ steps, %
and by induction on $n$ %
we construct a derivation of depth $n+1$ %
which concludes $\SessionTypeT \nasbut \SessionTypeS$. %
The only differences are the following:
\begin{itemize}
\item%
  in the base case $n=0$, %
  we observe that %
  if $\SessionTypeT$ and $\SessionTypeS$ cause an immediate derivation
  failure under $\ssubt$, %
  then they \emph{also} cause an immediate failure under $\asubt$ %
  --- \emph{except} %
  when $\SessionTypeT$ is a selection and $\SessionTypeS$ is a
  branching. %
  In this latter case, %
  we must consider that rule \rulename{sub-perm-async}
  \emph{might} %
  allow for a further derivation step under $\asubt$; %
  when this does \emph{not} happen %
  (i.e., the shapes of $\SessionTypeT$ and $\SessionTypeS$ %
  do not match the conclusion of \rulename{sub-perm-async}), %
  we construct a derivation of depth $1 = n+1$ %
  which concludes $\SessionTypeT \nasbut \SessionTypeS$, %
  by one of the axioms %
  \rulename{n-label-async}, %
  \rulename{n-bra-async} or \rulename{n-sel-async}; %
\item%
  in the inductive case $n = m+1$, %
  we must also consider the case %
  in which the shapes of $\SessionTypeT,\SessionTypeS$ %
  match the conclusion of rule \rulename{sub-perm-async}, %
  but there is some coinductive premise
  $\SessionTypeT' \asubt \SessionTypeS'$ %
  whose sub-derivation has a branch that fails after $m$ steps. %
  Then, by the induction hypothesis %
  there exists a derivation of depth $m+1$ %
  that concludes $\SessionTypeT' \nasbut \SessionTypeS'$; %
  using this as a premise, %
  by \rulename{n-exch-async} or \rulename{n-cont-async} %
  we construct a derivation of depth $(m+1)+1 = n+1$ %
  which concludes $\SessionTypeT \nasbut \SessionTypeS$.\qedhere %
\end{itemize}
\end{Proof}

\noindent Before proving completeness of subtyping, %
we need two more intermediate results %
on the predicates $\Branch \in \SessionTypeT$, 
$\Branch \not\in \SessionTypeT$, %
and 
$\oplus \in \SessionTypeT$ %
(defined 
on %
pages~\pageref{def:branch-in-t} and \pageref{def:sel-in-t}). %

Proposition~\ref{prop:branc-oplus-not-in-dual} can be easily proved %
by noticing that dualisation %
turns branchings in the continuation paths of $\tree\SessionTypeT$ %
into selections in the continuation paths of $\tree{\dualf\SessionTypeT}$,  and \emph{vice versa}, %
as remarked on page~\pageref{def:dualf}.

\begin{proposition}\label{prop:branc-oplus-not-in-dual}
  $\Branch \in \SessionTypeT$ %
  \;if and only if\; %
  $\oplus \in \dualf\SessionTypeT$.
\end{proposition}

We define duality of asynchronous contexts as expected:
$$\co{[ \; ]^{n}}=[ \; ]^{n}
\qquad
\co{\Branch_{i\in I} \In{\Tag_i}{\Type_i}.\AContext_i}=\Select_{i\in I} \Out{\Tag_i}{\Type_i}.\co{\AContext_i}
$$
We use $\BContext$ to range over duals of asynchronous contexts. We extend asynchronous subtyping to the duals of asynchronous contexts in the obvious way:
$$
 \begin{array}{@{}c@{}}
 \inferrule[\rulename{sub-empty}]{}
 {[ \; ]^{n }\subt[ \; ]^{n }}
 \qquad 
 \cinferrule[\rulename{sub-dual-cont}]{
   \forall i\in I: \TypeU'_i \subt \TypeU_i
   \quad \BContext_i \subt \BContext'_i
 }{
  \Select_{i\in I }
   \Out{\Tag_i}{\TypeU_i}.\BContext_i
     \subt
     \Select_{i\in I\cup J}
   \Out{\Tag_i}{\TypeU'_i}.\BContext'_i
 }
 \end{array}
 $$

The following lemma assures that, if there are continuation paths in the tree of a type $T$  which does not contain branchings, then we can find a type $S$ which is smaller than $T$ (according to the synchronous subtyping, and then also according to the asynchronous one) such that all continuation paths in the tree of $S$  do not contain branchings. Moreover $T$ and $S$ share similar structures.
\begin{lem}\label{last}
  If $\&\not \in \SessionTypeT$, %
  then there is $\Type\ssubt\SessionTypeT$  such that the continuation paths of $\tree\SessionTypeS$ do not contain branchings.
Moreover $\SessionTypeT=\BContext[\SessionTypeT_n]^{n\in N}$ and 
$\SessionTypeS=\BContext'[\SessionTypeT_n]^{n\in N'}$ with $\BContext'\subt\BContext$ 
and $N'\subseteq N$.
\end{lem}
\begin{Proof}
If $\Branch$ does not occur in the continuation paths of $\tree\SessionTypeT$ we can choose $\Type=\SessionTypeT$. Otherwise by definition $\tree\SessionTypeT$ contains some continuation paths 
with occurrences of $\Branch$ and other continuation paths without occurrences of $\Branch$. The continuation paths with occurrences of $\Branch$ must contain nodes labelled by selections. We can then choose $\Type$ as the session type whose 
tree is obtained
by pruning top-down from selection nodes the continuation paths containing $\Branch$ and the exchanged sub-trees with the same label in $\tree\SessionTypeT$. %
We then get $\SessionTypeS \ssubt \SessionTypeT$ through a derivation %
only composed by instances of rule \rulename{sub-sel}. This simple construction implies that
$\SessionTypeT=\BContext[\SessionTypeT_n]^{n\in N}$ and 
$\SessionTypeS=\BContext'[\SessionTypeT_n]^{n\in N'}$ with $\BContext'\subt\BContext$ 
and $N'\subseteq N$.
\end{Proof}

A last lemma connects tree representations of types and occurrences of channels as subjects in characteristic processes.

\begin{lem}\label{lastprop}
If the continuation paths of $\tree\SessionTypeT$ have no branchings, then $\ChannelA\not\in\sbnF( \cprocess(\ChannelA, \SessionTypeT))$.
\end{lem}
\begin{Proof}
  We first observe that, by Definition~\ref{async:charprocess}, %
  the exchanged types of $\SessionTypeT$ %
  do not influence whether $\ChannelA$ belongs to 
  $\sbnF(\cprocess(\ChannelA, \SessionTypeT))$. %
  Second, %
  we naturally extend type trees %
  to \emph{open} types (with \emph{closed} exchanged types) %
  by letting $\tree{\tvar} = \tvar$: %
  this allows us to prove the statement %
  by structural induction on $\SessionTypeT$, %
  neglecting its exchanged types. %
  We show that %
  if the continuation paths of $\tree{\SessionTypeT}$ %
  do not contain branchings %
  then %
  $\ChannelA \not\in\sbnD( \cprocess(\ChannelA, \SessionTypeT), \emptyset, \emptyset)$, %
  considering only the most interesting cases:
  \begin{itemize}
  \item%
    base case $\SessionTypeT = \tvar$.\quad %
    Then, %
    $\tree{\SessionTypeT}$ does not contain branchings %
    and, %
    by definition of $\sbnD$ (Table~\ref{sbnD}), %
    we also have %
    $\ChannelA \not\in %
     \sbnD(\invoke{\pvar_\tvar}{\ChannelA}, \emptyset, \emptyset)%
    = \emptyset$;
  \item%
    inductive case %
    $\SessionTypeT = \trec\tvar.\SessionTypeT'$.\quad %
    We observe that:%
    \begin{enumerate}
    \item%
      \label{item:lastprop:i}%
      the tree $\tree{\SessionTypeT}$ is obtained by recursively grafting %
      the tree $\tree{\SessionTypeT'}$ %
      on its own leaf nodes marked with $\tvar$. %
      Therefore, %
      the continuation paths of $\tree{\SessionTypeT}$ %
      do not contain branchings %
      if and only if %
      the continuation paths of $\tree{\SessionTypeT'}$ %
      do not contain branchings; %
    \item%
      \label{item:lastprop:ii}%
      assume that $\tree{\SessionTypeT}$ does not contain branchings. %
      From the previous item %
      and by the induction hypothesis, %
      we have %
      $\ChannelA \not\in \sbnD\left({%
        \cprocess(\ChannelA, \SessionTypeT'),%
        \emptyset,\emptyset%
      }\right)$;
    \item%
      \label{item:lastprop:iii}%
      by Definition~\ref{async:charprocess}, %
      $%
      \cprocess(\ChannelA, \SessionTypeT) = %
      \Def{\pvar_\tvar}{\var}{%
        \cprocess(\var, \SessionTypeT')%
      }{%
        \invoke{\pvar_\tvar}{\ChannelA}%
      }%
      $. %
      Therefore, we have %
      $\sbnD( \cprocess(\ChannelA, \SessionTypeT), \emptyset, \emptyset) = %
      \sbnD\left({%
        \cprocess(\ChannelA, \SessionTypeT'),\,%
        \left(%
        \pvar_\tvar(\var) = \cprocess(\var, \SessionTypeT')
        \right),\,%
        \emptyset%
      }\right)$, by definition of $\sbnD$. %
    \end{enumerate}
    Now, by contradiction, %
    assume $\ChannelA \in \sbnD( \cprocess(\ChannelA, \SessionTypeT), \emptyset, \emptyset)$. %
    By definition of $\sbnD$, %
    this means that $\ChannelA$ is yielded by a %
    syntactic occurrence of either %
    $\send\Expression\Tag{\ChannelA}.\ProcessP$ %
    or %
    $\sum_{i\in I} \receive\ChannelA{\Tag_i}{\var_i}.\Process_i$ %
    in $\cprocess(\ChannelA, \SessionTypeT')$. %
    The former is impossible, %
    because by Definition~\ref{async:charprocess}, %
    the parameter $\ChannelA$ of $\cprocess(\ChannelA, \SessionTypeT')$ %
    is never sent as an exchanged channel. %
    The latter, instead, %
    by item~\ref{item:lastprop:iii} above %
    implies: %
    $$\ChannelA \;\in\; \sbnD\left({%
      \cprocess(\ChannelA, \SessionTypeT'),\,%
      \left(%
      \pvar_\tvar(\var) = \cprocess(\var, \SessionTypeT')
      \right),\,%
      \emptyset%
    }\right)$$ %
    and thus %
    $\ChannelA  \in %
    \sbnD\left({%
        \cprocess(\ChannelA, \SessionTypeT'),%
        \emptyset,\emptyset%
    }\right) %
   $ %
    --- which contradicts 
    item~\ref{item:lastprop:ii} above.
  \end{itemize}
  Finally, assume that 
  the continuation paths of $\tree{\SessionTypeT}$ do
  not contain branchings: %
  we have proved that
  $\ChannelA \not\in\sbnD( \cprocess(\ChannelA, \SessionTypeT), \emptyset, \emptyset)$; %
  by Definition~\ref{def:sbn}, %
  we conclude %
  $\ChannelA\not\in\sbnF( \cprocess(\ChannelA, \SessionTypeT))$.
\end{Proof}

Completeness can now be shown:
\begin{thm} [Completeness for asynchronous subtyping]\label{thm:preciseness:async}
The asynchronous subtyping relation $\asubt$ is complete for the asynchronous calculus.
\end{thm}
\begin{Proof}
  We prove that $\SessionTypeT \nasbut \SessionTypeS$ implies that there are 
 $\SessionTypeT' \ssubt \SessionTypeT$ and $\SessionTypeS' \ssubt \co\SessionTypeS$, %
    with either $\SessionTypeT' = \SessionTypeT$ %
    or $\SessionTypeS' = \co\SessionTypeS$, %
  such that
\begin{equation}
  \label{eq:completeness-async:main}
  \new\ChannelA\ChannelB(
    \cprocess(\ChannelA,\SessionTypeT')
    \parop
    \cprocess(\ChannelB,\SessionTypeS')
    \parop \queue \ChannelB \ChannelA\EmptyQueue
    \parop \queue \ChannelA \ChannelB\EmptyQueue
  )
  \wreda
  \error
\end{equation}
where $\cprocess(\ChannelA,\SessionTypeT')$, $
\cprocess(\ChannelB,\SessionTypeS')$ are characteristic asynchronous
processes. %
Note that %
$\cprocess(\ChannelA,\SessionTypeT')$ %
and $\cprocess(\ChannelB,\SessionTypeS')$ %
play respectively the r\^oles of $\ProcessP$ and $\ProcessQ$ in
{\bf Step 3} (page~\pageref{step-iii}): %
in fact, %
we have %
${\wtp{}{\cprocess(\ChannelA,\SessionTypeT')}{\ASET{\ChannelA:\SessionTypeT}}}$ %
and %
${\wtp{}{\cprocess(\ChannelB,\SessionTypeS')}{\ASET{\ChannelB:\co\SessionTypeS}}}$, by using rule \rulename{t-sub} when $\SessionTypeT' \neq \SessionTypeT$ 
  or $\SessionTypeS' \neq \co\SessionTypeS$.
The proof is 
by induction on the derivation of $\SessionTypeT \nasbut
\SessionTypeS$. 
  In all inductive cases, we assume %
  $\SessionTypeT' \ssubt \SessionTypeT$ and %
  $\SessionTypeS' = \co\SessionTypeS$ %
 as induction hypothesis; %
  the proof assuming %
  $\SessionTypeT' = \SessionTypeT$ %
  and $\SessionTypeS' \ssubt \co\SessionTypeS$ %
  is symmetric.

We do not consider the rules
\rulename{n-label-sel}, \rulename{n-exch-sel} and \rulename{n-cont-sel}, since they are particular cases of the rules \rulename{n-label-async}, \rulename{n-exch-async} and \rulename{n-cont-async} when the asynchronous context is empty.

\smallskip

{\em Case} \rulename{n-end r}:
$\SessionTypeT = \End$ and $\SessionTypeS \not= \End$.
$$
\begin{array}{l}
\new\ChannelA\ChannelB
 (
 \cprocess(\ChannelA,\SessionTypeT)
 \parop
 \cprocess(\ChannelB,\co{\SessionTypeS})
 \parop 
 \queue \ChannelB \ChannelA \EmptyQueue 
 \parop
 \queue \ChannelA \ChannelB \EmptyQueue)  
=\new \ChannelA \ChannelB
(
 \idle \parop \cprocess(\ChannelB, \co\SessionTypeS)
 \parop 
 \queue \ChannelB \ChannelA \EmptyQueue 
 \parop
 \queue \ChannelA \ChannelB \EmptyQueue 
)
 \reda \error
\end{array}
$$
by rules $\rulename{err-in-async}$ and \rulename{err-context},
since $\ChannelA \not \in \fn(\idle)$.

\smallskip

{\em Case} \rulename{n-end l}:
$\SessionTypeT \not = \End$ and $\SessionTypeS = \End$.
The proof is as in the previous case. 

\smallskip

{\em Case} \rulename{n-brasel}: 
$\SessionTypeT = 
\Branch_{i\in I}\In{\Tag_i}{\SessionTypeT'_i}.\SessionTypeT_i$ and 
$\SessionTypeS = 
\Select_{j\in J}\Out{\Tag'_j}{\SessionTypeS'_j}.\SessionTypeS_j$.
$$
\begin{array}{l}
\new\ChannelA\ChannelB
 (
 \cprocess(\ChannelA,\SessionTypeT)
 \parop
 \cprocess(\ChannelB,\co{\SessionTypeS})
 \parop
 \queue \ChannelB \ChannelA \EmptyQueue 
 \parop
 \queue \ChannelA \ChannelB \EmptyQueue ) 
=\\
\new \ChannelA \ChannelB
(
 \sum_{i \in I} \cprocess[?](\ChannelA, \Tag_i, \SessionTypeT'_i, \SessionTypeT_i)
 \parop
 \sum_{j \in J} \cprocess[?](\ChannelB, \Tag'_j, \SessionTypeS'_j, \co{\SessionTypeS_j})
 \parop
 \queue \ChannelB \ChannelA \EmptyQueue 
 \parop
 \queue \ChannelA \ChannelB \EmptyQueue 
)
 \reda \error
\end{array}
$$
by rules $\rulename{err-in-in-async}$ and \rulename{err-context}.

\smallskip

{\em Case} \rulename{n-label-bra}:
$\SessionTypeT= \Branch_{i\in I} \In{\Tag_i}{\SessionTypeT'_i}.\SessionTypeT_i$, 
$\SessionTypeS = \Branch_{j\in J} \In{\Tag'_j}{\SessionTypeS'_j}.\SessionTypeS_j$, 
and $ \exists k\in J$ such that $\forall i\in I:\Tag'_k \ne \Tag_i$.
$$
\begin{array}{l}
\new\ChannelA\ChannelB
 (
 \cprocess(\ChannelA,\SessionTypeT)
 \parop
 \cprocess(\ChannelB,\co\SessionTypeS)
 \parop
 \queue \ChannelB \ChannelA \EmptyQueue 
 \parop
 \queue \ChannelA \ChannelB \EmptyQueue)
\reda \\
\new \ChannelA \ChannelB
(
 \sum_{i \in I}
 \cprocess[?](\ChannelA, \Tag_i, \SessionTypeT'_i, \SessionTypeT_i)
 \parop
 \new \ChannelC \ChannelD
 (
 \send{\ChannelB}{\Tag'_k}{\ChannelC}. \cprocess(\ChannelB, \co{\SessionTypeS_k}) \parop
 \cprocess(\ChannelD, \co{\SessionTypeS'_k}) \
 \parop \\\hfill
 \queue \ChannelD \ChannelC \EmptyQueue 
 \parop
 \queue \ChannelC \ChannelD \EmptyQueue 
 )
 \parop
 \queue \ChannelB \ChannelA \EmptyQueue 
 \parop
 \queue \ChannelA \ChannelB \EmptyQueue 
)
\reda\\

\ContextC[
 \sum_{i \in I}
 \cprocess[?](\ChannelA, \Tag_i, \SessionTypeT'_i, \SessionTypeT_i)
 \parop
 \queue \ChannelB \ChannelA \msg{\Tag'_k}{\ChannelC} 
 )
]
\end{array}
$$
where 
$
\ContextC
=\new \ChannelA \ChannelB
 \new \ChannelC \ChannelD
 (  \cprocess(\ChannelB, \co{\SessionTypeS_k})
\parop
 \cprocess(\ChannelD, \co{\SessionTypeS'_k})  \parop
 [ \; \; ] 
 \parop
 \queue \ChannelA \ChannelB \EmptyQueue 
 \parop
 \queue \ChannelD \ChannelC \EmptyQueue 
 \parop
 \queue \ChannelC \ChannelD \EmptyQueue 
 ).
$

\noindent
By rule $\rulename{err-mism-async}$
$$
\begin{array}{l}
 \sum_{i \in I}
 \cprocess[?](\ChannelA, \Tag_i, \SessionTypeT'_i, \SessionTypeT_i)
 \parop
 \queue \ChannelB \ChannelA \msg{\Tag'_k}{\ChannelC}
\reda
\error
\end{array}
$$
By rule \rulename{err-context}, we conclude
$$
\begin{array}{l}
\ContextC[  \sum_{i \in I}
 \cprocess[?](\ChannelA, \Tag_i, \SessionTypeT'_i, \SessionTypeT_i)
\parop
 \queue \ChannelB \ChannelA \msg{\Tag'_k}{\ChannelC} 
]
 \reda
\error
\end{array}
$$

\smallskip

{\em Case} \rulename{n-exch-bra}:
$\SessionTypeT =  
\Branch_{i\in I} \In{\Tag_i}{\SessionTypeT'_i}.\SessionTypeT_i$,
$\SessionTypeS = 
\Branch_{j\in J} \In{\Tag'_j}{\SessionTypeS'_j}.\SessionTypeS_j$, 
and $\exists k \in I ~ \exists k' \in J$ such that $\Tag_{k} = \Tag'_{k'}$
and $\SessionTypeT'_{k} \nasbut \SessionTypeS'_{k'}$. By induction there are 
$\SessionType^*\ssubt \SessionTypeT'_{k}$ and $\SessionTypeS^* = \co{\SessionTypeS'_{k'}}$
such that 
$$ \new \ChannelC \ChannelD
    (\cprocess(\ChannelC, \SessionTypeT^*)
    \parop
    \cprocess(\ChannelD, \SessionTypeS^*)
    \parop
    \queue \ChannelD \ChannelC \EmptyQueue 
    \parop
    \queue \ChannelC \ChannelD \EmptyQueue
   )
  \wreda \error$$ We can then choose $$\textstyle\SessionTypeT' =  \In{\Tag_k}{\SessionTypeT^*}.\SessionTypeT_k~\Branch~
\Branch_{i\in I, i\not=k} \In{\Tag_i}{\SessionTypeT'_i}.\SessionTypeT_i \quad\text{and}\quad%
\SessionTypeS' =  \Out{\Tag_{k'}}{\co{\SessionTypeS^*}}.\co{\SessionTypeS_{k'}}~\Select~
\Select_{j\in J,j\not=k'} \Out{\Tag'_j}{\SessionTypeS'_j}.\co{\SessionTypeS_j}.$$
By definition $\SessionTypeT' \ssubt \SessionTypeT$ and $\SessionTypeS' = \co\SessionTypeS$. We get
$$
\begin{array}{l}
\new \ChannelA \ChannelB
    (\cprocess(\ChannelA,\SessionTypeT')
    \parop
    \cprocess(\ChannelB, \SessionTypeS') 
    \parop
    \queue \ChannelB \ChannelA \EmptyQueue 
    \parop
    \queue \ChannelA \ChannelB \EmptyQueue  )
  \reda  \\ 
\new \ChannelA \ChannelB
(   
    \sum_{i \in I, i\not=k}
    \cprocess[?](\ChannelA,\Tag_i,\SessionTypeT'_i,\SessionTypeT_i) 
    +\receive{\ChannelA}{\Tag_k}{\var}.( \cprocess(\ChannelA, \SessionTypeT_k)\parop
    \cprocess(\var, \SessionTypeT^*)) \parop 
    \\ \qquad \qquad 
    \new {\ChannelC} {\ChannelD} (
    \send{\ChannelB}{\Tag_{k}}{{\ChannelC}}.
    \cprocess(\ChannelB,\co{\SessionTypeS_{k'}}) 
    \parop
    \cprocess({\ChannelD} ,\SessionTypeS^*)
   \parop
    \queue \ChannelD \ChannelC\EmptyQueue 
    \parop
    \queue \ChannelC \ChannelD \EmptyQueue)
    \parop
    \queue \ChannelB \ChannelA \EmptyQueue 
    \parop
    \queue \ChannelA \ChannelB \EmptyQueue   
) \wreda
\\ 
\Context
[ \new \ChannelC \ChannelD
    (\cprocess(\ChannelC, \SessionTypeT^*)
    \parop
    \cprocess(\ChannelD, \SessionTypeS^*)
    \parop
    \queue \ChannelD \ChannelC \EmptyQueue 
    \parop
    \queue \ChannelC \ChannelD \EmptyQueue
   )
]
\end{array}
$$
where\\ $\ContextC\hole = 
\new \ChannelA \ChannelB
( 
\cprocess(\ChannelA, \SessionTypeT_k) 
\parop
\cprocess(\ChannelB, \co{\SessionTypeS_{k'}})
\parop
\queue \ChannelB \ChannelA \EmptyQueue
\parop
\queue \ChannelA \ChannelB \EmptyQueue
)\parop \hole.
$
Then by rule \rulename{err-context} we conclude
$$
\Context
[ \new \ChannelC \ChannelD
    (\cprocess(\ChannelC, \SessionTypeT^*)
    \parop
    \cprocess(\ChannelD, \SessionTypeS^*)
    \parop
    \queue \ChannelD \ChannelC \EmptyQueue 
    \parop
    \queue \ChannelC \ChannelD \EmptyQueue
   )]
\wreda \error
$$

\smallskip

{\em Case} \rulename{n-cont-bra}:
$\SessionTypeT =  
\Branch_{i\in I} \In{\Tag_i}{\SessionTypeT'_i}.\SessionTypeT_i$,
$\SessionTypeS = 
\Branch_{j\in J} \In{\Tag'_j}{\SessionTypeS'_j}.\SessionTypeS_j$, 
and $\exists k \in I ~ \exists k' \in J$ such that $\Tag_{k} = \Tag'_{k'}$
and $\SessionTypeT_k \nasbut \SessionTypeS_{k'}$. By induction there are 
$\SessionType^*\ssubt \SessionTypeT_{k}$ and $\SessionTypeS^* = \co{\SessionTypeS_{k'}}$
 such that 
$$ \new \ChannelA \ChannelB
    (\cprocess(\ChannelA, \SessionTypeT^*)
    \parop
    \cprocess(\ChannelB, \SessionTypeS^*)
    \parop
    \queue \ChannelB \ChannelA \EmptyQueue 
    \parop
    \queue \ChannelA \ChannelB \EmptyQueue
   )
  \wreda \error$$ We can then choose $$\textstyle\SessionTypeT' =  \In{\Tag_k}{\SessionTypeT'_k}.\SessionTypeT^*~\Branch~
\Branch_{i\in I, i\not=k} \In{\Tag_i}{\SessionTypeT'_i}.\SessionTypeT_i \quad\text{and}\quad
\SessionTypeS' =  \Out{\Tag_{k'}}{\SessionTypeS'_{k'}}.\SessionTypeS^*~\Select~
\Select_{j\in J,j\not=k'} \Out{\Tag'_j}{\SessionTypeS'_j}.\co{\SessionTypeS_j}.$$
By definition $\SessionTypeT' \ssubt \SessionTypeT$ and $\SessionTypeS' = \co\SessionTypeS$. We get
$$
\begin{array}{l}
\new \ChannelA \ChannelB
    ( 
    \cprocess(\ChannelA,\SessionTypeT')
    \parop
    \cprocess(\ChannelB, \SessionTypeS') 
    \parop
    \queue \ChannelB \ChannelA \EmptyQueue 
    \parop
    \queue \ChannelA \ChannelB \EmptyQueue 
    )
  \reda \\ 
\new \ChannelA \ChannelB    
( \sum_{i \in  I,i\not=k}
  \cprocess[?](\ChannelA,\Tag_i,\SessionTypeT'_i,\SessionTypeT_i)
  +   \receive{\ChannelA}{\Tag_k}{\var}.(\cprocess(\ChannelA, \SessionTypeT^*)\parop
    \cprocess(\var, \SessionTypeT'_k)) \parop\\
   \qquad \new {\ChannelC} {\ChannelD} (
    \send{\ChannelB}{\Tag_{k}}{{\ChannelC}}.
    \cprocess(\ChannelB,\SessionTypeS^*) 
    \parop
    \cprocess({\ChannelD} ,\co{\SessionTypeS'_{k'}})
    \parop
    \queue \ChannelD \ChannelC\EmptyQueue 
    \parop
    \queue \ChannelC \ChannelD \EmptyQueue)
    \parop
    \queue \ChannelB \ChannelA\EmptyQueue 
    \parop
    \queue \ChannelA \ChannelB \EmptyQueue   
) 
 \reda \\
\new \ChannelA \ChannelB  
\new \ChannelC \ChannelD  
(   \sum_{i \in I, i\not=k}
    \cprocess[?](\ChannelA,\Tag_i,\SessionTypeT'_i,\SessionTypeT_i) +
    \receive{\ChannelA}{\Tag_k}{\var}.(\cprocess(\ChannelA, \SessionTypeT^*)\parop
    \cprocess(\var, \SessionTypeT'_k)) \parop\\
   \hfill \cprocess(\ChannelB, \SessionTypeS^*) 
    \parop \cprocess(\ChannelD, \co{\SessionTypeS'_{k'}})\parop
    \queue \ChannelB \ChannelA \msg{\Tag_{k}}{\ChannelC}  \parop
    \queue \ChannelA \ChannelB \EmptyQueue
    \parop \queue \ChannelD \ChannelC \EmptyQueue
\parop \queue \ChannelC \ChannelD \EmptyQueue
)
\reda\\
\Context[
\new \ChannelA \ChannelB    
( 
    \cprocess(\ChannelA,\SessionTypeT^*)
    \parop
    \cprocess(\ChannelB, \SessionTypeS^*) 
    \parop
    \queue \ChannelB \ChannelA \EmptyQueue 
    \parop
    \queue \ChannelA \ChannelB \EmptyQueue
    )
]
\end{array}
$$
where 
$$
\Context\hole = 
\new \ChannelC \ChannelD
( 
    \cprocess(\ChannelC, \SessionTypeT'_{k})   \parop
\cprocess(\ChannelD, \co{\SessionTypeS'_{k'}})
\parop \queue \ChannelD \ChannelC \EmptyQueue
\parop \queue \ChannelC \ChannelD \EmptyQueue
)\parop \hole
$$ 
Then by rule \rulename{err-context} we conclude
$$
\begin{array}{l}
\Context[
\new \ChannelA \ChannelB    
( 
    \cprocess(\ChannelA,\SessionTypeT^*)
    \parop
    \cprocess(\ChannelB, \SessionTypeS^*) 
    \parop
    \queue \ChannelB \ChannelA \EmptyQueue 
    \parop
    \queue \ChannelA \ChannelB \EmptyQueue
    )
] 
\wreda \error
\end{array}
$$

\smallskip

{\em Case} \rulename{n-label-async}: %
$\SessionTypeT = \Select_{i\in I} \Out{\Tag_i}{\SessionTypeT'_i}.\SessionTypeT_i$,
$\SessionTypeS = \AContextf{\Select_{j\in J_n} \Out{{\Tag'_j}^{n}}{{\SessionTypeS'_j}^{n}}.\SessionTypeS_j^{n}}^{n \in N}$
and there are $i_0 \in I, n_0\in N$ such that %
$\forall j\in J_{n_0}$ we get 
${\Tag'_j}^{n_0}\not=\Tag_{i_0}$. We show by induction on $\AContext$ that 
$\SessionTypeT \nasbut \SessionTypeS$ implies
$$ \new\ChannelA\ChannelB(
    \cprocess(\ChannelA,\SessionTypeT)
    \parop
    \cprocess(\ChannelB,\co\SessionTypeS)
    \parop \queue \ChannelB \ChannelA\Queue
    \parop \queue \ChannelA \ChannelB\EmptyQueue
  )
  \wreda
  \error$$
  for an arbitrary queue $\Queue$. 
  \begin{enumerate}
\item If $\AContext \!=\! \hole$, 
then
${\textstyle \SessionTypeS = 
\Select_{j\in J} \Out{\Tag'_j}{\SessionTypeS'_j}.\SessionTypeS_j}$, %
with $J = J_{n_0}$,
and $\co \SessionTypeS = 
\Branch_{j\in J} \In{\Tag'_j}{\SessionTypeS'_j}.\co{\SessionTypeS_j}$.
We get
$$
\begin{array}{l}
\new\ChannelA\ChannelB
 (
 \cprocess(\ChannelA,\SessionTypeT)
 \parop
 \cprocess(\ChannelB,\co\SessionTypeS)
 \parop \queue \ChannelB \ChannelA\Queue
 \parop \queue \ChannelA \ChannelB\EmptyQueue
 )
\reda  \\
\new \ChannelA \ChannelB
(
 \cprocess[!](\ChannelA, \Tag_{i_0}, \SessionTypeT'_{i_0}, \SessionTypeT_{i_0}) \parop
 \sum_{j \in J} \cprocess[?](\ChannelB, \Tag'_j, \SessionTypeS'_j, \co{\SessionTypeS_j})
\parop
\queue \ChannelB \ChannelA\Queue
 \parop \queue \ChannelA \ChannelB\EmptyQueue
)
\reda  \\
\Context[ \sum_{j \in J} 
 \receive{\ChannelB}{\Tag'_j}{\var}.(
 \cprocess(\ChannelB, \co{\SessionTypeS_j})\parop
 \cprocess(\var,\SessionTypeS'_j)) \parop\queue \ChannelA \ChannelB\msg{\Tag_{i_0}}{\ChannelC}]
\end{array}
$$
where
$$
\ContextC\hole =
\new \ChannelA \ChannelB
    \new\ChannelC\ChannelD(
    \cprocess(\ChannelA,\SessionTypeT_{i_0})
    \parop
    \cprocess(\ChannelD,\co{\SessionTypeT'_{i_0}})
    \parop\hole
    \parop
\queue \ChannelB \ChannelA\Queue\parop
    \queue \ChannelD \ChannelC\EmptyQueue
    \parop \queue \ChannelC \ChannelD\EmptyQueue
 )
$$

\noindent
By rule \rulename{err-mism-async}
$$
\sum_{j \in J} 
 \receive{\ChannelB}{\Tag'_j}{\var}.
 \cprocess(\ChannelB, \co{\SessionTypeS_j}) \parop\queue \ChannelA \ChannelB\msg{\Tag_{i_0}}{\ChannelC}
 \reda
\error$$
then by rule \rulename{err-context}, we conclude
$$
\Context[ \sum_{j \in J} 
 \receive{\ChannelB}{\Tag'_j}{\var}.
 \cprocess(\ChannelB, \co{\SessionTypeS_j}) \parop\queue \ChannelA \ChannelB\msg{\Tag_{i_0}}{\ChannelC}]
 \reda
\error$$

\item\label{case:complete:context:noempty}If $\& \in \AContext$, 
let $\AContext=
\Branch_{k\in K}\In{\Tag_k^*}{\SessionTypeS^*_k}. \AContext_k[\;]^{n\in N_k}$, where $\bigcup_{k\in K} N_k=N$. 
Then \begin{eqnarray} 
\co{\SessionTypeS}    &=&   
\Select_{k \in K} 
\Out{\Tag^*_{k}}{\SessionTypeS^*_{k}}.
\co{\AContext_{k}}
[\Branch_{j\in J_n} \In{{\Tag'_j}^{n}}{{\SessionTypeS'_j}^{n}}.\co{\SessionTypeS_j^{n}}]^{n \in N_{k}}
\nonumber
\end{eqnarray}
Let $k_0\in K$ be such that $n_0\in N_{k_0}$ and
$$\SessionTypeV= 
\co{\AContext_{k_0}}
[\Branch_{j\in J_n} \In{{\Tag'_j}^{n}}{{\SessionTypeS'_j}^{n}}.\co{\SessionTypeS_j^{n}}]^{n \in N_{k_0}}
$$
We get
$$\begin{array}{l}
\new \ChannelA \ChannelB
( 
 \cprocess(\ChannelA, \SessionTypeT) \parop
 \cprocess(\ChannelB, \co{\SessionTypeS}) \parop
 \queue{\ChannelB}{\ChannelA} \Queue \parop
 \queue{\ChannelA}{\ChannelB} \EmptyQueue
) \reda  \\ 
\new \ChannelA \ChannelB
( \cprocess(\ChannelA, \SessionTypeT) \parop
  \new{\ChannelC}{\ChannelD}
  ( 
  \send{\ChannelB}{\Tag^*_{k_0}}{\ChannelC}.  \cprocess(\ChannelB, \SessionTypeV) \parop 
  \cprocess(\ChannelD, \co{\SessionTypeS^*_{k_0}})
 \parop
  \queue{\ChannelD}{\ChannelC} \EmptyQueue \parop \queue{\ChannelC}{\ChannelD} \EmptyQueue 
  )  %
\parop
 \queue{\ChannelB}{\ChannelA} \Queue \parop
  \queue{\ChannelA}{\ChannelB} \EmptyQueue 
) \reda \\
\ContextC[\new \ChannelA \ChannelB(\cprocess(a,T)\parop \cprocess(b,V)\parop\queue{\ChannelB}{\ChannelA} \Queue \qconc\msg{\Tag^*_{k_0}}{\ChannelC} \parop
  \queue{\ChannelA}{\ChannelB} \EmptyQueue)]
 \end{array}$$ 
 where 
$$\ContextC=  \new{\ChannelC}{\ChannelD}
  ( 
   \cprocess(\ChannelD, \co{\SessionTypeS^*_{k_0}}) \parop[\;\;]
 \parop
  \queue{\ChannelD}{\ChannelC} \EmptyQueue \parop \queue{\ChannelC}{\ChannelD} \EmptyQueue 
  )$$
By induction $$\new \ChannelA \ChannelB(\cprocess(a,T)\parop \cprocess(b,V)\parop\queue{\ChannelB}{\ChannelA} \Queue \qconc\msg{\Tag^*_{k_0}}{\ChannelC} \parop
  \queue{\ChannelA}{\ChannelB} \EmptyQueue)\wreda \error$$
  then by rule \rulename{err-context}, we conclude
  $$\ContextC[\new \ChannelA \ChannelB(\cprocess(a,T)\parop \cprocess(b,V)\parop\queue{\ChannelB}{\ChannelA} \Queue \qconc\msg{\Tag^*_{k_0}}{\ChannelC} \parop
  \queue{\ChannelA}{\ChannelB} \EmptyQueue)]\wreda \error$$
\end{enumerate}

{\em Case} \rulename{n-exch-async}:
$
\begin{array}{ll}
\SessionTypeT = \Select_{i\in I} \Out{\Tag_i}{\SessionTypeT'_i}.\SessionTypeT_i,   &     
\SessionTypeS = \AContextf{\Select_{j\in J_n} \Out{{\Tag'_j}^{n}}{{\SessionTypeS'_j}^{n}}.\SessionTypeS_j^{n}}^{n \in N},
\end{array}
$
and there are $i_0 \in I,  n_0\in N, j_0 \in J_{n_0}$ such that  ${\Tag'_{j_0}}^{\!\!\!n_0}=\Tag_{i_0}$ and ${\SessionTypeS'_{j_0}}^{\!\!\!n_0} \nasbut \SessionType'_{i_0}$. 
We show by induction on $\AContext$ that 
 $\SessionTypeT \nasbut \SessionTypeS$ implies
$$
  \new\ChannelA\ChannelB(
    \cprocess(\ChannelA,\SessionTypeT')
    \parop
    \cprocess(\ChannelB,\SessionTypeS')
    \parop \queue \ChannelB \ChannelA\Queue
    \parop \queue \ChannelA \ChannelB\EmptyQueue
  )
  \wreda
  \error$$
%
for some 
    $\SessionTypeT'= \SessionTypeT$, $\SessionTypeS' \ssubt \co\SessionTypeS$ 
  and an arbitrary queue $\Queue$.
  \begin{enumerate}
\item\label{case:complete:context:empty2}
  If $\AContext \!=\! \hole$, then
$\SessionTypeS = 
\Select_{j\in J} \Out{\Tag'_j}{\SessionTypeS'_j}.\SessionTypeS_j$, where $J = J_{n_0}$,
and $\co \SessionTypeS = 
\Branch_{j\in J} \In{\Tag'_j}{\SessionTypeS'_j}.\co{\SessionTypeS_j}$ and ${\SessionTypeS'_{j_0}} \nasbut \SessionType'_{i_0}$. By induction there are  
$\SessionTypeT^* =\co{\SessionTypeT'_{i_0}}$ and $\SessionTypeS^* \ssubt  \SessionTypeS'_{j_0}$ such that
$$
\begin{array}{l}
 \new\ChannelC\ChannelD(
    \cprocess(\ChannelC,\SessionTypeS^*) \parop
    \cprocess(\ChannelD,\SessionTypeT^*)
    \parop \queue \ChannelD \ChannelC\EmptyQueue
    \parop \queue \ChannelC \ChannelD\EmptyQueue)
    \wreda \error
\end{array}
$$
We can choose %
$\SessionTypeT' = \Out{\Tag_{i_0}}{\co{\SessionTypeT^*}}.\SessionTypeT_{i_0}~\Select~\Select_{i\in I, i \not=i_0} \Out{\Tag_i}{\SessionTypeT'_i}.\SessionTypeT_i$ 
 (therefore, $\SessionTypeT' = \SessionTypeT$) 
and %
$\SessionTypeS'=\In{\Tag_{i_0}}{\SessionTypeS^*}.\co{\SessionTypeS_{j_0}}~\Branch~\Branch_{j\in J, j \not=j_0} \In{\Tag_j}{\SessionTypeS'_j}.\co{\SessionTypeS_j}$ %
  (therefore, $\SessionTypeS' \ssubt  \co{\SessionTypeS}$).
We get
$$
\begin{array}{l}
\new\ChannelA\ChannelB
 (
 \cprocess(\ChannelA,\SessionTypeT')
 \parop
 \cprocess(\ChannelB,\SessionTypeS')
 \parop \queue \ChannelB \ChannelA \Queue
 \parop \queue \ChannelA \ChannelB\EmptyQueue
 )
\reda\\
\new \ChannelA \ChannelB
    \new\ChannelC\ChannelD
   ( \cprocess(\ChannelA,\SessionTypeT_{i_0})
        \parop \receive{\ChannelB}{\Tag_{i_0}}{\var}.(
 \cprocess(\ChannelB, \co{\SessionTypeS_j}) \parop
 \cprocess(\var,\SessionTypeS^*))+
 \sum_{j \in J, j\not=j_0} 
 \receive{\ChannelB}{\Tag'_j}{\var}.(
 \cprocess(\ChannelB, \co{\SessionTypeS_j}) \parop
 \cprocess(\var,\SessionTypeS'_j)
  )
  \parop\\
  \hfill
    \cprocess(\ChannelD,\SessionTypeT^*)\parop
 \queue \ChannelB \ChannelA \Queue
 \parop \queue \ChannelA \ChannelB\msg{\Tag_{i_0}}{\ChannelC}
 \parop \queue \ChannelD \ChannelC\EmptyQueue
    \parop \queue \ChannelC \ChannelD\EmptyQueue
)
\reda
\\  
C[ \new\ChannelC\ChannelD(
    \cprocess(\ChannelC,\SessionTypeS^*) \parop
    \cprocess(\ChannelD,\SessionTypeT^*)
    \parop \queue \ChannelD \ChannelC\EmptyQueue
    \parop \queue \ChannelC \ChannelD\EmptyQueue)]
\end{array}
$$
where
$\ContextC=\new \ChannelA \ChannelB
( 
    \cprocess(\ChannelA,\SessionTypeT_{i_0})
    \parop
 \cprocess(\ChannelB, \co{\SessionTypeS_{j_0}}) \parop
 \queue \ChannelB \ChannelA \Queue
 \parop \queue \ChannelA \ChannelB \EmptyQueue
) \parop [\;\;]
.$
Then by \rulename{err-context}, we conclude
$$
\begin{array}{l}
C[ \new\ChannelC\ChannelD( 
    \cprocess(\ChannelC,\SessionTypeS^*) \parop
    \cprocess(\ChannelD,\SessionTypeT^*)
    \parop \queue \ChannelD \ChannelC\EmptyQueue
    \parop \queue \ChannelC \ChannelD\EmptyQueue)] \wreda
\error
\end{array}
$$
\item If $\& \in \AContext$, then the proof is as in case (\ref{case:complete:context:noempty}) of rule \rulename{n-label-async}.
\end{enumerate}

\smallskip

{\em Case} \rulename{n-cont-async}:
$\SessionTypeT = \Select_{i\in I} \Out{\Tag_i}{\SessionTypeT'_i}.\SessionTypeT_i$,
$\SessionTypeS = \AContextf{\Select_{j\in J_n} \Out{{\Tag'_j}^{n}}{{\SessionTypeS'_j}^{n}}.\SessionTypeS_j^{n}}^{n \in N}$, %
and for all $i\in I, n\in N: ~\exists j_{i,n}\in J_{n}$ such that
${\Tag'_{j_{i,n}}}^{\!\!\!\!\!\!\!n}=\Tag_{i}$ and
$\exists i_0\in I:
\SessionTypeT_{i_0} \nsubt
\AContextf{\SessionTypeS_{j_{i_0,n}}^{n}}^{n \in N}$. By
induction, there exist $\SessionTypeT^* \ssubt \SessionTypeT_{i_0}$ and
$\SessionTypeS^* =
\co{\AContextf{\SessionTypeS_{j_{i_0,n}}^{n}}^{n \in N}}$ such that:
\begin{equation}
  \label{eq:preciseness:n-cont-async-ind}
  \new\ChannelA\ChannelB(
  \cprocess(\ChannelA,\SessionTypeT^*)
  \parop
  \cprocess(\ChannelB,\SessionTypeS^*)
  \parop \queue \ChannelB \ChannelA\EmptyQueue
  \parop \queue \ChannelA \ChannelB\EmptyQueue
  )
  \wreda
  \error
\end{equation}
By Lemma~\ref{last}
$\SessionTypeS^*=\BContext[\SessionTypeS^n_*]^{n\in N'}$ with
$\BContext \subt \co\AContext$, $N' \subseteq N$ and
$\SessionTypeS^n_*=\co{\SessionTypeS^n_{j_{i_0,n}}}$.
We observe that \eqref{eq:preciseness:n-cont-async-ind} %
implies that %
in $\BContext$ %
there exists a continuation path of $m$ outputs %
$\Out{\Tag^{\flat}_1}{\SessionTypeS^{\flat}_1},%
\ldots,%
\Out{\Tag^{\flat}_m}{\SessionTypeS^{\flat}_m}$ %
reaching a $k$-indexed hole, %
such that if $\cprocess(\ChannelB,\SessionTypeS^*)$ fires
the $m$ outputs along such a path, %
we get%
\footnote{%
  Note that, %
  if there are errors in the exchanged types, %
  an $\error$ transition might be enabled %
  \emph{before} the
  whole continuation path is fired; %
  in this case, by Proposition~\ref{lem:async-errors-persistent}, %
  it will remain enabled until
  \eqref{eq:preciseness:n-cont-async-ind:ii} is reached.}: 
\begin{equation}
  \label{eq:preciseness:n-cont-async-ind:start-ii}
  \new\ChannelA\ChannelB(
  \cprocess(\ChannelA,\SessionTypeT^*)
  \parop
  \cprocess(\ChannelB,\SessionTypeS^*)
  \parop \queue \ChannelB \ChannelA\EmptyQueue
  \parop \queue \ChannelA \ChannelB\EmptyQueue
  )
  \;\overbrace{\reda \;\cdots\; \reda}^{\text{$m$ times}}%
  \end{equation}
  \begin{equation}
  \label{eq:preciseness:n-cont-async-ind:ii}
  \ContextC\left[%
  \new\ChannelA\ChannelB\left(
  \cprocess(\ChannelA,\SessionTypeT^*)
  \parop
  \cprocess(\ChannelB,\SessionTypeS^k_*)
  \parop \queue \ChannelB \ChannelA%
  \msg{\Tag^{\flat}_1}{\ChannelC^{\flat}_1}%
  \qconc\ldots\qconc%
  \msg{\Tag^{\flat}_m}{\ChannelC^{\flat}_m}%
  \parop \queue \ChannelA \ChannelB\EmptyQueue
  \right)%
  \right]
  \wreda\; \error%
\end{equation}
where $\ContextC\hole$ contains the restrictions and the characteristic
processes for the exchanged channels %
$\ChannelC^{\flat}_1,\ldots,\ChannelC^{\flat}_m$.

\noindent%
Now, we can choose\; %
$\SessionTypeT' =
\Out{\Tag_{i_0}}{\SessionTypeT'_{i_0}}.\SessionTypeT^*
~\Select~\Select_{i\in I,i\not=i_0} \Out{\Tag_i}{\SessionTypeT'_i}.\SessionTypeT_i$ 
  \;(therefore, $\SessionTypeT' \ssubt \SessionTypeT$) 
\;and %
$\SessionTypeS' = \BContext[\In{\Tag_{i_0}}{{\SessionTypeS^{'n}_{j_{i_0,n}}}}.\SessionTypeS^n_*~\Branch~{\Branch_{j\in J_n, j\not=j_{i_0,n}} \In{{\Tag'_j}^{n}}{{\SessionTypeS'_j}^{n}}.\co{\SessionTypeS_j^{n}}}]^{n \in N'}$ 
  (therefore, $\SessionTypeS' = \co{\SessionTypeS}$). 
We have:
\begin{equation}
  \label{eq:preciseness:n-cont-async-ind:start-iii}
  \new\ChannelA\ChannelB(
  \cprocess(\ChannelA,\SessionTypeT')
  \parop
  \cprocess(\ChannelB,\SessionTypeS')
  \parop \queue \ChannelB \ChannelA\EmptyQueue
  \parop \queue \ChannelA \ChannelB\EmptyQueue
  )
  \reda%
  \end{equation}
  \[
  \ContextC'\left[%
  \new\ChannelA\ChannelB(
  \cprocess(\ChannelA,\SessionTypeT^*)
  \parop
  \cprocess(\ChannelB,\SessionTypeS')
  \parop \queue \ChannelB \ChannelA\EmptyQueue
  \parop \queue \ChannelA \ChannelB\msg{\Tag_{i_0}}{\ChannelC_{i_0}}
  )\right]
  \;\overbrace{\reda \;\cdots\; \reda}^{\text{$m$ times}}%
 \]
  \[ 
  \ContextC'\Big[\ContextC\Big[%
  \new\ChannelA\ChannelB\Big(
  \cprocess(\ChannelA,\SessionTypeT^*)
  \parop
  \cprocess(\ChannelB,\In{\Tag_{i_0}}{{\SessionTypeS^{'k}_{j_{i_0,k}}}}.\SessionTypeS^k_*~\Branch~{\Branch_{j\in J_k, j\not=j_{i_0,k}}
  \In{{\Tag'_j}^{k}}{{\SessionTypeS'_j}^{k}}.\co{\SessionTypeS_j^{k}}})
  \]
  \[
  \phantom{x}
  \parop \queue \ChannelB \ChannelA%
  \Out{\Tag^{\flat}_1}{\ChannelC^{\flat}_1}%
  \qconc\ldots\qconc%
  \Out{\Tag^{\flat}_m}{\ChannelC^{\flat}_m}%
  \parop \queue \ChannelA \ChannelB\msg{\Tag_{i_0}}{\ChannelC_{i_0}}%
  \Big)\Big]\Big]
  \reda%
  \] 
  \begin{equation}\label{eq:preciseness:n-cont-async-ind:iii}
  \ContextC''\left[
  \ContextC\left[\new\ChannelA\ChannelB\left(
  \cprocess(\ChannelA,\SessionTypeT^*)
  \parop
  \cprocess(\ChannelB,\SessionTypeS^k_*)
  \parop \queue \ChannelB \ChannelA%
  \Tag^{\flat}_1{\ChannelC^{\flat}_1}%
  \qconc\ldots\qconc%
  \Tag^{\flat}_m{\ChannelC^{\flat}_m}%
  \parop \queue \ChannelA \ChannelB\EmptyQueue%
  \right)%
  \right]\right]
\end{equation}
where $\ContextC'\hole$ and $\ContextC''\hole$
contain the restrictions and the characteristic processes %
for the exchanged channel $\ChannelC_{i_0}$ and its co-channel. %
The reductions %
from \eqref{eq:preciseness:n-cont-async-ind:start-ii} and \eqref{eq:preciseness:n-cont-async-ind:start-iii}
perform the same communications, %
except for the enqueuing/dequeuing of $\msg{\Tag_{i_0}}{\ChannelC_{i_0}}$ on
$\ChannelA\ChannelB$. %
Moreover, the reached configurations
\eqref{eq:preciseness:n-cont-async-ind:ii} and
\eqref{eq:preciseness:n-cont-async-ind:iii} %
coincide, except for the surrounding context $\ContextC''\hole$. %
Thus, all errors reachable from
\eqref{eq:preciseness:n-cont-async-ind:ii} %
are also reachable from \eqref{eq:preciseness:n-cont-async-ind:iii}, %
by \rulename{err-context}. %
We conclude\; %
$\new\ChannelA\ChannelB(
  \cprocess(\ChannelA,\SessionTypeT')
  \parop
  \cprocess(\ChannelB,\SessionTypeS')
  \parop \queue \ChannelB \ChannelA\EmptyQueue
  \parop \queue \ChannelA \ChannelB\EmptyQueue
  ) \wreda \error$.

{\em Case} \rulename{n-bra-async}: 
$\Branch \not \in \SessionTypeT$
and $\SessionTypeS=\Branch_{i\in I} \In{\Tag_i}{\Type_i}.\SessionTypeT_i$. 
Then $\co\SessionTypeS=\Select_{i\in I} \Out{\Tag_i}{\Type_i}.\co{\SessionTypeT_i}$. By Lemma~\ref{last} %
there is $\SessionTypeT'\ssubt\SessionTypeT$ 
such that the continuation paths of $\tree{\SessionTypeT'}$ do not contain branchings. 
$$\begin{array}{c}
\new \ChannelA \ChannelB
( 
 \cprocess(\ChannelA, \SessionTypeT') \parop
 \cprocess(\ChannelB, \co{\SessionTypeS})\parop 
 \queue{\ChannelB}{\ChannelA} \EmptyQueue \parop
 \queue{\ChannelA}{\ChannelB} \EmptyQueue
)  \reda\\ 
 \new \ChannelA \ChannelB
( 
 \cprocess(\ChannelA, \SessionTypeT') \parop
 \cprocess[!]
 (\ChannelB, \Tag_{i_0}, \SessionTypeS_{i_0}, \co{\SessionTypeT_{i_0}})
 \parop 
 \queue{\ChannelB}{\ChannelA} \EmptyQueue \parop
 \queue{\ChannelA}{\ChannelB} \EmptyQueue ) \nonumber 
\reda \\
\Context[
 \new \ChannelA \ChannelB
( 
 \cprocess(\ChannelA, \SessionTypeT') \parop
 \cprocess(\ChannelB, \co{\SessionTypeT_{i_0}})
 \parop 
 \queue{\ChannelB}{\ChannelA} \msg{\Tag_{i_0}}{\ChannelC} \parop
 \queue{\ChannelA}{\ChannelB} \EmptyQueue  ) ] \nonumber 
\end{array}$$
where
$$
\ContextC[ \; ] = 
\new \ChannelC \ChannelD
(\cprocess(\ChannelD, \co{\SessionTypeS_{i_0}})
 \parop 
  [ \; \; ]
 \parop 
 \queue \ChannelD \ChannelC \EmptyQueue
 \parop \queue \ChannelC \ChannelD \EmptyQueue
 )
$$

By Lemma~\ref{lastprop} 
$\ChannelA\not\in\sbnF( \cprocess(\ChannelA, \SessionTypeT') \parop
 \cprocess(\ChannelB, \co{\SessionTypeT_{i_0}})\parop
 \queue{\ChannelA}{\ChannelB} \EmptyQueue)$, since the continuation paths of $\tree{\SessionTypeT'}$ do not contain branchings, and \\%
 \centerline{$\fpv(\cprocess(\ChannelA, \SessionTypeT') \parop
 \cprocess(\ChannelB, \co{\SessionTypeT_{i_0}})\parop
 \queue{\ChannelA}{\ChannelB} \EmptyQueue)=\emptyset$} so 
by 
\rulename{err-orph-mess-async} 
we get
\begin{eqnarray}
 \new \ChannelA \ChannelB
( 
 \cprocess(\ChannelA, \SessionTypeT') \parop
 \cprocess(\ChannelB, \co{\SessionTypeT_{i_0}})
 \parop 
  \queue{\ChannelB}{\ChannelA} \msg{\Tag_{i_0}}{\ChannelC} \parop
 \queue{\ChannelA}{\ChannelB} \EmptyQueue
 ) \nonumber 
 \reda 
\error \nonumber
\end{eqnarray}
Then by \rulename{err-context}, we conclude
\begin{eqnarray}
\Context[ 
 \new \ChannelA \ChannelB
( 
 \cprocess(\ChannelA, \SessionTypeT') 
 \parop
 \cprocess(\ChannelB, \co{\SessionTypeT_{i_0}})
 \parop
 \queue{\ChannelB}{\ChannelA} \msg{\Tag_{i_0}}{\ChannelC} 
 \parop
 \queue{\ChannelA}{\ChannelB} \EmptyQueue) ]
 {\reda}^{\ast}  
\error \hfill \nonumber
\end{eqnarray}

{\em Case} \rulename{n-sel-async}: 
$\SessionTypeT=\Select_{i\in I} \Out{\Tag_i}{\Type_i}.\SessionTypeT_i$
and $\oplus\not \in \SessionTypeS$. %
By Proposition~\ref{prop:branc-oplus-not-in-dual}, %
we have $\Branch \not \in \co{\SessionTypeS}$. %
By Lemma~\ref{last} there is $\SessionTypeS'\ssubt\co{\SessionTypeS}$ 
such that the continuation paths of $\tree{\SessionTypeS'}$ do not contain branchings. 
$$\begin{array}{c}
\new \ChannelA \ChannelB
( 
 \cprocess(\ChannelA, \SessionTypeT) \parop
 \cprocess(\ChannelB, \SessionTypeS')\parop 
 \queue{\ChannelB}{\ChannelA} \EmptyQueue \parop
 \queue{\ChannelA}{\ChannelB} \EmptyQueue
)  \reda\\ 
 \new \ChannelA \ChannelB
( 
 \cprocess(\ChannelA,  \Tag_{i_0}, \SessionTypeS_{i_0}, \SessionTypeT_{i_0}) \parop
 \cprocess[!]
 (\ChannelB,\SessionTypeS')
 \parop 
 \queue{\ChannelB}{\ChannelA} \EmptyQueue \parop
 \queue{\ChannelA}{\ChannelB} \EmptyQueue ) \nonumber 
\reda \\
\Context[
 \new \ChannelA \ChannelB
( 
 \cprocess(\ChannelA, \SessionTypeT_{i_0}) \parop
 \cprocess(\ChannelB, \SessionTypeS')
 \parop 
 \queue{\ChannelB}{\ChannelA} \msg{\Tag_{i_0}}{\ChannelC} \parop
 \queue{\ChannelA}{\ChannelB} \EmptyQueue  ) ] \nonumber 
\end{array}$$
where
$
\ContextC\hole = 
\new \ChannelC \ChannelD
(\cprocess(\ChannelD, \co{\SessionTypeS_{i_0}})
 \parop 
  \hole
 \parop 
 \queue \ChannelD \ChannelC \EmptyQueue
 \parop \queue \ChannelC \ChannelD \EmptyQueue
 )
$.\\
By Lemma~\ref{lastprop} 
$\ChannelB\not\in\sbnF( \cprocess(\ChannelA, \SessionTypeT_{i_0}) \parop
 \cprocess(\ChannelB, \SessionTypeS')\parop
 \queue{\ChannelA}{\ChannelB} \EmptyQueue)$, since the continuation paths of $\tree{\SessionTypeS'}$ do not contain branchings,  and \\%
 \centerline{$\fpv(\cprocess(\ChannelA, \SessionTypeT_{i_0}) \parop
 \cprocess(\ChannelB, \SessionTypeS')\parop
 \queue{\ChannelA}{\ChannelB} \EmptyQueue)=\emptyset$} so 
by 
\rulename{err-orph-mess-async} 
we get
\begin{eqnarray}
 \new \ChannelA \ChannelB
( 
 \cprocess(\ChannelA, \SessionTypeT_{i_0}) \parop
 \cprocess(\ChannelB, \SessionTypeS')
 \parop 
  \queue{\ChannelB}{\ChannelA} \msg{\Tag_{i_0}}{\ChannelC} \parop
 \queue{\ChannelA}{\ChannelB} \EmptyQueue
 ) \nonumber 
 \reda 
\error \nonumber
\end{eqnarray}
Then by \rulename{err-context}, we conclude
\begin{eqnarray}
\Context[ 
 \new \ChannelA \ChannelB
( 
 \cprocess(\ChannelA, \SessionTypeT_{i_0}) 
 \parop
 \cprocess(\ChannelB, \SessionTypeS')
 \parop
 \queue{\ChannelB}{\ChannelA} \msg{\Tag_{i_0}}{\ChannelC} 
 \parop
 \queue{\ChannelA}{\ChannelB} \EmptyQueue) ]
 {\reda}^{\ast}  
\error \hfill \nonumber
\end{eqnarray}
Summing up, we proved that $\SessionType\not\asubt\SessionTypeS$ implies 
that there are 
 $\SessionTypeT' \ssubt \SessionTypeT$ and $\SessionTypeS' \ssubt \co\SessionTypeS$, %
    with either $\SessionTypeT' = \SessionTypeT$ %
    or $\SessionTypeS' = \co\SessionTypeS$, %
  such that
\begin{equation*}
  \new\ChannelA\ChannelB(
    \cprocess(\ChannelA,\SessionTypeT')
    \parop
    \cprocess(\ChannelB,\SessionTypeS')
    \parop \queue \ChannelB \ChannelA\EmptyQueue
    \parop \queue \ChannelA \ChannelB\EmptyQueue
  )
  \wreda
  \error
\end{equation*}
Hence, the subtyping relation $\asubt$ %
is complete for the asynchronous calculus, %
according to Definition~\ref{def:preciseness}.
\end{Proof}

We end this section with two examples showing why, %
in the proof for Theorem~\ref{thm:preciseness:async}, %
for rules \rulename{n-bra-async} and \rulename{n-sel-async} we need to
build characteristic processes of \emph{subtypes} of the current types
$\SessionTypeT$ and $\dualf\SessionTypeS$.
By highlighting these two cases, %
  we will then discuss the existence %
  of other sound asynchronous subtypings and a further result on $\asubt$   %
  (Theorem~\ref{lem:asubt-largest}).%

\begin{example}\label{ex:n-bra-async-smaller-type}%
 {\em If we take:
  \[
  \SessionTypeT=\trec\tvar.\big(\Out{\Tag_1}{\Type_1}.\tvar\,\oplus\,\Out{\Tag_2}{\Type_2}.\In{\Tag_3}{\Type_3}.\End\big)%
  \qquad\text{and}\qquad%
  \SessionTypeS=\In{\Tag_3}{\Type_3}.\big(\trec\tvar.\Out{\Tag_1}{\Type_1}.\tvar\,\oplus\,\Out{\Tag_2}{\Type_2}.\End\big)%
  \]
  then $\SessionTypeT\nasbut\SessionTypeS$ %
  (by rule \rulename{n-bra-async}), %
  but $\new\ChannelA\ChannelB(
  \cprocess(\ChannelA,\SessionTypeT)
  \parop
  \cprocess(\ChannelB,\dualf\SessionTypeS)
  \parop \queue \ChannelB \ChannelA\EmptyQueue
  \parop \queue \ChannelA \ChannelB\EmptyQueue
  )$ does \emph{not} reduce to $\error$. 
  In fact, %
  the process $\cprocess(\ChannelA,\SessionTypeT)$ either sends an
  $\Tag_1$-labelled message and becomes
  $\cprocess(\ChannelA,\SessionTypeT)$ again, or it sends an
  $\Tag_2$-labelled message, receives an $\Tag_3$-labelled message and
  stops. The process $\cprocess(\ChannelB,\dualf\SessionTypeS)$
  sends one $\Tag_3$-labelled message, and then can receive either an
  $\Tag_1$-labelled message and become
  $\cprocess(\ChannelB,\dualf\SessionTypeS)$ again, or an
  $\Tag_2$-labelled message and stop. %
  Hence, %
    the $\Tag_3$-labelled message can always be potentially dequeued
    from $\ChannelB\ChannelA$ %
    (after $\cprocess(\ChannelA,\SessionTypeT)$ chooses to %
    output $\Tag_2$), %
    and this ensures that \rulename{err-orph-mess-async} never holds.

  Instead taking\; %
  $\SessionTypeT'=\trec\tvar.\Out{\Tag_1}{\Type_1}.\tvar$ %
  \;we get\;
  $\SessionTypeT'\asubt\SessionTypeT$, %
  \;and
  \[
  \new\ChannelA\ChannelB\big(
  \cprocess(\ChannelA,\SessionTypeT')
  \parop
  \cprocess(\ChannelB,\dualf\SessionTypeS)
  \parop \queue \ChannelB \ChannelA\EmptyQueue
  \parop \queue \ChannelA \ChannelB\EmptyQueue
  \big)\;\;\wreda\;\;\error
  \]
since $\ChannelA\not\in\sbnF( \cprocess(\ChannelA, \SessionTypeT'))$, i.e. $\cprocess(\ChannelA, \SessionTypeT')$ cannot read the $\Tag_3$-message sent by $\cprocess(\ChannelB,\dualf\SessionTypeS)$.}
\end{example}

\begin{example}\label{ex:n-sel-async-smaller-type}%
  {\em This example is the ``dual'' of Example~\ref{ex:n-bra-async-smaller-type}. %
  If we take:
  \[
  \SessionTypeT=\trec\tvar.\Out{\Tag_1}{\Type_1}.\big(\In{\Tag_2}{\Type_2}.\tvar\,\,\&\,\,\In{\Tag_3}{\Type_3}.\tvar\big)%
  \qquad\text{and}\qquad%
  \SessionTypeS=\trec\tvar.\big(\In{\Tag_2}{\Type_2}.\tvar\,\,\&\,\,\In{\Tag_3}{\Type_3}.\Out{\Tag_1}{\Type_1}.\tvar\big)%
  \]
  then $\SessionTypeT\nasbut\SessionTypeS$ %
  (by rule \rulename{n-sel-async}), %
  but $\new\ChannelA\ChannelB(
  \cprocess(\ChannelA,\SessionTypeT)
  \parop
  \cprocess(\ChannelB,\dualf\SessionTypeS)
  \parop \queue \ChannelB \ChannelA\EmptyQueue
  \parop \queue \ChannelA \ChannelB\EmptyQueue
  )$ does \emph{not} reduce to $\error$. %
  Instead taking\; %
  $\SessionTypeS'=\trec\tvar.\In{\Tag_2}{\Type_2}.\tvar$ %
  \;we get\; %
  $\dualf{\SessionTypeS'}\asubt\dualf{\SessionTypeS}$, %
  \;and %
  \[
  \new\ChannelA\ChannelB\big(
  \cprocess(\ChannelA,\SessionTypeT)
  \parop
  \cprocess(\ChannelB,\dualf{\SessionTypeS'})
  \parop \queue \ChannelB \ChannelA\EmptyQueue
  \parop \queue \ChannelA \ChannelB\EmptyQueue
  \big)\;\;\wreda\;\;\error%
  \]}
\end{example}

Examples~\ref{ex:n-bra-async-smaller-type} %
and~\ref{ex:n-sel-async-smaller-type} show that %
there exist sound asynchronous subtyping relations %
that are \emph{not} sub-relations of $\asubt$. %
For instance, %
take $\SessionTypeT$ and $\SessionTypeS$ from %
Example~\ref{ex:n-bra-async-smaller-type}, %
and let $\asubtB$ be the smallest reflexive relation between session types %
such that $\SessionTypeT \asubtB \SessionTypeS$. %
We can verify that $\asubtB$ is a sound subtyping, %
by Definition~\ref{def:preciseness}: %
if we take any $\SessionTypeT' \asubtB \SessionTypeS'$ %
and we compose two processes typed %
by $\SessionTypeT'$ and $\co{\SessionTypeS'}$, %
they will not reduce to error. %
Note, however, that $\asubtB$ is \emph{not} a sub-relation of $\asubt$, %
because $\SessionTypeT \asubtB \SessionTypeS$ %
but $\SessionTypeT \not\asubt \SessionTypeS$: %
this is unlike the synchronous calculus, %
where \emph{all} sound subtypings are sub-relations of $\ssubt$
(see proof of Corollary~\ref{lem:ssubt-largest}). %
However, $\asubt$ has an important property: %
if we only consider the asynchronous subtypings %
that extend $\ssubt$, %
then $\asubt$ is the unique precise one. %

\begin{thm}
  \label{lem:asubt-largest}%
  $\asubt$ is the unique 
  precise subtyping for the
  asynchronous calculus %
  that extends $\ssubt$.
\end{thm}
\begin{Proof}
  \def\asubti{\annotaterel{\sqsubseteq}{a}}%
  Take %
  a reflexive and transitive relation $\asubti$ %
  such that %
  $\mathord{\ssubt} \subseteq \mathord{\asubti}%
  \not\subseteq \mathord{\asubt}$ %
  --- i.e., there exist %
  $\SessionTypeT, \SessionTypeS$ %
  such that $\SessionTypeT \asubti \SessionTypeS$ %
  but $\SessionTypeT \not\asubt \SessionTypeS$. %
  We prove that %
  $\asubti$ is an unsound subtyping, %
  by showing that %
  there exist %
  $\SessionTypeT' \asubti \SessionTypeT$ %
  and $\SessionTypeS' \asubti \co\SessionTypeS$ such that
  \begin{equation*}
    \label{eq:proof-asubt-larger}%
    \new\ChannelA\ChannelB(
    \cprocess(\ChannelA,\SessionTypeT')
    \parop
    \cprocess(\ChannelB,\SessionTypeS')
    \parop \queue \ChannelB \ChannelA\EmptyQueue
    \parop \queue \ChannelA \ChannelB\EmptyQueue
    )
    \wreda
    \error
  \end{equation*}
  Since (by Lemma~\ref{pro:s:negation:async}) %
  $\SessionTypeT \not\asubt \SessionTypeS$ %
  implies $\SessionTypeT \nasbut \SessionTypeS$, %
  we proceed by induction %
  on the derivation of $\SessionTypeT \nasbut \SessionTypeS$, %
  similarly to the proof of Theorem~\ref{thm:preciseness:async}. %

  From such a proof, %
  we can see that in all cases %
  we can get 
  the error reduction above %
  for some %
  $\SessionTypeT',\SessionTypeS'$ %
  such that either $\SessionTypeT' \ssubt \SessionTypeT$
  and $\SessionTypeS' = \co{\SessionTypeS}$, or
  $\SessionTypeT' = \SessionTypeT$ and $\SessionTypeS' \ssubt
  \co{\SessionTypeS}$. %
  In the first case, we also have %
  $\SessionTypeT' \asubti \SessionTypeT$ %
  and $\SessionTypeS' \asubti \co\SessionTypeS$ %
  (because $\mathord{\ssubt} \subseteq \mathord{\asubti}$):
  hence, %
  we conclude that $\asubti$ is an unsound subtyping
  according to Definition~\ref{def:preciseness}. %
  The proof for the second case %
  ($\SessionTypeT' = \SessionTypeT$ and $\SessionTypeS' \ssubt
  \co{\SessionTypeS}$) %
  is dual.

  We conclude %
  that if $\asubti$ is sound %
  and %
  $\mathord{\ssubt} \subseteq \mathord{\asubti}$, %
  then $\mathord{\asubti} \subseteq \mathord{\asubt}$; %
  hence, among all subtypings that extend $\ssubt$, %
the subtyping $\asubt$ is the largest sound one %
  for the asynchronous calculus, %
  and therefore the unique precise one.
\end{Proof}

\section{Extensions}\label{ext}

In the original calculus~\cite{HVK} sessions are initialised using shared channels by request/accept prefixes and also expressions (including shared channels) can be communicated. In this section we show that preciseness is preserved when we augment the calculus, the types (adding sorts, following Honda \etal~\cite{HVK}) and the subtyping (following Demangeon and Honda~\cite{DemangeonH11}), both for the synchronous and for the asynchronous cases. The most challenging issue in this extension is the definition of characteristic processes. For the communication of expressions, taking inspiration from Ligatti \etal~\cite{BHLN12}, we add expression constructors distinguishing values of different sorts. To correctly deal with communication of shared channels the characteristic processes must contain both accepts and requests. If a shared channel carries a linear channel of type $S$, an accept can be typed with a linear channel of type $T \subt S$, while a request can be typed with a linear channel of type
$T \subt\co S$, see Table~\ref{tab:esync:typing}. Since $T \subt \co S$ is equivalent to $S \subt \co T$, typing the parallel composition of accept and request is invariant, as defined by Demangeon and Honda~\cite{DemangeonH11}.

\newpage

\subsection{Synchronous Communication}\label{esc}$\;$

\mypar{Syntax and operational semantics}

Table~\ref{tab:esync:syntax} shows the extended synchronous session calculus obtained by
adding session initialisations and communications of expressions (including shared channels) to the synchronous calculus of Table~\ref{tab:sync:syntax}.   We also add conditionals in order to get evaluation of expressions in reducing characteristic processes. 

The value $\val$ of expression $\e$ (notation $\eval\e\val$) is computed according to the rules of Table~\ref{tab:evaluation}.  We use $\valn$ to range over natural and $\valr$ to range over integer numbers. 
Evaluation contexts $\Econtext$ for expressions are defined by:
$$
\Econtext :: = [ \  ] \parop \neg(\Econtext) \parop \fneg\Econtext \parop 
\fsucc\Econtext  \parop 
\Econtext > \e \parop 
\val > \Econtext
$$
An expression $\e$ is {\em stuck} (notation $\neval\e$\,) if it does not evaluate to a value according to the rules of Table~\ref{tab:evaluation}. %
Note that $\succf$ reduces only if the argument is a natural number. We use $\eval{\tilde\e}{\tilde\val}$ and $\neval{\tilde\e}$ with the obvious meanings. 

We extend the structural congruence of synchronous processes (Table~\ref{tab:sync:congruence}) and the evaluation contexts (see page \pageref{pc}) in the obvious way. 
To reduce extended processes we add the rules of Table~\ref{tab:esync:red} to the rules of Table~\ref{tab:sync:red}. Table~\ref{tab:esync:red} takes into account session initialisations, communication of expressions, conditionals and expressions in definitions. 

A process containing a stuck expression reduces to $\error$, as well as a process with a value in a channel position, or with a shared channel in a linear channel position, or vice versa. This is prescribed by the rules of Table~\ref{tab:esync:ered}, which are added to the rules of Table~\ref{tab:sync:red:err}. 

\begin{table}
{\small%
\[
\begin{array}[t]{c}
\begin{array}[t]{@{}rcl@{\quad}l@{}@{}rcl@{\quad}l@{}}
  \Process \hspace{-3mm}& ::= \hspace{-3mm}& & \textbf{Process} \\
  &   & \vdots & \text{from Table~\ref{tab:sync:syntax}} &
  & | & \send\NameU\Tag{\e}.\Process & \text{(expression output)} \\
  & | & \acc\NameU\var\Process & \text{(session accept)} &
   & | &  \req\NameU\var \Process & \text{(session request)}\\
  & | & \res\s\Process & \text{(shared channel restriction)}&
   & | &   \cond\e\Process \Process & \text{(conditional)}\\
   &| & \invoke\pvar{\tilde\e\tilde\Expression} & \text{(variable with expressions)} 
 \end{array}
 \\\\
     \begin{array}[t]{@{}rcl@{\quad}l@{}}
\NameU & ::= & & \textbf{Identifiers} \\
 &   & \vdots & \text{from Table~\ref{tab:sync:syntax}} \\
   & | & \val & \text{(value)}
 \end{array}
\\\\
 \begin{array}[t]{@{}rcl@{\quad}l@{}@{}rcl}
  \val& ::= & \textbf{Value} \\
     &  & \s & \multicolumn{3}{c}{\text{(shared channel)}} \\ 
 & | &  \true&
 & | & \false\\
  & | & 0&
  & | & 1\\
  & | &-1&
  & | &\ldots
\end{array}
\qquad
\begin{array}[t]{@{}rcl@{\quad}l@{}@{}rcl}
  \e & ::= & \textbf{Expression} \\
  &   & \val & 
  & | & \var & \\
   & | & \mkkeyword{\neg} \e&
   & | & \fsucc \e\\
   & | & \fneg \e&
   & | & \e> \e
    \end{array}
\end{array}
\]
}
\caption{\label{tab:esync:syntax} Syntax of extended synchronous processes.}
\end{table}
\begin{table}
\centerline{$\begin{array}[t]{@{}c@{}}
\eval{\mkkeyword{\neg}\true}\false \quad \eval{\mkkeyword{\neg}\false}\true 
\qquad\qquad
\eval{\fsucc\valn}(\valn +1)
\qquad\qquad
\eval{\fneg\valr}(-\valr)
\\[5.5mm]
\eval{(\valr_1>\valr_2)}{\begin{cases}
 \true     & \text{if }\valr_1>\valr_2, \\
  \false    & \text{otherwise}
\end{cases}} \qquad\qquad
\inferrule[]{\eval{\e}{\val}\quad\eval{\Econtext(\val)}{\val'}}{\eval{\Econtext(\e)}{\val'}}
\qquad\qquad\eval\val\val
\end{array}
$}
\caption{\label{tab:evaluation}  Expression evaluation.}
\end{table}
\begin{table}
\[
\begin{array}[t]{@{}c@{}}
  \inferrule[\rulename{r-init-sync}]{ a, b\text{ fresh}}
  {\acc\s\var\Process\parop\req\s y\ProcessQ\red  \new \ChannelA \ChannelB (\Process \subst\ChannelA\var\parop\ProcessQ \subst\ChannelB y)}
 \\\\
  \inferrule[\rulename{r-com-sync-ext}]{
    k \in I\quad\eval\e\val
  }{
    \new \ChannelA \ChannelB (
      \send\ChannelA {\Tag_k} \e.\Process
      \parop
      \sum_{i\in I} \receive\ChannelB{\Tag_i}{\var_i}.\ProcessQ_i
    )
    \red
    \new \ChannelA \ChannelB (
      \Process
      \parop
      \ProcessQ_k \subst{\val}{\var_k}
    )
  }
  \\\\
   \inferrule[\rulename{r-t-cond}]{
   \eval{\e}{\true}}{
    \cond{\e}{\ProcessP}{\ProcessQ}  \red \ProcessP
   }  
  \qquad\qquad
    \inferrule[\rulename{r-f-cond}]{
   \eval{\e}{\false}}{
    \cond{\e}{\ProcessP}{\ProcessQ}  \red \ProcessQ
   }  
      \\\\
 \inferrule[\rulename{r-def-ext}]{\eval{\tilde\e}{\tilde\val}}
    {\Def\pvar{\tilde\var\tilde y}\ProcessP{
      (\invoke\pvar{\tilde\e\tilde \ChannelA}
      \parop
      \ProcessQ)
    }
    \red
    \Def\pvar{\tilde\var\tilde y}\ProcessP{
      (\ProcessP
      \subst{\tilde\val}{\tilde \var}\subst{\tilde\ChannelA}{\tilde y} 
      \parop
      \ProcessQ)
    }
  }

    \end{array}
\]
\caption{\label{tab:esync:red} 
Reduction of extended synchronous processes.
}
\end{table}
\begin{table}
\[
\begin{array}[t]{@{}c@{}}
\inferrule[\rulename{err-def}]{\neval{\tilde\e}{}}
{\invoke\pvar{\tilde\e\tilde \ChannelA}\red
    \error  }
  \qquad
  \inferrule[\rulename{err-com-ext}]{
    \neval\e
  }{
          \send\ChannelA {\Tag} \e.\Process
    \red
    \error  }
  \qquad
   \inferrule[\rulename{err-cond}]{
   \neval{\e}}{
    \cond{\e}{\ProcessP}{\ProcessQ}  \red \error
   }    \\\\
 \inferrule[\rulename{err-chan-in}]{} 
   {\sum_{i\in I} \receive\val{\Tag_i}{\var_i}.\Process_i  \red \error}
    \quad\quad
    \inferrule[\rulename{err-chan-out}]{}
   {\send\val\Tag{\Expression'}.\Process\red \error} 
   \quad\quad
    \inferrule[\rulename{err-acc}]{\NameU=\val\text{ or }  \NameU=\ChannelA} 
   {\acc\NameU\var\Process  \red \error}
    \quad\quad
    \inferrule[\rulename{err-req}]{\NameU=\val\text{ or }  \NameU=\ChannelA} 
   {\req\NameU\var\Process  \red \error}
  \end{array}
\]
\caption{\label{tab:esync:ered} 
Error reduction of extended synchronous processes. 
}
\end{table}
\mypar{Type system}
{\em Sorts}  (ranged over by $\B$) and {\em extended session types} (ranged over by $\T$) are defined by:
$$\begin{array}{lll} \B      & ::=   &               \tbool     \sep     \tnat \sep \tint \sep \sct\T\\
\SessionType & ::= & 
\Branch_{i\in I} \In{\Tag_i}{\U_i}.\SessionTypeT_i 
\ | \ 
\Select_{i\in I} \Out{\Tag_i}{\U_i}.\SessionTypeT_i
\ | \ 
\tvar 
\ | \ 
\trec\tvar.\SessionType
\ | \ 
 \End \\
 \U& ::= &\B\sep \T
\end{array}$$
where $\sct\T$ is the sort of shared channels binding linear channels of extended session type $\T$. 

{\em Subsorting }$\subs$ on sorts is the minimal reflexive and transitive closure of the relation induced by the rule:
  $\tnat \subs \tint$. Table~\ref{tab:te} gives the 
(expected) typing rules for expressions.

The synchronous subtyping rules for extended session types are obtained from the rules of Table~\ref {tab:sync:ssubt} by replacing $\SessionTypeS$ with
 $\U$, and by defining:%
\[
\U \,\subt\, \U' \;=\; \begin{cases}
\SessionTypeT \subt \SessionTypeT'      & \text{if }\U=\T\text{ and } \U'=\T', \\
\B'\subs\B     & \text{if }\U=\B\text{ and } \U'=\B',\\
\true     & \text{if }\U=\B\text{ and } \U'=\End,\\
\false    & \text{otherwise}.
\end{cases}
\]
Notice that processes do not contain occurrences of linear channels typed by $\End$, so any value can be sent to a process waiting for a  
linear channel typed by $\End$. This justifies the value $\true$ in
the definition of $\U\subt\U'$. Notice also that the extended session types and the sorts behave in opposite ways for inputs and outputs. 

{\em Shared environments} 
associate identifiers to sorts and process variables to sequences of sorts and extended session types
$$
\UEnv:: = 
\emptyset \mid \UEnv, \NameU:\B\mid 
\UEnv, \pbind{\pvar}{\tilde\B\tilde\SessionType}$$
 The typing rules for extended synchronous processes  are given in Tables~\ref{tab:sync:typing} and~\ref{tab:esync:typing}.
\begin{table}
\centerline{$
\begin{array}{c}
\der{\Gamma}{\true}{\tbool} \qquad \der{\Gamma}{\false}{\tbool}
  \qquad
  \der{\Gamma}{\valn}{\tnat} \qquad \der{\Gamma}{\valr}{\tint}
  \qquad
    \der{\Gamma,\NameU:\B}{\NameU}{\B}  
  \\ \\
  \myrule{\der{\Gamma}{\e}{\tbool}}{\der{\Gamma}{\mkkeyword{\neg}\e}{\tbool}}{}
   \qquad
  \myrule{\der{\Gamma}{\e}{\tnat}}{\der{\Gamma}{\fsucc\e}{\tnat}}{}
  \qquad
  \myrule{\der{\Gamma}{\e}{\tint}}{\der{\Gamma}{\fneg\e}{\tint}}{}
  \\ \\
  \myrule{\der{\Gamma}{\e_1}{\tint}\quad\der{\Gamma}{\e_2}{\tint}}{\der{\Gamma}{\e_1>\e_2}{\tbool}}{}
  \qquad
  \myrule{\der{\Gamma}{\e}{\B}\quad\B\subs\B'}{\der{\Gamma}{\e}{\B'}}{}
  \end{array}
  $}
  \caption{Typing rules for expressions.}
\label{tab:te}
 \end{table}

 \begin{table}
{\small%
\[\begin{array}{c}
  \inferrule[\rulename{t-acc}]{\wtp{\Gamma,\NameU:\sct\T}{\Process}{\LEnv,\bind{\var}{\SessionTypeT}}}
  {\wtp{\Gamma,\NameU:\sct\T}{\acc{\NameU}\var\Process}{\LEnv}}
   \qquad
    \inferrule[\rulename{t-req}]{\wtp{\Gamma,\NameU:\sct\T}{\Process}{\LEnv,\bind{\var}{\co\SessionTypeT}}}
  {\wtp{\Gamma,\NameU:\sct\T}{\req{\NameU}\var\Process}{\LEnv}}
   \qquad
    \inferrule[\rulename{t-res}]{\wtp{\Gamma,\s:\sct\T}{\Process}{\LEnv}}
  {\wtp{\Gamma}{\res{\s}\Process}{\LEnv}}
\\[5.5mm]
   \inferrule[\rulename{t-out-ext}]{\der\Gamma\e\B\quad  \wtp{\Gamma}{\Process}{\LEnv,\ASETT{\bind{\NameU}{\SessionTypeT}}}}{\wtp\Gamma{{\send {\NameU}  l \e  .\Process}}{\LEnv,
        \bind{\Name}{\Out{\Tag}{\B}.\SessionTypeT}}
}
   \qquad
 \inferrule[\rulename{t-cond}]{
  \der\Gamma\e\tbool\quad\wtp{\Gamma}{\Process_1}{\LEnv}\quad\wtp{\Gamma}{\Process_2}{\LEnv}}{\wtp\Gamma{{\cond{\e}{\Process_1}{\Process_2}}}{\LEnv}}
\\[5.5mm] 
 \inferrule[\rulename{t-var-ext}]{\der\UEnv{\tilde\e}{\tilde\B}}
    {\wtp{
    \UEnv,
    \pbind{\pvar}{\tilde\B\tilde\SessionType}
    }{
      \invoke\pvar{\tilde\e\tilde\Name}
    }{\ASET{\bind{\tilde \Name}{\tilde \SessionType}}}
  }
    \qquad
\inferrule[\rulename{t-def-ext}]{
    \wtp{
      \UEnv,\tilde\var:\tilde\B,
      \pbind{\pvar}{\tilde\B\tilde\SessionType}
    }{
      \ProcessP
    }{
        \ASET{\bind{\tilde y}{\tilde\SessionType}}
    }
    \\
    \wtp{
      \UEnv,
      \pbind{\pvar}{\tilde\B\tilde\SessionType}
    }{
      \ProcessQ
    }{
      \LEnv
    }
  }{
    \wtp{\UEnv}{
      \Def{\pvar}{\tilde\var\tilde y}{\ProcessP}{\ProcessQ}
    }{
      \LEnv
    }
  }
\end{array}\]
}
\caption{\label{tab:esync:typing} Typing rules for extended synchronous processes.}
\end{table}

\mypar{Preciseness}
The characteristic processes are  the processes of Definitions~\ref{s:cprocesses} and~\ref{s:ecprocesses}. For $\tbool,\tnat$ and $\tint$ we use conditionals and the constructors $\neg$, $\mkkeyword{succ}$ and $\mkkeyword{neg}$ in order to test values. For the sorts of shared channels both input and output characteristic processes contain accept and request constructors. If $\SessionTypeS$ is different from $\T$, then either $\SessionTypeS$ is not a subtype of $\T$ or vice versa. Therefore at least one of the session initialised by a shared channel of type $\sct\SessionTypeS$ and a shared channel of type $\sct\SessionTypeT$ will reduce to $\error$, see the last case of the proof of Theorem~\ref{thm:preciseness:easync}.
\begin{defi}[Characteristic extended synchronous processes]
\label{s:ecprocesses}$$\begin{array}{lll}
\cprocess[?](\Name,\Tag,\tbool,\SessionType)
 &\eqdef&
   \receive{\Name}{\Tag}{\var}.\cond{\mkkeyword{\neg}\var}
    {\cprocess(\Name,\SessionType)}
    {\cprocess(\Name,\SessionType)}
\\[1mm]
\cprocess[?](\Name,\Tag,\tnat,\SessionType)
 &\eqdef&
   \receive{\Name}{\Tag}{\var}.\cond{\fsucc\var>0}
    {\cprocess(\Name,\SessionType)}
    {\cprocess(\Name,\SessionType)}
\\[1mm]
\cprocess[?](\Name,\Tag,\tint,\SessionType)
 &\eqdef&
   \receive{\Name}{\Tag}{\var}.\cond{\fneg\var>0}
    {\cprocess(\Name,\SessionType)}
    {\cprocess(\Name,\SessionType)}
\\[1mm]
\cprocess[?](\Name,\Tag,\sct\SessionTypeS,\SessionType)
 &\eqdef&
   \receive{\Name}{\Tag}{\var}.(
    \cprocess(\Name,\SessionType)\parop\acc\var y
    \cprocess(y,\SessionTypeS)
    \parop\req\var z
    \cprocess(z,\co\SessionTypeS))
\\[1mm]
\cprocess[!](\Name,\Tag,\tbool,\SessionType)
 &\eqdef&
      \send{\Name}{\Tag}{\true}.\cprocess(\Name,\SessionType)
      \\[1mm]
\cprocess[!](\Name,\Tag,\tnat,\SessionType)
 &\eqdef&
      \send{\Name}{\Tag}{5}.\cprocess(\Name,\SessionType)
      \\[1mm]
\cprocess[!](\Name,\Tag,\tint,\SessionType)
 &\eqdef&
      \send{\Name}{\Tag}{-5}.\cprocess(\Name,\SessionType)
            \\[1mm]
\cprocess[!](\Name,\Tag,\sct\SessionTypeS,\SessionType)
 &\eqdef&\res\s(
      \send{\Name}{\Tag}{\s}.(\cprocess(\Name,\SessionType)
      \parop\acc\s y \cprocess(y,\SessionTypeS)\parop\req\s z \cprocess(z,\co{\SessionTypeS})))
    \end{array}
$$
\end{defi}

\bigskip

As for subtyping, 
the  negation of extended synchronous subtyping is obtained from the rules of Table~\ref{tab:negsubtype} by replacing $\SessionTypeS$ with $\U$, %
and by defining:
\[
\U \,\nsubt\, \U' \;=\; \begin{cases}
\SessionTypeT \nsubt \SessionTypeT'      & \text{if }\U=\T\text{ and } \U'=\T', \\
\B'\not\subs\B     & \text{if }\U=\B\text{ and } \U'=\B',\\
\false     & \text{if }\U=\B\text{ and } \U'=\End,\\
\true    & \text{otherwise}.
\end{cases}
\]
Lemmas~\ref{cpts} and~\ref{lem:s:negation} easily extend to these definitions, i.e. we get $\wtps{\;}{\cprocess(\Name,\SessionType)}{\ASET{\Name:\SessionType}}$ and if $S\ssubt T$ is not derivable, then  $S \nssubt T$ is derivable. We are now ready to show preciseness. 

\begin{thm}[Preciseness for extended synchronous subtyping]\label{thm:preciseness:esync}
The extended synchronous subtyping relation is precise for the extended synchronous calculus. 
\end{thm}
\begin{Proof}
As in previous cases soundness follows from subject reduction, which can be easily proved. For completeness the only new cases are applications of rules \rulename{n-exch-bra} and \rulename{n-exch-sel} with sorts. We consider two paradigmatic cases here and other two paradigmatic cases in the proof of preciseness for the extended asynchronous calculus (Theorem~\ref{thm:preciseness:easync}). 

\smallskip

{\em Case} \rulename{n-exch-bra}: 
$\SessionTypeT =  
\Branch_{i\in I} \In{\Tag_i}{\U_i}.\SessionTypeT_i$,
$\SessionTypeS = 
\Branch_{j\in J} \In{\Tag'_j}{\U'_j}.\SessionTypeS_j$, 
and $\exists k \in I~\exists k' \in J$ such that $\Tag_{k} = \Tag'_{k'}$
and $ \U_k \nssubt \U'_{k'}$. We only consider the case $ \U_k=\tnat$ and $\U'_{k'}=\tint$. 
$$
\begin{array}{l}
\new \ChannelA \ChannelB
    ( 
    \cprocess(\ChannelA,\SessionTypeT)
    \parop
    \cprocess(\ChannelB, \co \SessionTypeS) 
    )
\reds 
\\
\new \ChannelA \ChannelB
(   
    \sum_{i \in I \setminus \ASET{k}}
    \cprocess[?](\ChannelA,\Tag_i,\SessionTypeT'_i,\SessionTypeT_i)
  \ + \    
     \receive{\ChannelA}{\Tag_k}{\var}.\cond{\fsucc\var>0}
    {\cprocess(\ChannelA,\SessionType_k)}
    {\cprocess(\ChannelA,\SessionType_k)}
     \parop \\ \hfill
      \send{\ChannelB}{\Tag_k}{-5}.\cprocess(\ChannelB,\co{\SessionTypeS_{k'}}))
 \reds\\
\Context
[\cond{\fsucc{-5}>0}
    {\cprocess(\ChannelA,\SessionType_k)}
    {\cprocess(\ChannelA,\SessionType_k)}
]
\end{array}
$$
where $\Context\hole=\new \ChannelA \ChannelB    
(    \cprocess(\ChannelB, \co{\SessionTypeS_{k'}})\parop
   \hole
      )$.
Being $\fsucc{-5}>0$ stuck, by rule  \rulename{err-cond} \[\cond{\fsucc{-5}>0}
    {\cprocess(\ChannelA,\SessionType_k)}
    {\cprocess(\ChannelA,\SessionType_k)}\reds \error\] then by rule \rulename{err-context}, we conclude
$$
\Context
[ \cond{\fsucc{-5}>0}
    {\cprocess(\ChannelA,\SessionType_k)}
    {\cprocess(\ChannelA,\SessionType_k)}
]
\reds \error$$

\smallskip

{\em Case} \rulename{n-exch-sel}: 
$\SessionTypeT = \Select_{i\in I} \Out{\Tag_i}{\U_i}.\SessionTypeT_i$,
$\SessionTypeS =  \Select_{j\in J} \Out{\Tag'_j}{\U'_j}.\SessionTypeS_j$, 
and $\exists k \in I~\exists k' \in J$ such that $\Tag_{k} = \Tag'_{k'}$ and
$\U'_{k'} \nssubt \U_{k}$. We only consider the case $ \U_k=\tbool$ and $\U'_{k'}=\T'\not=\End$.
$$
\begin{array}{l}
\new \ChannelA \ChannelB
    (\cprocess(\ChannelA,\SessionTypeT)
    \parop
    \cprocess(\ChannelB, \co \SessionTypeS) )  \reds
\\ 
\new \ChannelA \ChannelB
(  
    \send{\ChannelA}{\Tag_{k}}{\true}.
    \cprocess(\ChannelA,\SessionTypeT_{k})
    \parop
    \sum_{j \in J \setminus \ASET{k'}}
    \cprocess[?](\ChannelB,\Tag'_j,\U'_j,\co{\SessionTypeS_j}) +
    \receive{\ChannelB}{\Tag_{k}}{\var}.
    (\cprocess(\ChannelB,\co{\SessionTypeS_{k'}}) \parop
    \cprocess(\var, \T') )
)  \reds
\\
\Context[
\cprocess( \true, \T'  )
]
\end{array}
$$
where $\Context\hole=\new \ChannelA \ChannelB
(
    \cprocess( \ChannelA,\SessionTypeT_{k} ) 
    \parop
    \cprocess(\ChannelB,\co{\SessionTypeS_{k'}} ) \parop
    \hole
    )$. 
By definition of characteristic process, being $\T'\not=\End$, the value $\true$ is used as a channel in $\cprocess( \true, \T'  )$, and this implies by rule \rulename{err-chan-in} or  \rulename{err-chan-out} 
$$ \cprocess( \true, \T'  )\reds \error$$ then by rule \rulename{err-context}, we conclude
$$
\Context
[ \cprocess( \true, \T'  )
]
\reds \error\vspace{-18 pt}$$
\end{Proof}

\subsection{Asynchronous Communication}\label{eac}$\;$

\mypar{Syntax and operational semantics}
The processes of the extended asynchronous session calculus are generated by the rules of Table~\ref{tab:async:syntax}, where messages can be of the form $\msg\Tag\val$, and by the rules of Table~\ref{tab:esync:syntax}. 

The rules for evaluating expressions remain those of Table~\ref{tab:evaluation}. Structural congruence and evaluation contexts are generalised in the obvious way. 
The reduction rules for the extended asynchronous processes are
obtained from the reduction rules of asynchronous processes  of
\S~\ref{sec:asynchronous_language} %
by adding the rules  \rulename{r-t-cond}, \rulename{r-f-cond}, \rulename{r-def-ext} of Table~\ref{tab:esync:red} and the rules of Table~\ref{tab:easync:red}.
The mapping $\sbnD$ is extended in the obvious way
wrt.~Table~\ref{sbnD} %
--- i.e., 
adding the cases:
$$
\begin{array}{rcl}
\sbnD(\invoke\pvar{\tilde\e\tilde\Expression},\tilde D,\chi)      & = &      \begin{cases}
\sbnD(\ProcessP\subst{\tilde\e\tilde\Expression}{\tilde\var \tilde y},\tilde D,\chi\cdot\invoke\pvar{\tilde\e\tilde\Expression})       &      \text{if } \invoke\pvar{\widetilde{\e_0}\tilde\Expression}\not\in\chi \text{ and }\\ &\invoke\pvar{\tilde\var \tilde y}=\ProcessP\in\tilde D\\
    \emptyset      &        \text{otherwise}
\end{cases}\end{array}$$
$$\begin{array}{lllll}
  \sbnD(\acc\NameU\var\Process,\, \tilde D,\, \chi) &=&\sbnD(\req\NameU\var\Process,\, \tilde D,\, \chi) &=&%
  \sbnD(\Process,\, \tilde D,\, \chi) \setminus \{\var\}\\
  \sbnD(\send\NameU\Tag{\e}.\Process,\, \tilde D,\, \chi) &=&%
  \sbnD(\res\s\Process,\, \tilde D,\, \chi) &=&%
  \sbnD(\Process,\, \tilde D,\, \chi)
  \end{array}
$$
  $$
\begin{array}{rcl}
  \sbnD(\cond{\e}{\Process_1}{\Process_2},\, \tilde D,\, \chi) &=&%
  \sbnD(\ProcessP_1,\, \tilde D,\, \chi) \;\cup\; \sbnD(\ProcessP_2,\, \tilde D,\,
  \chi)
\end{array}
$$
Notice that $\sbnD(\invoke\pvar{\tilde\e\tilde\Expression},\tilde D,\chi)=\emptyset$ if $\chi$ contains $\invoke\pvar{\widetilde{\e_0}\tilde\Expression}$ for some $\widetilde{\e_0}$, which can be different from $\tilde{\e}$.
\begin{table}
$$
\begin{array}[t]{@{}c@{}}
  \inferrule[\rulename{r-init-async}]{ a, b\text{ fresh}}
  {\acc\s\var\Process\parop\req\s y\ProcessQ\red  \new \ChannelA \ChannelB (\Process \subst\ChannelA\var\parop\ProcessQ \subst\ChannelB y
  \parop \queue \ChannelB \ChannelA \EmptyQueue 
    \parop
    \queue \ChannelA \ChannelB \EmptyQueue  )}
    \\\\
  \inferrule[\rulename{r-send-async-ext}]{\eval\e\val}{
      \queue\ChannelA\ChannelB \Queue
    \parop
    \send\ChannelA\Tag\e.\Process
        \red
      \queue\ChannelA\ChannelB \Queue \qconc \msg{\Tag}{\val}
    \parop
    \Process
      }
  \qquad
  \inferrule[\rulename{r-receive-async-ext}]{
    k\in I
  }{
    \textstyle
    \queue\ChannelA\ChannelB \msg{\Tag_k}{\val} \qconc \Queue
    \parop
    \sum_{i\in I} \receive\ChannelB{\Tag_i}{\var_i}.\Process_i
    \red
    \queue\ChannelA\ChannelB \Queue
    \parop
    \Process_k\subst\val{\var_k}
    }
\end{array}
$$
\caption{\label{tab:easync:red} Reduction of asynchronous processes.\strut}
\end{table}

The error reduction rules for extended asynchronous processes are obtained from the error reduction rules of asynchronous processes  of \S~\ref{sec:asynchronous_language} by adding the rules of Table~\ref{tab:esync:ered} and the rule
$$ \inferrule[\rulename{err-mism-async-ext}]
  { \forall i \in I : \Tag \not = \Tag_i
  }
  { 
        \queue\ChannelA\ChannelB \msg{\Tag}{\NameU} \qconc \Queue
        \parop
        \sum_{i\in I} \receive\ChannelB{\Tag_i}{\var_i}.\Process_i
        \red \error
  }$$
  
 \mypar{Type system} We define sorts, extended session types and shared environments as in \S~\ref{esc}. The typing rules for expressions remain those of Table~\ref{tab:te}. The queue types are those of \S~\ref{tap} by replacing $\SessionTypeS$ with $\U$, thus allowing $\tmsg{\Tag}\B$.   Also the definition of session remainder is obtained from that of \S~\ref{tap} by replacing $\SessionTypeS$ with $\U$.

 The asynchronous subtyping for extended session types is obtained by defining $\U\subt\U'$ as in \S~\ref{esc} and replacing $\SessionTypeS$ with $\U$  in the subtyping of \S~\ref{subsec:async}.
 The typing rules for extended asynchronous processes are obtained from the typing rules of asynchronous processes  of \S~\ref{tap} by adding the rules of Table~\ref{tab:esync:typing} and the following rule:
 $$ \inferrule[\rulename{t-message-q-v}]{\der{\Gamma}{\val}{\B}\quad
    \wtp{\UEnv}{
    \queue \ChannelB \ChannelA  \Queue
    }
    {
    \LEnv, \ASETT{\queuetype \ChannelB \ChannelA  \QueueType}
    }
  }{  
    \wtp{\UEnv}{
     \queue \ChannelB \ChannelA \Queue \qconc \msg{\Tag}{\val}
    }{ \LEnv,
       \queuetype \ChannelB \ChannelA 
       \QueueType  \qconc \tmsg{\Tag}{\B}}
    }
 $$
 
 The reduction of extended asynchronous session  environments is given by the rules of Table~\ref{tab:redtype} after the replacement of $\SessionTypeS$ with $\U$.
  \mypar{Preciseness} The characteristic processes are  the processes of Definition~\ref{async:charprocess} plus the processes of Definition~\ref{s:ecprocesses}.
   The  negation of asynchronous subtyping is obtained from the negation of Section~\ref{sec:a:completeness} by replacing $\SessionTypeS$ with $\U$  and by defining $\U\nsubt\U'$ as in~\ref{esc}.
   
   Lemmas~\ref{cpta} and~\ref{pro:s:negation:async} easily extend to these definitions, i.e. we get $\wtpa{\;}{\cprocess(\Name,\SessionType)}{\ASET{\Name:\SessionType}}$ and if $S\asubt T$ is not derivable, then  $S \nasbut T$ is derivable. The invariance of the types of shared channels is intriguing in the proof of completeness. We need  an auxiliary lemma.
 
 \begin{lem}\label{auxf}
 \begin{enumerate}
  \item\label{auxf1}
 If $T\nasbut S$ and $S \nasbut T$, then one of the two statements is derivable without using rules {\rm\rulename{n-bra-async}} and {\rm\rulename{n-sel-async}}.
 \item\label{auxf2}
 If $T\asubt S$ and $T\not=S$, then $S \nasbut T$ is derivable without using rules {\rm\rulename{n-bra-async}} and {\rm\rulename{n-sel-async}}.
 \end{enumerate}
 \end{lem}
 \begin{Proof} (\ref{auxf1}). Let assume we use \rulename{n-bra-async} in the proof of $T\nasbut S$. Then there  is a subderivation of $T\nasbut S$ such that \rulename{n-bra-async} is used to prove  
 $T'\nasbut S'$ or $S' \nasbut T'$. 
     Let $T'\nasbut S'$ be consequence of an application of
    rule  \rulename{n-bra-async}. We have two cases:
    \begin{itemize}
    \item%
      an application of rule \rulename{n-cont-async} with a
      non-empty asynchronous context occurs in the derivation branch 
      from $T'\nasbut S'$ to $T\nasbut S$.
      Let $T''\nasbut S''$ be the conclusion of the application of
      rule \rulename{n-cont-async} %
      (with a non-empty asynchronous context) %
      that is \emph{closest} to $T\nasbut S$. %
      Notice that the trees of $T'',S''$ are subtrees of those of
      $T,S$, respectively. We can then prove $S \nasbut T$ with a
      subderivation which shows $S'' \nasbut T''$ using rule
      \rulename{n-brasel};
    \item%
      otherwise, the derivation branch from $T'\nasbut S'$ to
      $T\nasbut S$ does \emph{not} contain any application of
      \rulename{n-cont-async} with a non-empty asynchronous context.
      In this case, the trees of $T',S'$ are subtrees of those of
      $T,S$, respectively. %
      Let $\&\not \in \SessionTypeT'$ and
      $S'=\Branch_{i\in I} \In{\Tag_i}{\Type_i}.\SessionTypeT_i$. Then
      $\SessionTypeT'$ is either $\End$ or a selection. We can prove
      $S\nasbut T$ with a subderivation which shows $S' \nasbut
      T'$.
      Taking into account the shapes of $T',S'$ the only applicable
      rules are either \rulename{n-end r} or \rulename{n-brasel}. 
    \end{itemize}
    The proof when we use \rulename{n-sel-async} %
    in the derivation of $T\nasbut S$  is similar.

 (\ref{auxf2}). The assumption $T\asubt S$ assures that $T$,$S$ have
 corresponding branchings and selections. Hence, rules {\rm\rulename{n-bra-async}} and {\rm\rulename{n-sel-async}} cannot be used in showing $S \nasbut T$.
 \end{Proof} 
   
  \begin{thm}[Preciseness for extended asynchronous subtyping]\label{thm:preciseness:easync}
The extended asynchronous subtyping relation is precise for the extended asynchronous calculus. 
\end{thm}
\begin{Proof}
As in previous cases soundness follows from subject reduction, which can be easily proved. For completeness let $\SessionTypeT\nasbut\SessionTypeS$. We show by induction on $\nasbut$: 
\begin{enumerate}
\item\label{i1} if $\SessionTypeT\nasbut\SessionTypeS$ is derivable without using rules {\rm\rulename{n-bra-async}} and {\rm\rulename{n-sel-async}}, then $$ \new {\ChannelA} {\ChannelB} (\cprocess(\ChannelA,\SessionType)\parop\cprocess(\ChannelB,\co\SessionTypeS)
 \parop
 \queue \ChannelB \ChannelA \EmptyQueue 
 \parop
 \queue \ChannelA \ChannelB \EmptyQueue) 
  \wreda \error$$
\item\label{i2} otherwise there are $\SessionTypeT'\ssubt\SessionTypeT$, $\SessionTypeS'\ssubt\co{\SessionTypeS}$ with either $\SessionTypeT'=\SessionTypeT$ or $\SessionTypeS'=\co{\SessionTypeS}$, such that $$ \new {\ChannelA} {\ChannelB} (\cprocess(\ChannelA,\SessionType')\parop\cprocess(\ChannelB,\SessionTypeS')
 \parop
 \queue \ChannelB \ChannelA \EmptyQueue 
 \parop
 \queue \ChannelA \ChannelB \EmptyQueue) 
  \wreda \error$$
\end{enumerate}
It is easy to check that the proof of Theorem~\ref{thm:preciseness:async}  
shows this stronger statement, since we need to consider subtypes of the current types only dealing with rules {\rm\rulename{n-bra-async}} and {\rm\rulename{n-sel-async}}. 

We consider here only two cases of applications of rules \rulename{n-exch-bra} and \rulename{n-exch-sel}.

\smallskip

{\em Case} \rulename{n-exch-bra}: 
$\SessionTypeT =  
\Branch_{i\in I} \In{\Tag_i}{\U_i}.\SessionTypeT_i$,
$\SessionTypeS = 
\Branch_{j\in J} \In{\Tag'_j}{\U'_j}.\SessionTypeS_j$, 
and $\exists k \in I~\exists k' \in J : \Tag_{k} = \Tag'_{k'}$
such that $ \U_k \nasbut \U'_{k'}$. We only consider the case 
 $\U'_{k}=\T' \not=\End$ and $\U'_{k'}=\sct{\SessionTypeS'}$. 
$$
\begin{array}{l}
\new \ChannelA \ChannelB
    (\cprocess(\ChannelA,\SessionTypeT)
    \parop
    \cprocess(\ChannelB, \co \SessionTypeS) 
    \parop
    \queue \ChannelB \ChannelA \EmptyQueue 
    \parop
    \queue \ChannelA \ChannelB \EmptyQueue  )
  \reda  \\ 
\new \ChannelA \ChannelB
( \cprocess(\ChannelA,\SessionTypeT)
\parop
    \res\s(
      \send{\ChannelB}{\Tag_k}{\s}.(\cprocess(\ChannelB,\co{\SessionTypeS_{k'}})
      \parop\acc\s y \cprocess(y,\SessionTypeS')\parop\req\s z \cprocess(z,\co{\SessionTypeS'})))
   \parop
    \queue \ChannelB \ChannelA \EmptyQueue 
   \parop
    \queue \ChannelA \ChannelB \EmptyQueue   
) \reda
\\ 
\new \ChannelA \ChannelB\res\s(
 \sum_{i \in I}
    \cprocess[?](\ChannelA,\Tag_i,\SessionTypeT'_i,\SessionTypeT_i) 
 \parop
     \cprocess(\ChannelB,\co{\SessionTypeS_{k'}})\parop\acc\s y \cprocess(y,\SessionTypeS')\parop\req\s z \cprocess(z,\co{\SessionTypeS'})
     \parop
    \queue \ChannelB \ChannelA \msg{\Tag_k}\s 
     \parop
    \queue \ChannelA \ChannelB \EmptyQueue   
)  \reda\\
\Context
[ \cprocess( \s, \T'  )]
\end{array}
$$
where $\Context\hole=\new \ChannelA \ChannelB\res\s(
\cprocess(\ChannelA,\SessionTypeT_{k}) \parop
     \cprocess(\ChannelB,\co{\SessionTypeS_{k'}})\parop\acc\s y \cprocess(y,\SessionTypeS')\parop\req\s z \cprocess(z,\co{\SessionTypeS'})  \parop \hole\parop
    \queue \ChannelB \ChannelA \EmptyQueue 
    \parop
    \queue \ChannelA \ChannelB \EmptyQueue   
)$.
By definition of characteristic process, being $\T'\not=\End$, the shared channel $\s$ is used as a linear channel in $\cprocess( \s, \T'  )$, and this implies by rule \rulename{err-chan-in} or  \rulename{err-chan-out} 
$$ \cprocess( \s, \T'  )\reda \error$$ then by rule \rulename{err-context}, we conclude 
$$
\Context
[ \cprocess( \s, \T'  )
]
\reda \error$$
{\em Case} \rulename{n-exch-sel}: $\SessionTypeT = \Select_{i\in I} \Out{\Tag_i}{\U_i}.\SessionTypeT_i$,
$\SessionTypeS =  \Select_{j\in J} \Out{\Tag'_j}{\U'_j}.\SessionTypeS_j$, 
and $\exists k \in I~\exists k' \in J$ such that $\Tag_{k} = \Tag'_{k'}$ and
$\U'_{k'} \nasbut \U_{k}$. We only consider the case 
 $\U_{k}=\sct{\T'}$ and $\U'_{k'}=\sct{\SessionTypeS'}$.
 $$
\begin{array}{l}
\new \ChannelA \ChannelB
    (\cprocess(\ChannelA,\SessionTypeT)
    \parop
    \cprocess(\ChannelB, \co \SessionTypeS) 
    \parop
    \queue \ChannelB \ChannelA \EmptyQueue 
    \parop
    \queue \ChannelA \ChannelB \EmptyQueue  )
  \reda  \\ 
\new \ChannelA \ChannelB
( \res\s(
      \send{\ChannelA}{\Tag_k}{\s}.(\cprocess(\ChannelA,\SessionType_k)
      \parop\acc\s y \cprocess(y,\SessionType')\parop\req\s z \cprocess(z,\co{\SessionType'})))\parop 
      \cprocess(\ChannelB, \co \SessionTypeS) 
    \parop
    \queue \ChannelB \ChannelA \EmptyQueue 
    \parop
    \queue \ChannelA \ChannelB \EmptyQueue  )
  \reda  \\ 
  \new \ChannelA \ChannelB\res\s(\cprocess(\ChannelA,\SessionType_k)
  \parop\acc\s y \cprocess(y,\SessionType')\parop\req\s z \cprocess(z,\co{\SessionType'})
  \parop 
      \sum_{j \in J \setminus \ASET{k'}}
    \cprocess[?](\ChannelB,\Tag'_j,\U'_i,\SessionTypeS_j) 
    \\ \hfill
    + \    
 \receive{\ChannelB}{\Tag_k}{\var}.(
    \cprocess(\ChannelB,\co{\SessionTypeS_{k'}})
    \parop\acc\var {y'} \cprocess(y',\SessionTypeS')\parop\req\var {z'}
    \cprocess(z',\co{\SessionTypeS'}))
     \parop
    \queue \ChannelB \ChannelA \EmptyQueue 
    \parop
    \queue \ChannelA \ChannelB \msg{\Tag_k}\s   )
  \reda  \\ 
   \new \ChannelA \ChannelB\res\s(\cprocess(\ChannelA,\SessionType_k)
   \parop\acc\s y \cprocess(y,\SessionType')\parop\req\s z \cprocess(z,\co{\SessionType'})
  \parop \\ \hfill
    \cprocess(\ChannelB,\co{\SessionTypeS_{k'}})\parop\acc\s {y'} \cprocess(y',\SessionTypeS')\parop\req\s {z'}
    \cprocess(z',\co{\SessionTypeS'})) \parop
    \queue \ChannelB \ChannelA \EmptyQueue 
    \parop
    \queue \ChannelA \ChannelB \EmptyQueue  ) \end{array}$$
    Since $\SessionType' $ and $\SessionTypeS'$ cannot be equal, then we can derive either  $\SessionTypeS' \nasbut\SessionType'$ or  $\SessionType' \nasbut\SessionTypeS'$ with a proof which does not use rules {\rm\rulename{n-bra-async}} and {\rm\rulename{n-sel-async}} by Lemma~\ref{auxf}.
    In the first case the obtained process can be written as  $C[\req\s z \cprocess(z,\co{\SessionType'})\parop\acc\s {y'}
    \cprocess(y',\SessionTypeS')]$ and 
$$
\begin{array}{l}
    C[\req\s z \cprocess(z,\co{\SessionType'})\parop\acc\s {y'}
    \cprocess(y',\SessionTypeS')]
  \reda   
C[ \new {\ChannelC} {\ChannelD} (\cprocess(\ChannelC,\co{\SessionType'})\parop\cprocess(\ChannelD,\SessionTypeS')
 \parop
 \queue \ChannelD \ChannelC \EmptyQueue 
 \parop
 \queue \ChannelC \ChannelD \EmptyQueue) ]
\end{array}$$
By above we are in case (\ref{i1}) of induction, therefore we get 
$$ \new {\ChannelC} {\ChannelD} (\cprocess(\ChannelC,\co{\SessionType'})\parop\cprocess(\ChannelD,\SessionTypeS')
 \parop
 \queue \ChannelD \ChannelC \EmptyQueue 
 \parop
 \queue \ChannelC \ChannelD \EmptyQueue) 
  \wreda \error$$
then by rule \rulename{err-context}
$$C[ \new {\ChannelC} {\ChannelD} (\cprocess(\ChannelC,\co{\SessionType'})\parop\cprocess(\ChannelD,\SessionTypeS')
 \parop
 \queue \ChannelD \ChannelC \EmptyQueue 
 \parop
 \queue \ChannelC \ChannelD \EmptyQueue) ]\wreda \error$$
 In the second case the obtained process can be written as  $C'[\acc\s y \cprocess(y,\SessionType')\parop\req\s {z'}
    \cprocess(z',\co{\SessionTypeS'})]$ and we conclude similarly.
\end{Proof}

\section{Denotational Preciseness}\label{sec:denotation}
In $\lambda$-calculus types are usually interpreted as subsets of the domains of $\lambda$-models~\cite{bcd83,H83}. {\em Denotational preciseness} of subtyping is then:
\begin{equation*}
T \subt S \quad \text{if and only if} \quad \dlsqb  T \drsqb \subseteq\dlsqb  S \drsqb
\end{equation*}
using $\dlsqb  \; \drsqb $ to denote type interpretation. 

In the present context let us interpret a session type $T$ as the set of processes with only one free channel typed by $T$, i.e.
$$\dlsqb  T \drsqb_*=\{P~ \mid ~~\vdash_* P \triangleright \{a:T\}\}$$
where $*\in \{\mathsf{s},\mathsf{a}\}$. We can then show that both the synchronous and the asynchronous subtypings are denotationally precise. Rule \rulename{t-sub} gives the denotational soundness. Denotational completeness follows from the following key property of characteristic processes:
\begin{equation}\label{key}
\vdash_* \cprocess{_*}(a,T) \triangleright \{a:S\} \quad \text{implies} \quad T\subt_* S
\end{equation}
where $ \cprocess{_*}(a,T)$ is the synchronous characteristic process if $*=\mathsf{s}$ and the asynchronous characteristic process if $*=\mathsf{a}$. 
The property (\ref{key}) can be shown by induction on $T$ using Inversion Lemmas for synchronous and asynchronous processes (Lemma~\ref{append:lem:inversion:pure} in Appendix~\ref{proof:soundness} and 
Lemma~\ref{append:lem:inversion:async} in Appendix~\ref{app:async_lang}).

  
  If $T\not\subt_* S$, then $\cprocess{_*}(a,T)\in \dlsqb  T \drsqb_*$, but $\cprocess{_*}(a,T)\not\in \dlsqb  S \drsqb_*$, which implies $\dlsqb  T \drsqb_*\not\subseteq\dlsqb  S \drsqb_*$. 
   Therefore we get 
  denotational completeness. 
  \begin{thm}[Denotational preciseness]\label{thm:dpreciseness}
The synchronous and the asynchronous subtyping relations are denotationally precise for the synchronous calculus and the asynchronous calculus, respectively. \end{thm}
To sum up, the existence of characteristic processes implies denotational preciseness when each session type $T$ is interpreted as the set of processes having only one channel of type $T$.

\section{Related Work}
\label{sec:related}
\mypar{Preciseness}
To the best of our knowledge, operational
preciseness was first defined by Ligatti \etal~\cite{BHLN12}, for a call-by-value $\lambda$-calculus 
with recursive functions, pairs and sums. In that paper 
the authors show that the iso-recursive subtyping induced by the Amber rules~\cite{LC86} is incomplete. 
They propose a new iso-recursive subtyping which they prove to be operationally precise. Denotational preciseness of this subtyping has been recently proved~\cite{DGJPY16}.

Both operational and denotational preciseness are shown by Dezani-Ciancaglini and Ghilezan~\cite{DG14} for the concurrent $\lambda$-calculus with intersection and union types introduced by Dezani-Ciancaglini \etal~\cite{SIAM}. In that paper divergence plays the r\^ole  of reduction to $\error$.

Preciseness in concurrency is more useful 
and challenging than in the functional setting,
since there are many interesting choices 
for the syntax, semantics, type errors of the calculi  and for the typing systems. 
A similar situation appears in the study of 
bisimulations, where many labelled transition relations 
can be defined.  
It is now common that researchers justify the correctness of labelled 
transition systems by proving  
that the bisimulation coincides with the contextual congruence~\cite{HondaKYoshida95,MiSa92}. 
Our claim is that preciseness should become a sanity check 
for subtypings. 

Recently preciseness has been shown for a synchronous multiparty session calculus without delegation~\cite{DGJPY15}.

\mypar{Choices of typing system and subtyping}
The first branching-selection subtyping for session types was proposed
by Gay and Hole~\cite{GH,GH05} and has been used in other works by various authors~\cite{CNV08,CGP08,castagna09foundations,Gay:2006b,Padovani11,P13,V09}. %
Their approach %
corresponds to safe substitutability of \emph{channels} %
(rather than \emph{processes}): %
as a consequence, %
their subtyping is the opposite of $\ssubt$, %
since branching is covariant and selection is controvariant %
in the set of labels; %
coherently, %
such a co/contra-variance is also embodied %
in their typing system and judgements. %
\\%
A subtyping relation with the opposite direction %
has been used by Honda, Yoshida, Mostrous and other authors~\cite{Carbone:2012:SCP:2220365.2220367,DemangeonH11,mostrous_phd,mostrous09sessionbased,MY15,
mostrous_yoshida_honda_esop09}:
their approach corresponds to safe substitutability of \emph{processes}. %
An insightful comparison between these subtypings is the argument of a recent paper~\cite{G16}.

In this work, %
we have adopted the subtyping direction and typing system %
of Honda \emph{et al.}, %
since they directly match a definition of preciseness %
based on process substitution (Definition~\ref{def:preciseness}), %
and thus allow for direct reasoning on characteristic processes. %
Establishing the preciseness of the Gay and Hole subtyping %
with respect to their typing system is less immediate, %
since the notion of characteristic process (and their substitution) %
needs to be adapted to their setting. %
However, since their typing rules ``mirror'' ours, %
and their subtyping ``mirrors'' $\ssubt$, %
similar preciseness results can be proved %
by reversing the
ordering both in the preciseness definition and in the extension of
subtyping to session environments.

\mypar{Other completeness results}
Subtyping of recursive types requires algorithms for checking subtype relations, 
as discussed by Pierce~\cite[\S~21]{PierceBC:typsysfpl}. 
These algorithms need to be proved sound and complete with respect to the definition of the corresponding subtyping, 
as done in several works~\cite{castagna09foundations,GH05,PierceSangiorgi95}. Synchronous subtyping can be easily decided, see for example~\cite{GH05}.
We leave the development of an algorithm for asynchronous subtyping, %
and the proof of its soundness and completeness, as future work.\\
Several works on subtyping formulate
the errors using typed reductions or type environments~\cite{HR02,PierceSangiorgi95}, and they prove
soundness with respect to the typed reductions and their erasure theorems. 
In contrast with these approaches, 
our error definitions in Tables \ref{tab:sync:red:err} and  \ref{tab:async:red:err} 
do not rely on any type-case construct or explicit 
type information, but are defined 
{\em syntactically} over untyped terms.
Note that once the calculus is annotated by type information or 
equipped with type case, completeness 
becomes trivial, since any two processes of incomparable types can be operationally distinguished. 

\mypar{Semantic subtyping}
In the semantic subtyping approach by Frisch \etal~\cite{FCB08}, each type is interpreted as the set of values having that type and subtyping is subset inclusion between type interpretations.
This gives a precise subtyping as soon as the calculus allows to distinguish operationally values of different types. Semantic subtyping has been studied by Castagna, De Nicola \etal~\cite{CNV08} for a $\pi$-calculus with a patterned input, and by Castagna, Dezani-Ciancaglini \etal~\cite{castagna09foundations} for a session calculus with internal and external choices and typed input. Types are built using a rich set of type constructors including union, intersection and negation: they extend IO-types in Castagna, De Nicola \etal~\cite{CNV08}, and 
session types in Castagna, Dezani-Ciancaglini \etal~\cite{castagna09foundations}. Semantic subtyping is precise for the calculi of all such works, %
thanks to the type case constructor in the work by Frisch \etal~\cite{FCB08},  and to the blocking of inputs for values of  ``wrong'' types in the works by Castagna \etal~\cite{CNV08,castagna09foundations}.

\mypar{Subtyping of Mostrous PhD thesis} Our subtyping relation differs from that defined in Mostrous thesis~\cite{mostrous_phd,MY15} for the premises $\& \in \AContext
$ and $\&\in\SessionTypeT_i$ in rule \rulename{sub-perm-async}. 
As a consequence in that thesis
$T$ is a subtype of $S$ when 
$T=\mu \tvar.\Out{\Tag}{T'}.\tvar$   
and $S=\mu \tvar.\Out{\Tag}{T'}.\In{\Tag'}{S'}.\tvar$ 
(see p.~116 of Mostrous thesis). This subtyping is not sound in our system: intuitively 
$T$ accumulates infinite orphan messages in a queue, 
while  $S$ ensures that the messages are eventually received. 
The subtyping relation in Mostrous thesis 
unexpectedly allows an unsound process (typed by $T$) 
to act as if it were a sound process (typed by $S$).  
Let 
$\Context=
(\nu ab)([ \ ]  \ | \ Q
\ | \ \queue\ChannelA\ChannelB \EmptyQueue \ | \  \queue\ChannelB\ChannelA\EmptyQueue )$
where
$$Q =\DefD{\pvarY(x) = \send{\ChannelB}{\Tag}{x}.{\ChannelB}\In{\Tag'}{y}.\invoke{\pvarY}{x}}{\new{\ChannelC}{\ChannelC'}(\invoke{\pvarY}{\ChannelC})}$$
Then we can derive
$\wtp{}{\Context[\ChannelA:\SessionTypeS]}{\EmptyEnv}$. 
Let 
$$\Process =\DefD{\pvarZ(z) =\send{\ChannelA}{\Tag}{z}.\invoke{\pvarZ}{z}}{\new{\ChannelD}{\ChannelD'}(\invoke{\pvarZ}{\ChannelD})}$$
Then $\wtpa{}{\Process}{\ASET{a:\SessionTypeT}}$. We get 
$$C[\Process]\wreda (\nu \ChannelA\ChannelB)\new{\ChannelC}{\ChannelC'}(P \ | \ Q
\ | \ \queue\ChannelA\ChannelB\EmptyQueue \ | \  \queue\ChannelB\ChannelA\msg{\Tag}{\ChannelC} )\reda\error$$ 
by rule \rulename{err-orph-mess-async}, since 
$a\not\in  \sbnF(P \ | \ Q
\ | \  \queue\ChannelA\ChannelB\EmptyQueue )$ and $\fpv(P \ | \ Q
\ | \  \queue\ChannelA\ChannelB\EmptyQueue )=\emptyset$.

The subtyping of Mostrous thesis is sound for the session calculus
defined there, which does not consider orphan messages as errors.
However, the subtyping of Mostrous thesis
is not complete, an example being
$\trec\tvar.\Out{\Tag}{\SessionTypeT} . \tvar \not \subt \trec \tvar. \In{\Tag'}{\SessionTypeS} . \tvar$. 
There is no context $\Context$ which is safe for all processes
with one channel typed by $\trec \tvar. \In{\Tag'}{\SessionTypeS} . \tvar$
and no process $\Process$ with one channel typed by $\trec\tvar.\Out{\Tag}{\SessionTypeT} . \tvar$ such that $\Context[\Process]$ deadlocks.

\section{Conclusion}
\label{sec:conclusion}
This article gives, as far as we know, the first formulation and
proof techniques for the preciseness of subtyping in mobile processes. 
We consider the synchronous and asynchronous session calculi 
to investigate the preciseness of the existing subtypings. 
While 
the well-known branching-selection subtyping 
\cite{Carbone:2012:SCP:2220365.2220367,DemangeonH11,GH05} is precise for the
synchronous calculus,  the subtyping in Mostrous thesis~\cite{mostrous_phd,MY15} 
turns out to be not sound for the asynchronous calculus. 
We propose a simplification of previous asynchronous subtypings
\cite{mostrous09sessionbased,mostrous_yoshida_honda_esop09} 
and prove its preciseness. 
As a matter of fact only soundness is a consequence of subject reduction, while completeness can fail also when subject reduction holds.

The formulation of preciseness along with
the proof methods and techniques could be useful to examine 
other subtypings and calculi. 
Our future work includes the applications to higher-order processes 
\cite{mostrous_phd,mostrous09sessionbased,MY15},   
polymorphic types~\cite{CPPT13,Gay:2006b,Goto12}, 
fair subtypings~\cite{Padovani11,P13} and contract subtyping~\cite{BL10}.  We plan to use the characteristic processes in typecheckers for session types. More precisely the error messages can show processes of given types when type checking fails.
One interesting problem is to find the necessary
and sufficient conditions to obtain completeness of the generic
subtyping by Igarashi and Kobayashi~\cite{igarashi.kobayashi04:gentypPi}. Such a characterisation would give preciseness for the many type systems which are instances of generic
subtyping~\cite{igarashi.kobayashi04:gentypPi}. The notion of subtyping is clearly connected with that of type duality. Various definitions of dualities are compared by Bernardi \etal~\cite{BDGK14}, 
and we plan to investigate if completeness of subtyping can be used in finding the largest safe duality. 

The recent study on the Curry-Howard isomorphism between 
session types and Linear Logic~\cite{CPPT13,CF10,PCPT12,Wadler2012}
gives a logic basis to session type duality. Also 
the one-to-one correspondence between session types and a class of 
deadlock-free communicating automata~\cite{GMY84} 
has shown that session types have solid roots~\cite{DY13,Villard11}. 

The preciseness result for the synchronous calculus 
in \S~\ref{sec:synchronous_language} and \S~\ref{sec:s:completeness} 
shows a rigorousness of 
the branching-selection subtyping, which is 
implemented (as a default) in most of session-based programming languages 
and tools 
\cite{Carbone:2012:SCP:2220365.2220367,DemangeonH11,HNHYH13,scribble10,event,HU07TYPE-SAFE,scribsite} 
for enlarging typeability. 
For the asynchronous calculus, preciseness is more debatable 
since it depends on the choice of type safety properties, see \S~\ref{sec:asynchronous_language} and \S~\ref{sec:a:completeness}.  
But in this case 
preciseness plays a more important r\^ole, since 
a programmer can adjust a subtyping relation
to loosen or tighten subtypings with respect to 
the type safety properties which she wishes to guarantee.   
Once preciseness has been proved, she can be sure that 
her safety specifications 
and the subtyping have an exact match with respect to 
both static and dynamic semantics.

\mypar{Acknowledgments} 
We are grateful to the anonymous reviewers for their useful
suggestions, which led to substantial improvements. We are indebted to Jovanka Pantovi\'c for pointing out a subtle mistake 
in a previous version of the completeness proof. 

\bibliographystyle{plain}
\bibliography{session}


\appendix
\section{Proofs of Section~\ref{sec:synchronous_language}}
\label{proof:soundness}
\begin{thm} \label{thm:trans:ssubt}
The relation $\ssubt$ is transitive. 
\end{thm}
\begin{Proof}
Let $\SessionTypeT \ssubtTrans \SessionTypeS$ if there exists $\SessionTypeV$ such that 
$\SessionTypeT \ssubt \SessionTypeV$ and $\SessionTypeV \ssubt \SessionTypeS$. It suffices to show $\ssubtTrans \subseteq \ssubt$,
i.e. that $\ssubtTrans$ satisfies all the rules of Table \ref{tab:sync:ssubt}.
We consider all the possible shapes of 
$\SessionTypeT$, $\SessionTypeV$, and $\SessionTypeS$
based on the rules of Table \ref{tab:sync:ssubt}:
\begin{enumerate}
\item $\SessionTypeT = \End$ and $\SessionTypeV = \End$ and $\SessionTypeS = \End$.
Then $\SessionTypeT \ssubtTrans \SessionTypeS$ agrees with rule \rulename{sub-end}.

\item $\SessionTypeT = \Branch_{i\in I\cup J\cup H} \In{\Tag_i}{\TypeU_i}.\SessionTypeT_i$
and $\SessionTypeV = \Branch_{i\in I\cup J} \In{\Tag_i}{\TypeU'_i}.\SessionTypeT'_i$
and $\SessionTypeS = \Branch_{i\in I} \In{\Tag_i}{\TypeU''_i}.{\SessionTypeT''_i}$
and $\SessionTypeS_i \ssubt \SessionTypeS'_i$ and $\SessionTypeT_i \ssubt \SessionTypeT'_i$ for all $i \in I\cup J$,
and $\SessionTypeS'_i \ssubt \SessionTypeS''_i$ and $\SessionTypeT'_i \ssubt \SessionTypeT''_i$ for all $i \in I$.
By the definition of $\ssubtTrans$, we get 
$\SessionTypeS_i \ssubtTrans \SessionTypeS''_i$ and $\SessionTypeT_i \ssubtTrans \SessionTypeT''_i$ for all $i \in I$.
Then 
$\SessionTypeT \ssubtTrans \SessionTypeS$ agrees with rule \rulename{sub-bra}.

\item $\SessionTypeT = \Select_{i\in I} \Out{\Tag_i}{\TypeU_i}.\SessionTypeT_i$
and $\SessionTypeV =\Select_{i\in I \cup J} \Out{\Tag_i}{\TypeU'_i}.\SessionTypeT'_i$
and 
$\SessionTypeS = \Select_{i\in I \cup J \cup H} \Out{\Tag_i}{\TypeU''_i}.\SessionTypeT''_i$
and $\SessionTypeS'_i \ssubt \SessionTypeS_i$ and $\SessionTypeT_i \ssubt \SessionTypeT'_i$ for all $i \in I$,
and $\SessionTypeS''_i \ssubt \SessionTypeS'_i$ and $\SessionTypeT'_i \ssubt \SessionTypeT''_i$ for all $i \in I \cup J$.
By the definition of $\ssubtTrans$, we get
$\SessionTypeS''_i \ssubtTrans \SessionTypeS_i$ and $\SessionTypeT_i \ssubtTrans \SessionTypeT''_i$ for all $i \in I$.
Then 
$\SessionTypeT \ssubtTrans \SessionTypeS$ agrees with rule \rulename{sub-sel}.\qedhere
\end{enumerate}
\end{Proof}

\noindent The remainder of this section is devoted to the
proof of the subject reduction theorem and of the soundness of the synchronous subtyping.

\begin{lem}[Inversion lemma for synchronous processes]
\label{append:lem:inversion:pure} $\;$ 
\begin{enumerate}
\item\label{inver:end} If $\wtps{\UEnv}{\idle}{\LEnv}$, 
         then $\LEnv$ is end-only.        
\item\label{inver:var} If 
         $\wtps{\UEnv}
         {\invoke\pvar{\tilde\Name}}{\LEnv}$,
         then $\UEnv = \UEnv', \pbind{\pvar}{\tilde\SessionType}$ and
         $\ASET{\bind{\tilde \Name}{\tilde \SessionType}} \ssubt \LEnv$.     
\item\label{inver:bra} If 
          $\wtps{\UEnv}{\sum_{i\in I} \receive{\Name}{\Tag_i}{\var_i}.\Process_i}{\LEnv}$,
          then $\LEnv = \LEnv', 
                        \ASETT{
                          \bind{\Name}{\Branch_{j\in J}
                         \In{\Tag_j}{\Type_j}.\SessionTypeT_j}}$ and  $J \subseteq I$ and\\
          $\forall j \in J:\wtps{\UEnv}{\Process_j}{\LEnv',\ASETT{
                                                             \bind{\Name}{\SessionTypeT_j},
                                                             \bind{\var_j}{\SessionTypeS_j}}}$.
\item\label{inver:sel} If 
         $\wtps{\UEnv}{\send{\NameU}{\Tag}{\Name'}.\Process}{\LEnv}$,
         then $\LEnv = 
         {\LEnv', \ASETT{\bind{\Name}{\SessionTypeT},
          \bind{\Name'}{\SessionTypeS}}}$ and
         $\Out{\Tag}{\SessionTypeS}.\SessionTypeT' \ssubt \SessionTypeT$,
         and
         \mbox{$\wtps{\UEnv}{\Process}{\LEnv', \ASETT{\bind{\NameU}{\SessionTypeT'}}}$.}      

\item\label{inver:par} If $\wtps{\UEnv}{\Process_1 \parop \Process_2}{\LEnv}$,
          then $\LEnv = \LEnv_1, \LEnv_2$ and $\wtps{\UEnv}{\Process_1}{\LEnv_1}$ and
          $\wtps{\UEnv}{\Process_2}{\LEnv_2}$.
          
\item\label{inver:choice} If $\wtps{\UEnv}{\Process_1 \choice \Process_2}{\LEnv}$, 
          then $\wtps{\UEnv}{\Process_1}{\LEnv}$ and 
          $\wtps{\UEnv}{\Process_2}{\LEnv}$.       
\item\label{inver:def} If $\wtps{\UEnv}
        {\Def{\pvar}{\tilde\varY}{\ProcessP}{\ProcessQ}}{\LEnv}$,
         then $\wtps{\UEnv,\pbind{\pvar}{\tilde\SessionType}}{\ProcessP}{
                   \ASET{\bind{\tilde\varY}{\tilde\SessionType}}}$ and  
        $\wtps{\UEnv,\pbind{\pvar}{\tilde\SessionType}}{
          \ProcessQ}{\LEnv}$.
\item\label{inver:new:sync} If $\wtps{\UEnv}{\new{\ChannelA}{\ChannelB}\Process}{\LEnv}$,
         then 
         $\wtps{\UEnv}{\Process}
         {\LEnv,
          \ASETT{\bind{\ChannelA}{\SessionType},
          \bind{\ChannelB}{\co\SessionType}}
         }$. 
\end{enumerate}
\end{lem}
\begin{Proof}
By induction on derivations.
\end{Proof}

\begin{lem}[Substitution lemma for synchronous processes]\label{app:lem:subs} 
If $\wtps{\UEnv}{\Process}{\LEnv, \ASETT{\bind{y}{\SessionType}}}$
and $\ChannelA \not \in \dom(\LEnv)$, 
then $\wtps{\UEnv}{\Process \subst{a}{y}}{\LEnv, \ASETT{\bind{a}{\SessionType}}}$.
\end{lem}
\begin{Proof}
Standard.
\end{Proof}

\begin{lem} \label{thm:preserve:sync} 
If $\wtps{\UEnv}{\Process}{\LEnv}$ and $\Process \equiv \Process'$, then
$\wtps{\UEnv}{\Process'}{\LEnv}$.
\end{lem}
\begin{Proof}
The proof by induction on $\equiv$ is easy.
\end{Proof}

\paragraph{\bf Theorem \ref{thm:sbr:sync}.} 
(Subject reduction for synchronous processes) 
\label{app:thm:sbr:sync}
If $\wtps{\UEnv}{\Process}{\LEnv}$ and 
$\Process \wreds \ProcessQ$, then
\mbox{$\wtps{\UEnv}{\ProcessQ}{\LEnv}$.}
\begin{Proof}
  We first prove that %
  if $\wtps{\UEnv}{\Process}{\LEnv}$ and 
  $\Process \reds \ProcessQ$, then
  \mbox{$\wtps{\UEnv}{\ProcessQ}{\LEnv}$.} %
  The proof is by induction on the derivation %
  of $\Process \reds \ProcessQ$. %
  We consider some interesting rules of Table \ref{tab:sync:red}.

\smallskip

\begin{enumerate}
\item Case $\rulename{r-com-sync}$:
\begin{eqnarray} \label{eq:sbred:sync:r-com-syn:1}
\wtps{\UEnv}{\new \ChannelA \ChannelB
( \send\ChannelA {\Tag_k} \ChannelC.\Process \parop
  \sum_{i\in I} \receive\ChannelB{\Tag_i}{\var_i}.\ProcessQ_i)}{\LEnv}
\end{eqnarray} where $k \in I$.
By applying Lemma \ref{append:lem:inversion:pure}.\ref{inver:new:sync} to 
 (\ref{eq:sbred:sync:r-com-syn:1}),
we have
\begin{eqnarray} \label{eq:sbred:sync:r-com-syn:2}
\wtps{\UEnv}{\send\ChannelA {\Tag_k} \ChannelC.\Process \parop
  \sum_{i\in I} \receive\ChannelB{\Tag_i}{\var_i}.\ProcessQ_i} 
  {\LEnv, \ASETT{\bind{\ChannelA}{\SessionTypeT}, 
  \bind{\ChannelB}{\co{\SessionTypeT}}}}
\end{eqnarray}
By applying Lemma \ref{append:lem:inversion:pure}.\ref{inver:par} to 
 (\ref{eq:sbred:sync:r-com-syn:2}),
we get
\begin{eqnarray} \label{eq:sbred:sync:r-com-syn:3}
\wtps{\UEnv}{\send\ChannelA {\Tag_k} \ChannelC.\Process}{\LEnv_1}
\end{eqnarray}
\begin{eqnarray} \label{eq:sbred:sync:r-com-syn:4}
\wtps{\UEnv}{\sum_{i\in I} \receive\ChannelB{\Tag_i}{\var_i}.\ProcessQ_i}{\LEnv_2}
\end{eqnarray}
where
$\LEnv_1, \LEnv_2=
\LEnv, 
\ASETT{\bind{\ChannelA}{\SessionTypeT}, 
\bind{\ChannelB}{\co{\SessionTypeT}}}$.
By applying Lemma \ref{append:lem:inversion:pure}.\ref{inver:sel}
to  (\ref{eq:sbred:sync:r-com-syn:3}), we have
$$
\LEnv_1 = 
          \LEnv'_1,
          \ASETT{
          \bind{\ChannelA}{\SessionTypeT}, 
          \bind{\ChannelC}{\SessionTypeS}}
$$
\begin{eqnarray} \label{eq:sbred:sync:r-com-syn:6}
\Out{\Tag_k}{\SessionTypeS}.\SessionTypeT' \ssubt \SessionTypeT
\end{eqnarray}
\begin{eqnarray} \label{eq:sbred:sync:r-com-syn:7}
\wtps{\UEnv}{\Process}{\LEnv'_1, \ASETT{\bind{\ChannelA}{\SessionTypeT'}}}
\end{eqnarray}
By applying Lemma \ref{append:lem:inversion:pure}.\ref{inver:bra}
to  (\ref{eq:sbred:sync:r-com-syn:4}), we have
\begin{eqnarray} \label{eq:sbred:sync:r-com-syn:8}
\LEnv_2 = \LEnv'_2, \ASETT{\bind{\ChannelB}{\co{\SessionTypeT}}} \nonumber
\end{eqnarray}
\begin{eqnarray} \label{eq:sbred:sync:r-com-syn:9}
\co{\SessionTypeT} =
\Branch_{j\in J}
\In{\Tag_j}{\SessionTypeS_j}.\SessionTypeT_j, \quad
J \subseteq I
\end{eqnarray}
\begin{eqnarray} \label{eq:sbred:sync:r-com-syn:10}
\forall j \in J :
\wtps{\UEnv}{\ProcessQ_j}{\LEnv'_2, 
\ASETT{\bind{\ChannelB}{\SessionTypeT_j},
\bind{\var_j}{\SessionTypeS_j}}}
\end{eqnarray}
By (\ref{eq:sbred:sync:r-com-syn:9}) and duality,
\begin{eqnarray} \label{eq:sbred:sync:r-com-syn:10-1}
\SessionTypeT
= \co{\Branch_{j\in J} \In{\Tag_j}{\SessionTypeS_j}.\SessionTypeT_j}
= \Select_{j \in J} \Out{\Tag_j}{\SessionTypeS_j}.\co{\SessionTypeT_j}
\end{eqnarray}
Then by (\ref{eq:sbred:sync:r-com-syn:6}) $k\in J$ and
\begin{eqnarray} \label{eq:sbred:sync:r-com-syn:11}
\SessionTypeS = \SessionTypeS_k
\end{eqnarray}
which together with  (\ref{eq:sbred:sync:r-com-syn:10}) imply
\begin{eqnarray} \label{eq:sbred:sync:r-com-syn:12}
\wtps{\UEnv}{\ProcessQ_k}{\LEnv'_2, 
\ASETT{\bind{\ChannelB}{\SessionTypeT_k},
\bind{\var_k}{\SessionTypeS_k} }}
\end{eqnarray}
By (\ref{eq:sbred:sync:r-com-syn:11}) and by applying  
Lemma \ref{app:lem:subs} to (\ref{eq:sbred:sync:r-com-syn:12}),
we have
\begin{eqnarray} \label{eq:sbred:sync:r-com-syn:13}
\wtps{\UEnv}{\ProcessQ_k \subst{\ChannelC}{\var_k}}{
                               \LEnv'_2, 
                               \ASETT{\bind{\ChannelB}{\SessionTypeT_k},
                               \bind{\ChannelC}{\SessionTypeS}}}
\end{eqnarray}
From  (\ref{eq:sbred:sync:r-com-syn:6}) and  (\ref{eq:sbred:sync:r-com-syn:10-1}), 
we get
\begin{eqnarray} \label{eq:sbred:sync:r-com-syn:14}
\SessionTypeT' \ssubt \co \SessionTypeT_k
\end{eqnarray}
Applying \rulename{t-sub} to (\ref{eq:sbred:sync:r-com-syn:7}) 
and  (\ref{eq:sbred:sync:r-com-syn:14}),
we derive
\begin{eqnarray} \label{eq:sbred:sync:r-com-syn:15}
\wtps{\UEnv}{\ProcessP}{\LEnv'_1, \ASETT{\bind{\ChannelA}{\co \SessionTypeT_k}}}
\end{eqnarray}
By applying $\rulename{t-par}$ to (\ref{eq:sbred:sync:r-com-syn:13}) and  (\ref{eq:sbred:sync:r-com-syn:15}),
we derive
\begin{eqnarray} \label{eq:sbred:sync:r-com-syn:16}
\wtps{\UEnv}{\Process \parop \ProcessQ_k \subst{\ChannelC}{\var_k}}
{\LEnv_1', 
\LEnv'_2,
\ASETT{
\bind{\ChannelA}{\co\SessionTypeT_k},
\bind{\ChannelB}{\SessionTypeT_k},
\bind{\ChannelC}{\SessionTypeS}}}
\end{eqnarray}
By applying \rulename{t-new-sync} to (\ref{eq:sbred:sync:r-com-syn:16}), we derive
$\wtps{\UEnv}{\new \ChannelA \ChannelB
(\Process \parop \ProcessQ_k \subst{\ChannelC}{\var_k})}
{\LEnv'_1, \LEnv'_2, \ASETT{\bind{\ChannelC}{\SessionTypeS}}}.$
Since $\LEnv'_1, \LEnv'_2, \ASETT{\bind{\ChannelC}{\SessionTypeS}} = \LEnv$,
we conclude
$$\wtps{\UEnv}{\new \ChannelA \ChannelB
(\Process \parop \ProcessQ_k \subst{\ChannelC}{\var_k})}{\LEnv}$$
\item Case $\rulename{r-def}$:
      \begin{eqnarray} \label{eq:sbred:sync:r-def:1}     
      \wtps{\UEnv}
      {\Def\pvar{\tilde \varY}\ProcessP{
      (\invoke\pvar{\tilde\Channel}
      \parop \ProcessQ)}}
      {\LEnv}
      \end{eqnarray}   
Applying Lemma \ref{append:lem:inversion:pure}.\ref{inver:def} 
to  (\ref{eq:sbred:sync:r-def:1}),
we have
\begin{eqnarray} \label{eq:sbred:sync:r-def:2}
\wtps{\UEnv,\pbind{\pvar}{\tilde\SessionType}}{\ProcessP}{
                   \ASET{\bind{\tilde\varY}{\tilde\SessionType}}}
\end{eqnarray}
\begin{eqnarray} \label{eq:sbred:sync:r-def:3}
 \wtps{\UEnv,\pbind{\pvar}{\tilde\SessionType}}{
          \invoke\pvar{\tilde\Channel}
          \parop\ProcessQ}{\LEnv}
\end{eqnarray} 
Applying Lemma \ref{append:lem:inversion:pure}.\ref{inver:par} to (\ref{eq:sbred:sync:r-def:3}), 
we have
\begin{eqnarray} \label{eq:sbred:sync:r-def:4}
\wtps{\UEnv,\pbind{\pvar}{\tilde\SessionType}}{\invoke\pvar{\tilde\Channel}}{\LEnv_1}
\end{eqnarray}
\begin{eqnarray} \label{eq:sbred:sync:r-def:5}
\wtps{\UEnv, \pbind{\pvar}{\tilde\SessionType}}{\ProcessQ}{\LEnv_2}
\end{eqnarray}
where $\LEnv = \LEnv_1, \LEnv_2$.
By Lemma \ref{append:lem:inversion:pure}.\ref{inver:var}
and  (\ref{eq:sbred:sync:r-def:4}), we have
\begin{eqnarray} \label{eq:sbred:sync:r-def:6}
\ASET{\bind{\tilde \Channel}{\tilde \SessionType}} \ssubt \LEnv_1
\end{eqnarray}
Applying Lemma \ref{app:lem:subs} to (\ref{eq:sbred:sync:r-def:2}),
we get
\begin{eqnarray} \label{eq:sbred:sync:r-def:7}
\wtps{\UEnv,\pbind{\pvar}{\tilde\SessionType}}{
\Process \subst{\tilde \Channel}{\tilde \varY}}
\ASET{\bind{\tilde \Channel}{\tilde \SessionType}}
\end{eqnarray}
By applying $\rulename{t-sub}$ to (\ref{eq:sbred:sync:r-def:6}) and  (\ref{eq:sbred:sync:r-def:7}), we derive
\begin{eqnarray} \label{eq:sbred:sync:r-def:8}
\wtps{\UEnv,\pbind{\pvar}{\tilde\SessionType}}{
\Process \subst{\tilde \Channel}{\tilde \varY}}{\LEnv_1}
\end{eqnarray}
By applying $\rulename{t-par}$ to (\ref{eq:sbred:sync:r-def:5}) and  (\ref{eq:sbred:sync:r-def:8}),
we derive
\begin{eqnarray} \label{eq:sbred:sync:r-def:9}
\wtps{\UEnv, \pbind{\pvar}{\tilde\SessionType}}
{\Process \subst{\tilde \Channel}{\tilde \varY} \parop \ProcessQ}
{\LEnv_1,\LEnv_2}
\end{eqnarray}
By applying $\rulename{t-def}$ to (\ref{eq:sbred:sync:r-def:9}),
we conclude
$$\wtps{\UEnv}{
\Def\pvar{\tilde \varY}
\ProcessP{
(\Process \subst{\tilde \Channel}{\tilde \varY} \parop \ProcessQ)}
}{\LEnv_1, \LEnv_2}$$ where $\LEnv_1, \LEnv_2 = \LEnv$.

\smallskip

\item Case \rulename{r-context}:
We consider the context $\new \ChannelA \ChannelB \Context$.
The proofs for other cases are similar and simpler.  
Let
\begin{eqnarray} \label{eq:sbred:sync:r-context:1}
\wtps{\UEnv}{\new \ChannelA \ChannelB \Context[\Process]}{\LEnv}
\end{eqnarray}
By applying \ref{append:lem:inversion:pure}.\ref{inver:new:sync} to (\ref{eq:sbred:sync:r-context:1}), 
we have
\begin{eqnarray} \label{eq:sbred:sync:r-context:2}
\wtps{\UEnv}{\Context[\Process]}{\LEnv, 
\ASETT{\bind{\ChannelA}{\SessionType}, 
\bind{\ChannelB}{\co{\SessionType}}}} \nonumber
\end{eqnarray}
By induction, we derive
\begin{eqnarray} \label{eq:sbred:sync:r-context:3}
\wtps{\UEnv}{\Context[\ProcessQ]}{\LEnv, 
\ASETT{\bind{\ChannelA}{\SessionType}, 
\bind{\ChannelB}{\co{\SessionType}}}}
\end{eqnarray}
By applying $\rulename{t-new-sync}$ to (\ref{eq:sbred:sync:r-context:3}),
we conclude
$$\wtps{\UEnv}{\new \ChannelA \ChannelB \Context[\ProcessQ]}{\LEnv}$$
\end{enumerate}
We have proved that if $\wtps{\UEnv}{\Process}{\LEnv}$ and 
$\Process \reds \ProcessQ$, then
\mbox{$\wtps{\UEnv}{\ProcessQ}{\LEnv}$}. %
Now, the main statement of the theorem %
can be easily proved %
by induction on the length of the sequence of transitions %
$\Process \wreds \ProcessQ$. %
\end{Proof}

As a consequence of subject reduction we get a substitution lemma for process variables, 
which can be also proved independently by induction on reduction contexts. 
\begin{lem} \label{lem:substitution:context}
If $\wtps{\UEnv, \pbind{\pvar}{\SessionType}}{\Context[\invoke\pvar{\ChannelA}]}{\LEnv}$ and $\wtps{\UEnv}{\Process}{\ASET{\bind{\ChannelA}{\SessionType}}}$ 
and $\pvar$ does not occur free in $\Context$,
then $\wtps{\UEnv}{\Context[\Process]}{\LEnv}$.
\end{lem}
\begin{Proof}
The proof is  by structural induction on reduction contexts.

\smallskip

\begin{enumerate}
\item\label{case:context:sync:empty} If $\ContextC$ is empty, i.e. $\Context = [ \; ]$,
by Lemma \ref{append:lem:inversion:pure}.\ref{inver:var},
$\ASET{\bind{ \ChannelA}{ \SessionType}} \ssubt \LEnv$.
By \rulename{t-sub}, we derive 
$\wtps{\UEnv}{\Process}{\LEnv}$.

\smallskip

\item\label{case:context:sync:new} If $\Context = \new \ChannelC \ChannelD \Context'$,
let
\begin{eqnarray} \label{eq:context:sync:new:1}
\wtps{\UEnv, \pbind{\pvar}{\SessionType}}
{\new \ChannelC \ChannelD \Context'[\invoke\pvar{\ChannelA}]}{\LEnv}
\end{eqnarray}
\begin{eqnarray} \label{eq:context:sync:process}
\wtps{\UEnv}{\Process}{\ASET{\bind{ \ChannelA}{ \SessionType}}}
\end{eqnarray}
By applying Lemma \ref{append:lem:inversion:pure}.\ref{inver:new:sync} to (\ref{eq:context:sync:new:1}),
we have
\begin{eqnarray} \label{eq:context:sync:new:3}
\wtps{\UEnv, \pbind{\pvar}{\SessionType}}{\Context'[\invoke\pvar{\ChannelA}]}{
\LEnv, \ASETT{\bind{\ChannelC}{\SessionType}, \bind{\ChannelD}{\co\SessionType}}}
\end{eqnarray}
By induction,  (\ref{eq:context:sync:process}) and  (\ref{eq:context:sync:new:3}) together imply
\begin{eqnarray} \label{eq:context:sync:new:3.1}
\wtps{\UEnv}{\Context'[\Process]}{
\LEnv, \ASETT{\bind{\ChannelC}{\SessionType}, \bind{\ChannelD}{\co\SessionType}}}
\end{eqnarray}
By applying \rulename{t-new-sync} to (\ref{eq:context:sync:new:3.1}), 
we derive
$$
\wtps{\UEnv}{\new \ChannelC \ChannelD \Context'[\Process]}{\LEnv}
$$
Notice that this proof holds also if  
$\ChannelA=  \ChannelC$ or $\ChannelA=  \ChannelD$. 
\item\label{case:context:sync:par}
If $\Context = \Context' \parop \ProcessQ$,
let
\begin{eqnarray} \label{eq:context:sync:par:1}
\wtps{\UEnv, \pbind{\pvar}{\SessionType}}{\Context'[\invoke\pvar{\ChannelA}] \parop \ProcessQ}{\LEnv}
\end{eqnarray}
and  (\ref{eq:context:sync:process}).
By applying Lemma \ref{append:lem:inversion:pure}.\ref{inver:par} to 
 (\ref{eq:context:sync:par:1}) and by
the assumption that $\pvar$ does not occur free in $\ProcessQ$, we have
\begin{eqnarray} \label{eq:context:sync:par:2}
&&\wtps{\UEnv, \pbind{\pvar}{ \SessionType}}{\Context'[\invoke\pvar{\ChannelA}]}{\LEnv_1}
\end{eqnarray}
\begin{eqnarray} \label{eq:context:sync:par:2.1}
&&\wtps{\UEnv}{\ProcessQ}{\LEnv_2}
\end{eqnarray}
where $\LEnv_1, \LEnv_2 = \LEnv$.
By induction, 
 (\ref{eq:context:sync:process}) and  (\ref{eq:context:sync:par:2}) together imply
\begin{eqnarray} \label{eq:context:sync:par:3}
\wtps{\UEnv}{\Context'[\Process]}{\LEnv_1}
\end{eqnarray}
By applying \rulename{t-par} to (\ref{eq:context:sync:par:2.1}) and  (\ref{eq:context:sync:par:3}),
we derive
$$
\wtps{\UEnv}{\Context'[\Process] \parop \ProcessQ}{\LEnv}
$$


\smallskip
\item\label{case:context:sync:def} If $\Context = \DefD{D}{\Context'}$, 
let
\begin{eqnarray} \label{eq:context:sync:def:1}
\wtps{\UEnv, \pbind{\pvar}{\tilde\SessionType}}
{\DefD{D}{\Context'[\invoke\pvar{\tilde\ChannelA}]}}{\LEnv}
\end{eqnarray}
where $D = (\pvarY (\tilde \varY) = \Process')$ 
and $\pvarY \not = \pvar$,
and  (\ref{eq:context:sync:process}).
By applying Lemma \ref{append:lem:inversion:pure}.\ref{inver:def} to (\ref{eq:context:sync:def:1})
and by the assumption that $\pvar$ does not occur free in $D$,
we have
\begin{eqnarray} \label{eq:context:sync:def:2:1}
\wtps{\UEnv,\pbind{\pvarY}{\tilde\SessionType'}}{\Process'}{
                   \ASET{\bind{\tilde\varY}{\tilde\SessionType'}}}
\end{eqnarray}
\begin{eqnarray} \label{eq:context:sync:def:2}
\wtps{\UEnv,\pbind{\pvar}{\SessionType},\pbind{\pvarY}{\tilde\SessionType'}}{\Context'[\invoke\pvar{\tilde\ChannelA}]}{\LEnv}
\end{eqnarray}
By induction, 
 (\ref{eq:context:sync:process}) and  (\ref{eq:context:sync:def:2}) imply
\begin{eqnarray} \label{eq:context:sync:def:3}
\wtps{\UEnv, \pbind{\pvarY}{\tilde\SessionType'}}{\Context'[\Process]}{\LEnv}
\end{eqnarray}
By applying \rulename{t-def} to  (\ref{eq:context:sync:def:2:1}) and  (\ref{eq:context:sync:def:3}), 
we derive\\[3pt]
\centerline{$
\wtps{\UEnv}{
\DefD{
\pvarY (\tilde \varY) = \ProcessP'}{\ContextC'[\Process]}
}{\LEnv}
$}\vspace{-5pt}
\end{enumerate}
\end{Proof}

\paragraph{\bf Theorem \ref{thm:s:sound}.} 
 The synchronous subtyping relation is sound  for the synchronous calculus.
\label{app:thm:s:sound}
\begin{Proof}
Suppose  
$\SessionTypeT \ssubt \SessionTypeS$
and 
$\wtps{\UEnv}{\Process}{\ASET{\bind{\ChannelA}{\SessionTypeT}}}$.
By applying \rulename{t-sub} 
we have
\mbox{$\wtps{\UEnv}{\Process}{\ASET{\bind{\ChannelA}{\SessionTypeS}}}$.}
By definition of $\Context[\bind{\ChannelA}{\SessionTypeS}]$, 
we have 
$\wtps{\UEnv, \pbind{\pvar}{\SessionTypeS}}{\Context[\invoke\pvar\ChannelA]}{\LEnv}$.
By Lemma \ref{lem:substitution:context}, 
we can get
$\wtps{\UEnv}{\Context[\Process]}{\LEnv}$.
By Corollary \ref{pro:comsafe:sync}, we have 
$\Context[\Process] \not \reds^\ast \error$. 
\end{Proof}

\section{Proofs of Section~\ref{sec:asynchronous_language}} 
\label{app:async_lang}

\begin{proposition}
  \label{lem:branch-in-t:unfolding}%
  $\& \in \trec\tvar.\SessionType$ \;if and only if\; %
  $\& \in \SessionType \subst{\trec\tvar.\SessionType}\tvar$.
\end{proposition}
\begin{Proof}
  \def\someDeriv{\mathcal{D}}%
  ($\implies$). %
  Assume $\& \in \trec\tvar.\SessionType$, %
  for some derivation $\someDeriv$. %
  We observe that $\someDeriv$ can only conclude %
  by the rule for 
  recursion on page~\pageref{def:branch-in-t}, %
  and from its premise %
  we have a derivation $\someDeriv'$ proving $\& \in \SessionType$. %
  We observe that $\someDeriv'$ cannot reach $\tvar$ in $\SessionType$, %
  where no rule is defined: %
  hence, if we inductively rewrite $\someDeriv'$ %
  substituting each occurrence of $\tvar$ with $\trec\tvar.\SessionType$, %
  we obtain a derivation 
  proving %
  $\& \in \SessionType \subst{\trec\tvar.\SessionType}\tvar$.
  
  ($\impliedby$). %
  Assuming $\& \in \SessionType \subst{\trec\tvar.\SessionType}\tvar$, %
  for some derivation $\someDeriv$, %
  we have two cases. %
  If $\someDeriv$ %
  traverses a substitution %
  of $\tvar$ in $\SessionType$ with $\trec\tvar.\SessionType$, %
  then we have a corresponding sub-derivation $\someDeriv'$ proving %
  $\& \in \trec\tvar.\SessionType$, %
  which is the thesis\footnote{%
    \label{footnote:proof:branch-in-t:unfolding}%
    We could also prove that this case is actually absurd, %
    but this detail is not relevant for the main proof.
  }. %
  Otherwise, %
  $\someDeriv$ reaches its axioms \emph{without} traversing any substitution %
  of $\tvar$ in $\SessionType$ with $\trec\tvar.\SessionType$: %
  hence, by inductively rewriting $\someDeriv$ %
  removing such a substitution %
  and restoring $\tvar$ in all terms, %
  we obtain a derivation $\someDeriv'$ proving $\& \in \SessionType$. %
  By using $\someDeriv'$ as a premise %
  for the rule for 
  recursion on page~\pageref{def:branch-in-t}, %
  we obtain a derivation 
  proving $\& \in \trec\tvar.\SessionType$.
\end{Proof}

 To show the transitivity of asynchronous subtyping (Theorem~\ref{asubtrans})
 we extend the asynchronous subtyping to asynchronous contexts (Table \ref{tab:cont:asubt}). 
 It is easy to verify that 
 if $\AContext\asubt\AContext'$, and $N$ and $N'$
 are the set of hole indices of $\AContext$ and $\AContext'$, respectively, 
 then $N' \subseteq N$.
 If previous conditions hold and $\SessionTypeT_n \asubt \SessionTypeT'_n$ for all $n\in N'$,  
 then $\AContext[\SessionTypeT_n]^{n\in N}\asubt\AContext'[\SessionTypeT'_n]^{n\in N'}$.
 \begin{table}[h]
 $$
 \begin{array}{@{}c@{}}
 \inferrule[\rulename{sub-empty}]{}
 {[ \; ]^{n }\subt[ \; ]^{n }}
 \qquad 
 \cinferrule[\rulename{sub-cont}]{
   \forall i\in I: \TypeU_i \subt \TypeU'_i
   \quad \AContext_i \subt \AContext'_i
 }{
  \Branch_{i\in I \cup J}
   \In{\Tag_i}{\TypeU_i}.\AContext_i
     \subt
     \Branch_{i\in I}
   \In{\Tag_i}{\TypeU'_i}.\AContext'_i
 }
 \end{array}
 $$
 \caption{Subtyping for asynchronous contexts.}\label{tab:cont:asubt}
 \end{table}
Two more lemmas on properties of $\asubt$ are handy.
\begin{lem} \label{brancontain}
If $\SessionTypeT \asubt \SessionTypeS$ and $\& \in \SessionTypeS$, 
then $\& \in \SessionTypeT$.
\end{lem}
\begin{Proof}
By cases on the rules defining $\asubt$.
\end{Proof}

\begin{lem}\label{asubc}
If\; %
$\AContext[\Select_{i\in I_n} \Out{\Tag^{n}_i}{\Type_i^{n}}.\SessionTypeT_i^{n} ]^{n\in N}\asubt\SessionType$, %
\;then %
there exists\; %
$\AContext'[\,]^{n \in N'}$ %
\;such that:
\begin{enumerate}
\item%
  \label{item:asubc:i}%
  $\AContext \asubt \AContext'$ %
  \;and\; %
  $N' \subseteq N$
\item%
  \label{item:asubc:ii}%
  for all $n \in N'$, %
  there exists\; %
  $\AContext_n[\,]^{m \in M_n}$ %
  \;such that
  \begin{itemize}
  \item%
    for all $i \in I_n$ and $m \in M_n$,\; %
    there exist\; %
    $H_{n,m}$,\; %
    $\Type_i^{n,m}$ %
    \;and\; %
    $\SessionTypeT_i^{n,m}$ %
    \;such that:
    \begin{enumerate}
    \item%
      \label{item:asubc:ii:a}%
      $\Type_i^{n,m}\asubt \Type_i^{n}$
    \item%
      \label{item:asubc:ii:b}%
      $\SessionType_i^{n}\asubt\AContext_n[\SessionType_i^{n,m}]^{m \in M_n}$
    \item%
      \label{item:asubc:ii:c}%
      $\SessionType=\AContext'[\AContext_n[\Select_{i\in I_n\cup H_{n,m}} \Out{\Tag_i^{n}}{\Type_i^{n,m}}.\SessionTypeT_i^{n,m}]^{m\in M_n} ]^{n\in N'}$
    \end{enumerate}
  \end{itemize}
\end{enumerate}
\end{lem}
\begin{Proof}
By structural induction on $\AContext$. %

In the base case where $\AContext$ is just one hole %
(i.e., $N$ is a singleton), %
then we choose $\AContext'$ to be just one hole, too; %
hence, we get $\AContext \asubt \AContext'$ (by rule \rulename{sub-empty}) %
and $N' = N=\{ 1\}$, %
thus satisfying item~\ref{item:asubc:i} of the statement. %
Then, according to the rules of Table \ref{tab:sync:ssubt}, 
we have either: %
\begin{itemize}
\item%
  by rule \rulename{sub-sel},\; %
$\SessionType=\Select_{i\in I_1\cup J} \Out{\Tag_i}{\Type^1_i}.\SessionTypeT^1_i$ \;and
  \; %
  $\Type^1_i\asubt\Type_i$ 
  \;and\; $\SessionTypeT_i\asubt\SessionTypeT^1_i$ for all $i\in I_1$. %
  Then we let:
 \begin{itemize}
  \item%
    $\AContext_1$ be just one hole %
    (i.e., $M_1=\{ 1 \}$ is a singleton); %
  \item%
       $H_{1,1} = J$,\; %
    and for all $i \in I_1$,\; %
    $S_i^{1,1} = S_i^1$ %
    \;and\; %
    $\SessionTypeT_i^{1,1} = T_i^1$;
  \end{itemize}
\item%
  otherwise, by rule \rulename{sub-perm-async},\; %
  there exists\; $\AContext''[\,]^{m \in M}$ \;such that $\& \in \AContext''$ %
  and %
  $\SessionType=\AContext''[\Select_{i\in I\cup J_m} \Out{\Tag_i}{\Type_i^{m}}.\SessionTypeT_i^{m} ]^{m\in M}$, %
  \;and for all $i\in I_1$,\; %
  $\SessionTypeT_i\asubt\AContext''[\SessionTypeT_i^{m}]^{m \in M}$ %
  \;and for all $m \in M$,\;
  $\TypeU_i^{m}\asubt\TypeU_i$.\; %
  Then we let:
  \begin{itemize}
  \item%
    $\AContext_1 = \AContext''$ (and thus, $M_1 = M$); %
  \item%
    for all $m \in M_1$,\; %
    $H_{1,m} = J_m$,\; %
    and for all $i \in I_1$,\; %
    $S_i^{1,m} = S_i^m$ %
    \;and\; %
    $\SessionTypeT_i^{1,m} = T_i^m$.
  \end{itemize}
\end{itemize}
In both cases, we satisfy items~\ref{item:asubc:ii:a}, \ref{item:asubc:ii:b}
and \ref{item:asubc:ii:c} of the statement.

For the induction step,
let $\AContext = \Branch_{j\in J} \In{\Tag_j}{\Type_j}.\AContext_j$
with $N_j$ being the set of hole indices occurring in $\AContext_j$ %
and $N = \bigcup_{j \in J} N_j$. 
Since by assumption
$$
\begin{array}{l}
\Branch_{j\in J} \In{\Tag_j}{\Type_j}.
\AContext_j [ \Select_{i\in I_{n}} \Out{\Tag^{n}_i}{\Type_i^{n}}.\SessionTypeT_i^{n} ]^{n \in N_j} \;\asubt\; \SessionType
\end{array}
$$
and since such a relation can only hold by rule \rulename{sub-bra}, 
the shape of $\SessionTypeT$ must be
\begin{equation}
  \label{eq:asubc:t}%
  \SessionTypeT \;=\; \Branch_{j\in J'} \In{\Tag_j}{\Type'_j}.\SessionTypeT'_j
\end{equation}
where $J' \subseteq J$ and, %
for all $j \in J'$,\; %
$\Type_j \asubt \Type'_j$ %
\;and\; %
$\AContext_j [ \Select_{i\in I_{n}} \Out{\Tag^{n}_i}{\Type_i^{n}}.\SessionTypeT_i^{n} ]^{n \in N_j} \asubt \SessionTypeT'_j$. %
By the induction hypothesis, for all $j \in J'$, %
there exists\; $\AContext'_j[\,]^{n \in N'_j}$ \;such that %
(by item~\ref{item:asubc:ii:c}): %
\begin{equation}
  \label{eq:asubc:tj}%
  \SessionTypeT'_j
  \;=\;\AContext'_j[ 
    \AContext_{n}[{\textstyle%
        \Select_{i\in I_{n} \cup H_{n, m}} \Out{\Tag^{n}_i}{\Type_i^{n, m}}.\SessionTypeT_i^{n, m}%
    }]^{m \in M_{n}}%
  ]^{n \in N'_j}
\end{equation}
where:
\begin{itemize}
\item%
  $\AContext_j \asubt \AContext'_j$ \;and\; $N'_j \subseteq N_j$ %
  (from item~\ref{item:asubc:i} of the statement); %
\item%
  for all $n \in N'_j$, $i \in I_{n}$ and $m \in M_{n}$,\; %
  we have: %
  \begin{enumerate}
  \item%
    \label{item:proof:asubc:ii:a}%
    $\SessionTypeS_i^{n, m} \asubt \SessionTypeS_i^{n}$ %
    (from item~\ref{item:asubc:ii:a}), %
    \;and\; %
  \item%
    \label{item:proof:asubc:ii:b}%
    $\SessionTypeT_i^{n} \asubt \AContext_{n} [\SessionTypeT_i^{n, m}]^{m \in M_{n}}$ %
    (from item~\ref{item:asubc:ii:b}). %
  \end{enumerate}
\end{itemize}

Now, let %
$\AContext' = \Branch_{j\in J'} \In{\Tag_j}{\Type'_j}.{\AContext'_j}$
\;and\; $N' = \bigcup_{j \in J'} N'_j$. %
Since for all $j \in J' \subseteq J$ we have\; %
$N'_j \subseteq N_j$,\;%
we also get \;$N' \subseteq N$;\; %
moreover, %
since for all $j \in J'$ %
we have\; %
$\SessionTypeS_j \asubt \SessionTypeS'_j$ %
\;and\; %
$\AContext_j \asubt \AContext'_j$,\; %
by \rulename{sub-cont} %
we also get\; $\AContext \asubt \AContext'$:\; %
hence, we satisfy item~\ref{item:asubc:i} of the statement. %
Furthermore, %
from items~\ref{item:proof:asubc:ii:a} and \ref{item:proof:asubc:ii:b} above %
we satisfy respectively items~\ref{item:asubc:ii:a} and
\ref{item:asubc:ii:b} %
of the statement. %
Finally, %
from (\ref{eq:asubc:t}) and (\ref{eq:asubc:tj}) %
we obtain:
$$
\begin{array}{lll}
\SessionTypeT &  =  & 
\Branch_{j\in J'} \In{\Tag_j}{\Type'_j}.{\AContext'_j[ 
\AContext_{n}[\Select_{i\in I_{n} \cup H_{n, m}} \Out{\Tag^{n}_i}{\Type_i^{n, m}}.\SessionTypeT_i^{n, m}]^{m \in M_{n}} ]^{n \in N'_j}}
\end{array}
$$
from which we get:
$$
\begin{array}{lll}
\SessionTypeT &  =  &%
\AContext'[\AContext_n[\Select_{i\in I_n\cup H_{n,m}} \Out{\Tag_i^{n}}{\Type_i^{n,m}}.\SessionTypeT_i^{n,m}]^{m\in M_n} ]^{n\in N'}
\end{array}
$$
thus satisfying item \ref{item:asubc:ii:c} of the statement.
\end{Proof}

\paragraph{\bf Theorem \ref{asubtrans}}
The relation $\asubt$ is transitive.
\begin{Proof}
It suffices to show $\asubtTrans \subseteq \asubt$,
where $\SessionTypeT \asubtTrans \SessionTypeS$
if there exists $\SessionTypeV$ such that 
$\SessionTypeT \asubt \SessionTypeV$ and $\SessionTypeV \asubt \SessionTypeS$.
We proceed by cases on the rules concluding %
$\SessionTypeT \asubt \SessionTypeV$ and $\SessionTypeV \asubt \SessionTypeS$.

If $\SessionTypeT \asubt \SessionTypeV$ %
by rule \rulename{sub-sel} (respectively \rulename{sub-bra}), %
and %
$\SessionTypeV \asubt \SessionTypeS$ %
by rule \rulename{sub-sel} (respectively \rulename{sub-bra}),\quad %
then the proof is straightforward, %
and $\SessionTypeT \asubt \SessionTypeS$ %
holds again by rule \rulename{sub-sel} (respectively \rulename{sub-bra}).

If $\SessionTypeT \asubt \SessionTypeV$ %
by rule \rulename{sub-sel} (respectively \rulename{sub-perm-async}), %
and %
$\SessionTypeV \asubt \SessionTypeS$ %
by rule \rulename{sub-perm-async} (respectively \rulename{sub-bra}),\quad %
then we have\; %
$\SessionTypeT = \Select_{i\in I} \Out{\Tag_i}{\Type_i}.\SessionTypeT_i$ %
\;and\;
$\SessionTypeV =\AContext[\Select_{i\in I\cup J_n} \Out{\Tag_i}{\Type_i^{n}}.\SessionTypeT_i^{n} ]^{n\in N}$\; %
(where $\AContext$ is just one hole %
when $\SessionTypeT \asubt \SessionTypeV$ holds by \rulename{sub-sel}). %
From $\SessionTypeV \asubt \SessionTypeS$, %
by Lemma~\ref{asubc} (item~\ref{item:asubc:ii:c}) %
we have: %
\[
  \SessionTypeS\;\;=\;\;\AContext'[\AContext_n[{\textstyle%
    \Select_{i\in I \cup J_n\cup H_{n,m}} \Out{\Tag_i}{\Type_i^{n,m}}.\SessionTypeT_i^{n,m}%
    }]^{m\in M_n} ]^{n\in N'}
  \]
\noindent
where:
\begin{enumerate}
\item%
  \label{item:proof:asubtrans:i}%
  $\AContext \asubt \AContext'$ \;and\; $N' \subseteq N$ %
  \;(by item~\ref{item:asubc:i} of Lemma~\ref{asubc});
\item%
  \label{item:proof:asubtrans:ii}%
  for all $n\in N'$, $m \in M_n$ and $i\in I \cup J_n$,\; %
  we have %
  $\Type_i^{n,m}\asubt \Type_i^{n}$ %
  \;and\; %
  $\SessionType_i^{n}\asubt\AContext_n[\SessionType_i^{n,m}]^{m \in M_n}$ %
  (resp.~from items~\ref{item:asubc:ii:a} and \ref{item:asubc:ii:b} %
  of Lemma~\ref{asubc}).
\end{enumerate}
Moreover, from $\SessionTypeT \asubt \SessionTypeV$, we also have:
\begin{enumerate}[resume]
\item%
  \label{item:proof:asubtrans:iii}%
  for all $n \in N$,\; %
  $\TypeU_i^{n}\asubt\TypeU_i$;
\item%
  \label{item:proof:asubtrans:iv}%
  for all $i\in I$,\; %
  $\SessionTypeT_i\asubt\AContext[\SessionTypeT_i^{n}]^{n \in N}$.
\end{enumerate}
By the definition of $\asubtTrans$, %
from items~\ref{item:proof:asubtrans:ii} %
and \ref{item:proof:asubtrans:iii} above, %
we have \;$\Type_i^{n,m}\asubtTrans \Type_i$\; %
for all $i \in I$, $m \in M_n$ and $n \in N'$. %
Furthermore, %
from\; $\AContext \asubt \AContext'$ %
(item~\ref{item:proof:asubtrans:i}) %
\;and\; %
$\SessionType_i^{n}\asubt\AContext_n[\SessionType_i^{n,m}]^{m \in M_n}$ %
(item~\ref{item:proof:asubtrans:ii}), %
for all $i \in I \cup J_n$ and $n \in N'$
we have:
\begin{eqnarray} \label{pf:async:trans:eq1}
\AContext[\SessionTypeT_i^{n}]^{n \in N}
\;\asubt\; \AContext'[\AContext_n[\SessionType_i^{n,m}]^{m \in M_n}]^{n \in N'}
\end{eqnarray}

Now, %
according to the definition of $\asubtTrans$,
(\ref{pf:async:trans:eq1}) and
$\SessionTypeT_i\asubt\AContext[\SessionTypeT_i^{n}]^{n \in N}$
(from item~\ref{item:proof:asubtrans:iv})
\;imply\; %
$\SessionTypeT_i \asubtTrans \AContext'[\AContext_n[\SessionType_i^{n,m}]^{m \in M_n}]^{n \in N'}$,
for all $i \in I$. %
Moreover, %
we observe that %
since 
$\SessionTypeV \asubt \SessionTypeS$ %
holds by rule \rulename{sub-perm-async} (resp.~\rulename{sub-bra}), %
then we have $\& \in \SessionTypeS$; %
and since $\SessionTypeV \asubt \SessionTypeS$ and $\SessionTypeT \asubt \SessionTypeV$, %
by applying Lemma \ref{brancontain} twice %
we obtain %
$\&\in\SessionTypeV$ %
and %
$\&\in\SessionTypeT$, %
i.e.~%
$\&\in\SessionTypeT_i$ for all $i\in I$.
Thus 
$\SessionTypeT \asubtTrans \SessionTypeS$ %
agrees with rule \rulename{sub-perm-async}.
\end{Proof}

The remaining of this section is devoted to the
proof of the subject reduction theorem and of the soundness of the asynchronous subtyping.

\begin{lem}[Inversion lemma for asynchronous processes]
\label{append:lem:inversion:async}$\;$ 
\begin{enumerate}
\item\label{inver:end:async} 
         If $\wtpa{\UEnv}{\idle}{\LEnv}$, 
         then $\LEnv$ is end-only.        
\item\label{inver:var:async} If 
         $\wtpa{\UEnv}
         {\invoke\pvar{\tilde\Name}}{\LEnv}$,
         then $\UEnv = \UEnv', \pbind{\pvar}{\tilde\SessionType}$ and
         $\ASET{\bind{\tilde \Name}{\tilde \SessionType}} \asubt \LEnv$.         
\item\label{inver:bra:async} If 
          $\wtpa{\UEnv}{\sum_{i\in I} \receive{\Name}{\Tag_i}{\var_i}.\Process_i}{\LEnv}$,
          then $\LEnv = \LEnv', \ASETT{\bind{\Name}{\Branch_{j\in J}
                         \In{\Tag_j}{\Type_j}.\SessionTypeT_j}}$ and $J \subseteq I$ and\\
          $\forall j \in J:\wtpa{\UEnv}{\Process_j}{\LEnv',\ASETT{\bind{\Name}{\SessionTypeT_j},
                                                             \bind{\var_j}{\SessionTypeS_j}}}$.
\item\label{inver:sel:async} If $\wtpa{\UEnv}{\send{\NameU}{\Tag}{\Name'}.\Process}{\LEnv}$,
         then $\LEnv = 
         {\LEnv',\ASETT{\bind{\Name}{\SessionTypeT},
          \bind{\Name'}{\SessionTypeS}}}$ and         
         $\Out{\Tag}{\SessionTypeS}.\SessionTypeT' \asubt \SessionTypeT$
         and
         \mbox{$\wtpa{\UEnv}{\Process}{\LEnv', \ASETT{\bind{\NameU}{\SessionTypeT'}}}$.}      
\item\label{inver:par:async} If $\wtpa{\UEnv}{\Process_1 \parop \Process_2}{\LEnv}$,
          then $\LEnv = \LEnv_1, \LEnv_2$ and $\wtpa{\UEnv}{\Process_1}{\LEnv_1}$ and
          $\wtpa{\UEnv}{\Process_2}{\LEnv_2}$.         
\item\label{inver:choice:async} If $\wtpa{\UEnv}{\Process_1 \choice \Process_2}{\LEnv}$, 
          then $\wtpa{\UEnv}{\Process_1}{\LEnv}$ and 
          $\wtpa{\UEnv}{\Process_2}{\LEnv}$.
\item\label{inver:def:async} If $\wtpa{\UEnv}
        {\Def{\pvar}{\tilde\varY}{\ProcessP}{\ProcessQ}}{\LEnv}$,
         then $\wtpa{\UEnv,\pbind{\pvar}{\tilde\SessionType}}{\ProcessP}{
                   \ASET{\bind{\tilde\varY}{\tilde\SessionType}}}$ and 
        $\wtpa{\UEnv,\pbind{\pvar}{\tilde\SessionType}}{
          \ProcessQ}{\LEnv}$.          
\item\label{inver:new:async} 
         If $\wtpa{\UEnv}{\new{\ChannelA}{\ChannelB}\Process}{\LEnv}$,
         then $\wtpa{\UEnv}{\Process}
         {\LEnv,
          \ASETT{
          \bind{\ChannelA}{\SessionType_1},
          \bind{\ChannelB}{\SessionType_2},
          \queuetype \ChannelB \ChannelA \QueueType_1,
          \queuetype \ChannelA \ChannelB \QueueType_2}
          }
          $          
         and
         $\remainder{\SessionType_1}{\QueueType_1}
         \dual
         \remainder{\SessionType_2}{\QueueType_2}$. 
   \item\label{inver:q:empty} If 
   $\wtpa{\UEnv}{\queue \ChannelB\ChannelA \EmptyQueue}{\LEnv}$,
then $\ASET{\queuetype \ChannelB \ChannelA \EmptyQueueT} \asubt\LEnv$.
\item\label{inver:q} If $\wtpa{\UEnv}{
                     \queue \ChannelB \ChannelA 
                     \Queue \qconc \msg{\Tag}{\ChannelC}}
                     {\LEnv}$,
                 then
                $\LEnv = \LEnv', 
                 \ASETT{
                 \bind{\ChannelC}{\SessionTypeS},
                 \queuetype \ChannelB \ChannelA
                 \QueueType \qconc \tmsg{\Tag}{\SessionTypeS'}
                 }$
                 and $\SessionTypeS'\asubt\SessionTypeS$ and\\ 
                 $\wtpa{\UEnv}{\queue \ChannelB \ChannelA  \Queue}
                 {\LEnv',
                 \ASETT{\queuetype \ChannelB \ChannelA  \QueueType}}$.
\end{enumerate}
\end{lem}
\begin{Proof}
By induction on derivations.
\end{Proof}

\begin{lem}[Substitution lemma  for asynchronous processes]
\label{app:lem:subs:async}
If $\wtpa{\UEnv}{\Process}{\LEnv, \ASETT{\bind{y}{\SessionType}}}$
and $\ChannelA \not \in \dom(\LEnv)$, 
then $\wtpa{\UEnv}{\Process \subst{a}{y}}{\LEnv, \ASETT{\bind{a}{\SessionType}}}$.
\end{lem}
\begin{Proof}
The proof is by induction on the derivation of $\wtpa{\UEnv}{\Process}{\LEnv, \ASETT{\bind{y}{\SessionType}}}.$
The only interesting case is:
$$\inferrule{
     \wtpa{\UEnv}{\Process}
      {
      \LEnv,\bind{y}{\SessionType},
      \ASETT{
      \bind{\ChannelB}{\SessionType_1},
      \bind{\ChannelC}{\SessionType_2},
      \queuetype \ChannelC \ChannelB \QueueType_1,
      \queuetype \ChannelB \ChannelC \QueueType_2}
     }
      \quad
      \remainder{\SessionType_1}{\QueueType_1}
      \dual
      \remainder{\SessionType_2}{\QueueType_2}
    }{
      \wtpa{\UEnv}{\new{\ChannelB}{\ChannelC}\Process}{\LEnv,\bind{y}{\SessionType}}
  }$$
By induction
     $\wtpa{\UEnv}{\Process \subst{\ChannelA}{y}}
      {
      \LEnv, \bind{\ChannelA}{\SessionType},
      \ASETT{
      \bind{\ChannelB}{\SessionType_1},
      \bind{\ChannelC}{\SessionType_2},
      \queuetype \ChannelC \ChannelB \QueueType_1,
      \queuetype \ChannelB \ChannelC \QueueType_2}.
     }$
     Thus by \rulename{t-new-async}, 
     we conclude
     $\wtpa{\UEnv}{\new \ChannelB \ChannelC\ProcessP \subst{\ChannelA}{y}}{\LEnv,\bind{\ChannelA}{\SessionType}}.$
\end{Proof}
\begin{lem}[Types of queues] \label{lem:queue:type}
If $\wtpa{\UEnv}{
                     \queue \ChannelB \ChannelA 
                     \msg{\Tag}{\ChannelC} \qconc\Queue}
                     {\LEnv}$, 
then $\LEnv=\LEnv', 
\ASETT{\bind{\ChannelC}{\SessionTypeS},\queuetype \ChannelB \ChannelA
\tmsg{\Tag}{\SessionTypeS'} \qconc\QueueType}$, 
and $\SessionTypeS'\asubt\SessionTypeS$ and 
$\wtpa{\UEnv}{
                     \queue \ChannelB \ChannelA 
                     \Queue}
                     {\LEnv', \ASETT{\queuetype \ChannelB \ChannelA \QueueType}}$.
\end{lem}
\begin{Proof} By induction on $n$ we show:\\
\centerline{If $\wtpa{\UEnv}{
                     \queue \ChannelB \ChannelA 
                     \msg{\Tag_1}{\ChannelC_1} \qconc\ldots \qconc\msg{\Tag_n}{\ChannelC_n}}
                     {\LEnv}$, 
then 
$\LEnv=\LEnv', 
\ASETT{\bind{\ChannelC_1}{\SessionTypeS_1},\ldots, 
\bind{\ChannelC_n}{\SessionTypeS_n}, 
\queuetype \ChannelB \ChannelA
\tmsg{\Tag_1}{\SessionTypeS_1'} \qconc\ldots\qconc \tmsg{\Tag_n}{\SessionTypeS'_n}},$} 
where  $\LEnv'$ is end-only and $\SessionTypeS_i'\asubt\SessionTypeS_i$ for $1\leq i\leq n$.
The first step follows from Lemma \ref{append:lem:inversion:async}.\ref{inver:q:empty}. 
The induction step follows from Lemma \ref{append:lem:inversion:async}.\ref{inver:q}. 
\end{Proof}

\begin{thm} \label{thm:sbr:async:struct}
If $\wtpa{\UEnv}{\Process}{\LEnv}$ and $\Process \equiv \Process'$, then
$\wtpa{\UEnv}{\Process'}{\LEnv}$.
\end{thm}
\begin{Proof}
The proof is by induction on $\equiv$. The most interesting case is rule 
\rulename{s-queue-equiv}. Let 
$\Queue \equiv \Queue'$  and 
$\wtpa{\UEnv}{\queue \ChannelA \ChannelB \Queue}{\LEnv}.$
The equivalence $\Queue \equiv \Queue'$ should come from one of the following cases:
$\Queue \equiv \EmptyQueue \qconc \Queue = \Queue'$, or
$\Queue \equiv \Queue \qconc \EmptyQueue = \Queue'$, or
$\Queue  = \Queue_1 \qconc ( \Queue_2 \qconc \Queue_3) 
\equiv (\Queue_1 \qconc \Queue_2) \qconc \Queue_3 = \Queue'$.
For all cases, 
the messages in $\Queue$ and $\Queue'$ are the same {and they are in the same order}. 
Therefore, by Lemma \ref{lem:queue:type}, if
$\wtpa{\UEnv}{\queue \ChannelA \ChannelB \Queue'}{\LEnv'}$, then
$\LEnv' = \LEnv$.
\end{Proof}

We extend the session remainder to type contexts in the expected way.

\begin{lem} \label{lem:context:remainder} 
$$
\remainder{\AContextf{\SessionType_n}^{n \in N}}\QueueType=\begin{cases}
\AContextfp{\SessionType_n}^{n \in N'}    & 
\text{if }\remainder{\AContextf{\;}^{n \in N}}\QueueType=\AContextfp{\;}^{n \in N'}, \\
\remainder{\SessionType_{n_0}}{\QueueType_0}  & 
\text{if }\remainder{\AContextf{\;}^{n \in N}}\QueueType=\remainder{[ \; ]^{n_0}}{\QueueType_0}.
\end{cases}
$$
\end{lem}
\begin{Proof}
The proof by cases is easy.
\begin{itemize}
\item If $\remainder{\AContextf{\;}^{n \in N}}{\QueueType}=\AContextfp{\;}^{n \in N'}$,
we have $\remainder{\AContextf{\SessionType_n}^{n \in N}}{\QueueType}
=\AContextfp{\SessionType_n}^{n \in N'}$.
\item If $\remainder{\AContextf{\;}^{n \in N}}\QueueType= \remainder{[ \; ]^{n_0}}{\QueueType_0}$,
we have $\remainder{\AContextf{\SessionType_n}^{n \in N}}{\QueueType}= 
\remainder{\SessionType_{n_0}}{\QueueType_0}$.\qedhere
\end{itemize}
\end{Proof}

\paragraph{\bf Lemma \ref{lem:redsessiontype:async}.}
\label{app:lem:redsessiontype:async}
{\em If $ 
\LEnv\Red\LEnv'$
and $\LEnv$ is  balanced, then $\LEnv'$ is balanced.}
\begin{Proof} By cases on the definition of $\Red$.
\begin{enumerate}
\item 
Let
$
\LEnv,
\ASETT{
\bind{\ChannelA}{\SessionType}, 
\queuetype \ChannelB \ChannelA \QueueType}
\Red
\LEnv',
\ASETT{
\bind{\ChannelA}{\SessionType'}, 
\queuetype \ChannelB \ChannelA \QueueType'}
$
and $\remainder{\SessionType}{\QueueType}$ be defined.
\begin{enumerate}
\item If the applied rule is \rulename{tr-out}:
$\SessionTypeT = 
\AContextf{\Select_{i\in I_n} \Out{\Tag_i^{n}}{\Type_i^{n}}.\SessionTypeT_i^{n}}^{n \in N}$
and\\ 
$
\forall n \in N ~ \exists i_n\in I_n : 
\Tag_{i_n}^{n}=\Tag$ and 
$\Type_{i_n}^{n} \asubt \Type $,
then 
we get $\SessionTypeT'=\AContextf{\SessionTypeT^{n}_{i_n}}^{n \in N}$ 
and $\QueueType'=\QueueType$. 
By Lemma \ref{lem:context:remainder} we have the following subcases:
\begin{enumerate}
\item\label{redbalanced:case1} If $\remainder{\AContextf{\;}^{n \in N}}{\QueueType}=\AContextfp{\;}^{n \in N'}$, then $
\begin{array}{l}
\remainder{\SessionTypeT'}{\QueueType'} 
=
\remainder{\AContextf{\SessionTypeT^{n}_{i_n}}^{n \in N}}{\QueueType}
=
\AContext' [ \SessionTypeT_{i_n}^{n} ]^{n \in N'},
\end{array}
$
which is defined.
\item\label{redbalanced:case1:a} If $\remainder{\AContextf{\;}^{n \in N}}{\QueueType}=\remainder{[ \; ]^{n_0}}{\QueueType_0},$
 then $
\remainder{\SessionTypeT'}{\QueueType'} 
=
\remainder{\AContextf{\SessionTypeT^{n}_{i_n}}^{n \in N}}{\QueueType}
=\remainder{\SessionTypeT^{n_0}_{i_{n_0}}}{\QueueType_0} 
$, which is defined because we know that
$\remainder{
\Select_{i\in I_{n_0}} \Out{\Tag_i^{n_0}}{\Type_i^{n_0}}.\SessionTypeT_i^{n_0}}{\QueueType_0}$ 
is defined and, therefore, $\forall i \in I_{n_0}:\remainder{\SessionTypeT_i^{n_0}}{\QueueType_0}$ is defined.
\end{enumerate}
\item\label{redbalanced:case1:b} If the applied rule is \rulename{tr-in}:
$\SessionTypeT = \Branch_{i\in I} \In{\Tag_i}{\Type_i}.\SessionTypeT_i$
and $\QueueType = \tmsg{\Tag_k}{\SessionTypeS} \qconc\QueueType'$ and 
$\SessionTypeS_k\asubt\SessionTypeS$, 
and $T'=T_k$ and $k \in I$.
We get
$
\remainder{\Branch_{i\in I} \In{\Tag_i}{\Type_i}.\SessionTypeT_i}{\tmsg{\Tag_k}{\SessionTypeS} \qconc\QueueType'}
=\remainder{\SessionTypeT_k}{\QueueType'}.
$
Therefore $\remainder{\SessionTypeT_k}{\QueueType'}$ is defined because it is equal to
$\remainder{\SessionTypeT}{\tmsg{\Tag_k}{\SessionTypeS} \qconc\QueueType'}$, which
is defined by assumption.
\end{enumerate}

\item
Let
$
\LEnv,
\ASETT{ 
\bind{\ChannelA}{\SessionType_1}, 
\bind{\ChannelB}{\SessionType_2},
\queuetype \ChannelB \ChannelA \QueueType_1,
\queuetype \ChannelA \ChannelB \QueueType_2}
\Red 
\LEnv',
\ASETT{
\bind{\ChannelA}{\SessionType'_1}, 
\bind{\ChannelB}{\SessionType'_2},
\queuetype \ChannelB \ChannelA \QueueType'_1,
\queuetype \ChannelA \ChannelB \QueueType'_2}
$ and $
\remainder{\SessionType_1}{ \QueueType_1}
\dual \remainder{\SessionType_2}{\QueueType_2}.
$
\begin{enumerate}
\item\label{redbalanced:case2:a}  
If the applied rule is \rulename{tr-out}:
$\SessionTypeT_1 = 
\AContextf{\Select_{i\in I_n} \Out{\Tag_i^{n}}{\Type_i^{n}}.\SessionTypeT_i^{n}}^{n \in N}$
and 
$
\forall n \in N ~ \exists k_n\in I_n : 
\Tag_{k_n}^{n}=\Tag$ and $\Type_{k_n}^{n} \asubt\Type
$
and
$ \SessionTypeT'_1= \AContextf{\SessionTypeT^{n}_{k_n}}^{n \in N}, \SessionTypeT'_2=\SessionTypeT_2, \QueueType'_1=\QueueType_1,
       \QueueType_2'=\QueueType_2 \qconc \tmsg{\Tag}{\SessionTypeS}.
$
By Lemma \ref{lem:context:remainder} we have the following subcases: 
\begin{enumerate}
\item If $\remainder{\AContextf{\;}^{n \in N}}\QueueType_1=\AContextfp{\;}^{n \in N'}$, then we have
$
\remainder{\SessionType_1}{\QueueType_1} 
=
\AContextfp{\Select_{i\in I_n} \Out{\Tag_i^{n}}{\Type_i^{n}}.{\SessionTypeT_i}^{n}}^{n \in N'}$ and \mbox{$
\remainder{\SessionType'_1}{\QueueType'_1} 
= \remainder{\SessionType'_1}{\QueueType_1} 
= \AContext' [ \SessionTypeT_{k_n}^{n} ]^{n \in N'}.
$} 
By $\remainder{\SessionType_1}{ \QueueType_1}
\dual \remainder{\SessionType_2}{ \QueueType_2}$, we get
$$
\remainder{\SessionTypeT_2}{\QueueType_2}
~ = ~\co{\AContextfp{\textstyle \Select_{i\in I_n} \Out{\Tag_i^{n}}{\Type_i^{n}}.{\SessionTypeT_i}^{n}}^{n \in N'}}
~ = ~\co{\AContext'}[
\Branch_{i \in I_n} \In{\Tag_i^{n}}{\Type_i^{n}}.{\co{\SessionTypeT_i}}^{n}
]^{n \in N'}
$$
which implies\\[3pt]
$
\remainder{\SessionTypeT'_2}{\QueueType'_2} 
~= ~\remainder{\SessionTypeT_2}{\QueueType_2 \qconc \tmsg{\Tag}{\SessionTypeS}}
~=~ \remainder{\co{\AContext'}[
\Branch_{i \in I_n} \In{\Tag_i^{n}}{\Type_i^{n}}.{\co{\SessionTypeT_i}}^{n}
]^{n \in N'}}{}~~{\tmsg{\Tag}{\SessionTypeS}}
~=~
\co{\AContext'}[\co{\SessionTypeT_{k_n}}^{n} ]^{n \in N'}.\\[3pt]
$
We conclude $\remainder{\SessionType'_1}{ \QueueType'_1}
\dual \remainder{\SessionType'_2}{ \QueueType'_2}$.  
\item If $\remainder{\AContextf{\;}^{n \in N}}{\QueueType_1}=\remainder{[ \; ]^{n_0}}{\QueueType_0}$,
 then we have
$
\remainder{\SessionType_1}{ \QueueType_1} 
=
\Select_{i\in I_{n_0}} \Out{\Tag_i^{n_0}}{\Type_i^{n_0}}.\remainder{\SessionTypeT_i^{n_0}}
{\QueueType_0}$ and \mbox{$
\remainder{\SessionType'_1}{ \QueueType'_1}$} 
$=\remainder{\SessionTypeT_{k_{n_0}}^{n_0}}{\QueueType_0}.
$  By $\remainder{\SessionType_1}{ \QueueType_1}
\dual \remainder{\SessionType_2}{ \QueueType_2}$, we get
$$
\remainder{\SessionTypeT_2}{\QueueType_2}
~ = ~\co{\textstyle \Select_{i\in I_{n_0}} \Out{\Tag_i^{n_0}}{\Type_i^{n_0}}.\remainder{\SessionTypeT_i^{n_0}}
{\QueueType_0}}
~ = ~
\Branch_{i \in I_{n_0}} \In{\Tag_i^{n_0}}{\Type_i^{n_0}}.{\co{\remainder{\SessionTypeT_i^{n_0}}
{\QueueType_0}}}
$$
which implies 
$$
\remainder{\SessionTypeT'_2}{\QueueType'_2} 
~=~ \remainder{\Branch_{i \in I_{n_0}} \In{\Tag_i^{n_0}}{\Type_i^{n_0}}.{\co{\remainder{\SessionTypeT_i^{n_0}}
{\QueueType_0}}}}{\tmsg{\Tag}{\SessionTypeS}}\\
~=~ \co{\remainder{\SessionTypeT_{k_{n_0}}^{n_0}}{\QueueType_0}}.
$$
We conclude $\remainder{\SessionType'_1}{ \QueueType'_1}
\dual \remainder{\SessionType'_2}{ \QueueType'_2}$. 
\end{enumerate}
\item\label{redbalanced:case2:b}  If the applied rule is \rulename{tr-in}:
$\SessionTypeT_2 = \Branch_{i\in I} \In{\Tag_i}{\Type_i}.\SessionTypeT_i$
and $\QueueType_2 = \tmsg{\Tag_k}{\SessionTypeS} \qconc\QueueType$ and 
$\SessionTypeS_k\asubt\SessionTypeS$, 
and  $k \in I$. 
As in the proof of case~(\ref{redbalanced:case1:b}),
we can shown that
$
\remainder{\SessionTypeT_2}{\QueueType_2}
=\remainder{\SessionTypeT'_2}{\QueueType'_2}.
$
Since $\remainder{\SessionTypeT'_1}{\QueueType'_1} = 
\remainder{\SessionTypeT_1}{\QueueType_1}$ 
we conclude $\remainder{\SessionType'_1}{ \QueueType'_1}
\dual \remainder{\SessionType'_2}{ \QueueType'_2}$.\qedhere
\end{enumerate}
\end{enumerate}
\end{Proof}

 \paragraph{\bf Theorem \ref{thm:sbr:async}.}
(Subject reduction for asynchronous processes)  
\label{app:thm:sbr:async}
If $\wtpa{\UEnv}{\Process}{\LEnv}$
and $\LEnv$ is balanced and $\Process \wreda \ProcessQ$, then there is
$\LEnv'$ such that
$\LEnv \Red^\ast \LEnv'$ 
and
$\wtpa{\UEnv}{\ProcessQ}{\LEnv'}$.
\begin{Proof}
  We first prove that %
  if $\wtpa{\UEnv}{\Process}{\LEnv}$ and %
  $\LEnv$ is balanced and %
  $\Process \reda \ProcessQ$, then %
  there is $\LEnv'$ such that $\LEnv \Red \LEnv'$ and %
  \mbox{$\wtpa{\UEnv}{\ProcessQ}{\LEnv}$.} %
  The proof is by induction on the derivation %
  of $\Process \reda \ProcessQ$. %

We only consider some interesting rules of Table~\ref{tab:async:red}.

\smallskip

\begin{enumerate}
\item Case \rulename{r-send-async}:
\begin{eqnarray}\label{eq:sbreduction:async:r-send:1}
\wtpa{\UEnv}{
    \queue\ChannelA\ChannelB  \Queue
    \parop
    \send\ChannelA\Tag\ChannelC.\Process 
    }{\LEnv}
\end{eqnarray}
By applying 
Lemma \ref{append:lem:inversion:async}.\ref{inver:par:async} to (\ref{eq:sbreduction:async:r-send:1}),
we get
\begin{eqnarray} \label{eq:sbreduction:async:r-send:2}
\wtpa{\UEnv}{\queue\ChannelA\ChannelB \Queue}{\LEnv_1}
\end{eqnarray}
\begin{eqnarray} \label{eq:sbreduction:async:r-send:3}
\wtpa{\UEnv}{\send\ChannelA\Tag\ChannelC.\Process}
{\LEnv_2}
\end{eqnarray}
where  
$\LEnv = \LEnv_1, \LEnv_2$. By applying Lemma \ref{append:lem:inversion:async}.\ref{inver:sel:async} to (\ref{eq:sbreduction:async:r-send:3}), 
we have 
\begin{eqnarray} \label{eq:sbreduction:async:r-send:5}
\LEnv_2 = \LEnv'_2, 
\ASETT{
\bind{\ChannelA}{\SessionTypeT},
\bind{\ChannelC}{\SessionTypeS}} \qquad  \Out\Tag\SessionTypeS . \SessionTypeT' \asubt \SessionTypeT \nonumber
\\
\wtpa{\UEnv}{\Process}
{\LEnv'_2, \ASETT{\bind{\ChannelA}{\SessionTypeT'}}}
\end{eqnarray}
By applying 
Lemma \ref{append:lem:inversion:async}.\ref{inver:q:empty} and \ref{append:lem:inversion:async}.\ref{inver:q}
and rule \rulename{t-message-q} to (\ref{eq:sbreduction:async:r-send:2}), 
we get
\begin{eqnarray} \label{eq:sbreduction:async:r-send:4}
&&\LEnv_1=\LEnv'_1, \ASETT{\queuetype \ChannelA \ChannelB \QueueType} \nonumber
\\
&&\wtpa{\UEnv}
{\queue \ChannelA \ChannelB \Queue \qconc \msg{\Tag}{\ChannelC}}
{\LEnv'_1, 
\ASETT{\queuetype \ChannelA \ChannelB \QueueType \qconc \tmsg{\Tag}{\SessionTypeS}, 
\bind{\ChannelC}{\SessionTypeS}}} 
\end{eqnarray}
By rule \rulename{sub-sel} if $\AContext$ is just one hole and by rule \rulename{sub-perm-async} 
if $\& \in \AContext$,
$$\SessionTypeT=
\AContextf{\textstyle \Select_{i\in I_n} \Out{\Tag_i^{n}}{\Type_i^{n}}.\SessionTypeT_i^{n}}^{n \in N}$$ 
and for all $n\in N$ there is  $i_n\in I_n$
such that $\Tag_{i_n}^{n}=\Tag$, 
$\Type_{i_n}^{n}\asubt\Type$, and
$\SessionTypeT' \asubt\AContextf{\SessionTypeT_{i_n}^{n}}^{n \in N}$.
By applying \rulename{t-sub} to (\ref{eq:sbreduction:async:r-send:5}), 
we derive
\begin{eqnarray} \label{eq:sbreduction:async:r-send:6}
\wtpa{\UEnv}{\Process}
{\LEnv'_2, \ASETT{\bind{\ChannelA}{\AContextf{\SessionTypeT_{i_n}^{n}}^{n \in N}}}}
\end{eqnarray}

By applying \rulename{t-par} to (\ref{eq:sbreduction:async:r-send:4}) and  (\ref{eq:sbreduction:async:r-send:6}), 
we derive
$$
\begin{array}{c}
\wtpa{\UEnv}
{\queue \ChannelA \ChannelB \Queue \qconc \msg{\Tag}{\ChannelC} \parop
\ProcessP}{\LEnv'}
\end{array}
$$
where
$
\LEnv'=
\LEnv'_1, \LEnv'_2, 
\ASETT{
\queuetype \ChannelA \ChannelB \QueueType \qconc \tmsg{\Tag}{\SessionTypeS},
\bind{\ChannelC}{\SessionTypeS}, 
\bind{\ChannelA}{\AContextf{\SessionTypeT_{i_n}^{n}}^{n \in N}}}.$ 
By \rulename{tr-out}, we have
$$\bind{\ChannelA}{\SessionTypeT}, \queuetype \ChannelA \ChannelB \QueueType\Red
\bind{\ChannelA}{\AContextf{\SessionTypeT_{i_n}^{n}}^{n \in N}},\queuetype \ChannelA \ChannelB \QueueType\qconc \tmsg{\Tag}{\SessionTypeS}$$
which implies
$
\LEnv
\Red
\LEnv'_1, 
\LEnv'_2,
\ASETT{
\queuetype \ChannelA \ChannelB \QueueType \qconc \tmsg{\Tag}{\SessionTypeS},
\bind{\ChannelC}{\SessionTypeS}, 
\bind{\ChannelA}{\AContextf{\SessionTypeT_{i_n}^{n}}^{n \in N}}}.$
\item Case \rulename{r-receive-async}:
\begin{eqnarray} \label{eq:sbreduction:async:r-receive:1}
\wtpa{\UEnv}{
    \queue\ChannelA\ChannelB \msg{\Tag_k}{\ChannelC} \qconc \Queue 
    \parop
    \sum_{i\in I} \receive\ChannelB{\Tag_i}{\var_i}.\Process_i
    }{\LEnv}
    \end{eqnarray}
By applying Lemma \ref{append:lem:inversion:async}.\ref{inver:par:async} to (\ref{eq:sbreduction:async:r-receive:1}),
we derive
\begin{eqnarray} \label{eq:sbreduction:async:r-receive:2}
\wtpa{\UEnv}
{\queue\ChannelA\ChannelB \msg{\Tag_k}{\ChannelC} \qconc  \Queue
 }
{\LEnv_1}
\end{eqnarray}
\begin{eqnarray} \label{eq:sbreduction:async:r-receive:3}
\wtpa{\UEnv}{
\sum_{i\in I} \receive\ChannelB{\Tag_i}{\var_i}.\Process_i
}{\LEnv_2}
\end{eqnarray}
where $\LEnv = \LEnv_1, \LEnv_2$.
By applying Lemma \ref{lem:queue:type} to (\ref{eq:sbreduction:async:r-receive:2}), 
we have
\begin{eqnarray} \label{eq:sbreduction:async:r-receive:4}
&&\LEnv_1=
\LEnv'_1,
\ASETT{
\queuetype \ChannelA \ChannelB 
\tmsg{\Tag_k}{\SessionTypeS'} \qconc\QueueType,
\bind{\ChannelC}{\SessionTypeS}} \nonumber
\\
&&
\SessionTypeS' \asubt \SessionTypeS
\label{eq:sbreduction:async:r-receive:asubt1}
\\
&&
\wtpa{\UEnv}{\queue\ChannelA\ChannelB  \Queue}
{\LEnv'_1,
\ASETT{
\queuetype \ChannelA \ChannelB 
\QueueType}} \label{eq:sbreduction:async:r-receive:asubt12}
\end{eqnarray}
By applying Lemma \ref{append:lem:inversion:async}.\ref{inver:bra:async} to (\ref{eq:sbreduction:async:r-receive:3}),
we have
\begin{eqnarray} \label{eq:sbreduction:async:r-receive:5}
&&\LEnv_2 = 
\LEnv'_2, \ASETT{\bind{\ChannelB}{\SessionTypeT}}  \qquad
\SessionTypeT =  
\Branch_{j\in J}\In{\Tag_j}{\Type_j}.\SessionTypeT_j \qquad
J \subseteq I  \nonumber
\\
&& 
\forall j \in J: 
\wtpa{\UEnv}{\Process_j}
{\LEnv'_2, \ASETT{\bind{\ChannelB}{\SessionTypeT_j}, \bind{\var_j}{\Type_j}}}
\end{eqnarray}
Since $\LEnv$ is balanced, $\remainder\SessionTypeT{\tmsg{\Tag_k}{\SessionTypeS'} \qconc\QueueType}$ is defined, and this implies $k\in J$ 
and 
\begin{eqnarray} \label{eq:sbreduction:async:r-receive:asubt2}
\SessionTypeS_k \asubt \SessionTypeS'
\end{eqnarray}

By applying Lemma \ref{app:lem:subs:async} to (\ref{eq:sbreduction:async:r-receive:5}), 
we have
\begin{eqnarray} \label{eq:sbreduction:async:r-receive:6}
\wtpa{\UEnv}{\ProcessP_k \subst{\ChannelC}{\var_k}}
{\LEnv'_2, 
\ASETT{\bind{\ChannelB}{\SessionTypeT_k}, \bind{\ChannelC}{\SessionTypeS_k}}}
\end{eqnarray}
By (\ref{eq:sbreduction:async:r-receive:asubt1}) and 
(\ref{eq:sbreduction:async:r-receive:asubt2}),   
we have $\SessionTypeS_k \asubt \SessionTypeS' \asubt \SessionTypeS$.
Hence 
by applying rule  \rulename{t-sub} to (\ref{eq:sbreduction:async:r-receive:6}), we derive  
\begin{eqnarray}  \label{eq:sbreduction:async:r-receive:61}
\wtpa{\UEnv}{\ProcessP_k \subst{\ChannelC}{\var_k}}
{\LEnv'_2, \ASETT{\bind{\ChannelB}{\SessionTypeT_k}, \bind{\ChannelC}{\SessionTypeS}}}
\end{eqnarray}
By applying rule \rulename{t-par} to (\ref{eq:sbreduction:async:r-receive:asubt12}) 
and  (\ref{eq:sbreduction:async:r-receive:61}) we derive
\begin{eqnarray} \label{eq:sbreduction:async:r-receive:7}
\wtpa{\UEnv}
{\queue \ChannelA \ChannelB\Queue \parop
\ProcessP_k \subst{\ChannelC}{\var_k}}
{\LEnv'_1,
\LEnv'_2,
\ASETT{
\queuetype \ChannelA \ChannelB 
\QueueType, \bind{\ChannelB}{\SessionTypeT_k}, \bind{\ChannelC}{\SessionTypeS}}} \nonumber
\end{eqnarray}
By rule \rulename{tr-in},
we derive
$
\queuetype \ChannelA \ChannelB \tmsg{\Tag_k}{\SessionTypeS'} \qconc
\QueueType,  \bind{\ChannelB}{\SessionTypeT}\Red
\queuetype \ChannelA \ChannelB \QueueType,  \bind{\ChannelB}{\SessionTypeT_k}
$
which implies $$\LEnv\Red\LEnv'_1, \LEnv'_2,
\ASETT{
\queuetype \ChannelA \ChannelB 
\QueueType,  \bind{\ChannelB}{\SessionTypeT_k}, \bind{\ChannelC}{\SessionTypeS}}$$
\item Case \rulename{r-context}:
We only illustrate 
the case when the context is a channel restriction. 
Let $\wtpa{\UEnv}{\new{\ChannelA}{\ChannelB}\Process}{\LEnv}$,
         then $$\wtpa{\UEnv}{\Process}
         {\LEnv,
         \ASETT{
          \bind{\ChannelA}{\SessionType_1},
          \bind{\ChannelB}{\SessionType_2},
          \queuetype \ChannelB \ChannelA \QueueType_1,
          \queuetype \ChannelA \ChannelB \QueueType_2}}
          $$        
         and 
         $\remainder{\SessionType_1}{\QueueType_1}
         \dual
         \remainder{\SessionType_2}{\QueueType_2}$ by Lemma \ref{append:lem:inversion:async}.\ref{inver:new:async}. If $\Process\wreda\Process'$, then by induction there is $\LEnv'$ such that 
$\wtpa{\UEnv}{\Process}{\LEnv'}$ and 
$$\LEnv,
\ASETT{
          \bind{\ChannelA}{\SessionType_1},
          \bind{\ChannelB}{\SessionType_2},
          \queuetype \ChannelB \ChannelA \QueueType_1,
          \queuetype \ChannelA \ChannelB \QueueType_2} \Red^* \LEnv'$$ 
where $\LEnv'$ is balanced.          
          This implies $\LEnv'=\LEnv'',\ASETT{\bind{\ChannelA}{\SessionType'_1},
          \bind{\ChannelB}{\SessionType'_2},
          \queuetype \ChannelB \ChannelA \QueueType'_1,
          \queuetype \ChannelA \ChannelB \QueueType'_2}$ and $\remainder{\SessionType'_1}{\QueueType'_1}
         \dual
         \remainder{\SessionType'_2}{\QueueType'_2}$ by Lemma \ref{lem:redsessiontype:async}. By rule \rulename{t-new-async} we derive$$\wtpa{\UEnv}{\new{\ChannelA}{\ChannelB}\Process}{\LEnv''}$$
\end{enumerate}

  \noindent We have proved that %
  if $\wtpa{\UEnv}{\Process}{\LEnv}$ and %
  $\LEnv$ is balanced and %
  $\Process \reda \ProcessQ$, then %
  there is $\LEnv'$ such that $\LEnv \Red \LEnv'$ and %
  \mbox{$\wtpa{\UEnv}{\ProcessQ}{\LEnv}$.} %
  Now, the main statement of the theorem %
  can be easily proved %
  by induction on the length of the sequence of transitions %
  $\Process \wreda \ProcessQ$. %
\end{Proof}

As in the case of synchronous subtyping we can show:
\begin{lem} \label{lem:substitution:context:a}
If $\wtpa{\UEnv, \pbind{\pvar}{\SessionType}}{\Context[\invoke\pvar{\ChannelA}]}{\LEnv}$,
$\wtpa{\UEnv}{\Process}{\ASET{\bind{\ChannelA}{\SessionType}}}$, 
and $\pvar$ does not occur free in $\Context$,
then $\wtpa{\UEnv}{\Context[\Process]}{\LEnv}$.
\end{lem}\newpage

\begin{Proof}
The proof is similar to the proof of Lemma \ref{lem:substitution:context}.
The only case in which the proof differs is 
$\Context = \new \ChannelC \ChannelD \Context'$.
Let
\begin{eqnarray} \label{eq:context:async:new:1}
\wtpa{\UEnv, \pbind{\pvar}{\SessionType}}
{\new \ChannelC \ChannelD \Context'[\invoke\pvar{\ChannelA}]}{\LEnv}
\end{eqnarray}
\begin{eqnarray} \label{eq:context:async:process}
\wtpa{\UEnv}{\Process}{\ASET{\bind{ \ChannelA}{ \SessionType}}}
\end{eqnarray}
By applying Lemma \ref{append:lem:inversion:async}.\ref{inver:new:async} to  (\ref{eq:context:async:new:1}), 
we have
\begin{eqnarray} \label{eq:context:async:new:3}
\wtpa{\UEnv, \pbind{\pvar}{\SessionType}}{\Context'[\invoke\pvar{\ChannelA}]}{
\LEnv, \ASETT{
\bind{\ChannelC}{\SessionType}, \bind{\ChannelD}{\SessionType'},
\queuetype \ChannelC \ChannelD \QueueType, 
\queuetype \ChannelD \ChannelC \QueueType'}}\text{ and } 
\remainder{\SessionType}{\QueueType} \dual \remainder{\SessionType'}{\QueueType'}
\end{eqnarray}
By induction,  (\ref{eq:context:async:process}) and  (\ref{eq:context:async:new:3}) together imply
$
\wtpa{\UEnv}{\Context'[\Process]}{
\LEnv, 
\ASETT{
\bind{\ChannelC}{\SessionType}, \bind{\ChannelD}{\SessionType'},
\queuetype \ChannelC \ChannelD \QueueType, 
\queuetype \ChannelD \ChannelC \QueueType'}}$ and $
\remainder{\SessionType}{\QueueType} \dual 
\remainder{\SessionType'}{\QueueType'}$. By applying \rulename{t-new-async}, 
we derive
$$
\wtpa{\UEnv}{\new \ChannelC \ChannelD \Context'[\Process]}{\LEnv}\vspace{-18 pt}
$$
\end{Proof}

\paragraph{\bf Theorem \ref{thm:a:sound}.}
The asynchronous subtyping relation is sound  for the asynchronous calculus.\vspace{-\baselineskip}
\label{app:thm:a:sound}
\begin{Proof}
The proof is similar to that of  Theorem \ref{thm:s:sound},
using  
Lemma \ref{lem:substitution:context:a} and Corollary \ref{pro:comsafe:async}.
\end{Proof}

\section{Proofs of Section~\ref{sec:a:completeness}}\label{ac}

\begin{proposition}
  \label{lem:branch-not-in-t:unfolding}%
  $\& \not\in \trec\tvar.\SessionType$ \;if and only if\; %
  $\& \not\in \SessionType \subst{\trec\tvar.\SessionType}\tvar$.
\end{proposition}
\begin{Proof}
  \def\someDeriv{\mathcal{D}}%
  ($\implies$) %
  Assume $\& \not\in \trec\tvar.\SessionType$, %
  for some derivation $\someDeriv$. %
  We observe that $\someDeriv$ can only conclude %
  by the rule for 
  recursion on page~\pageref{def:branch-not-in-t}, %
  and from its premise %
  we have a derivation $\someDeriv'$ proving $\& \not\in \SessionType$. %
  We can inductively rewrite $\someDeriv'$ %
  into a derivation 
  proving %
  $\& \not\in \SessionType \subst{\trec\tvar.\SessionType}\tvar$, %
  by replacing %
  \emph{(i)} each occurrence of $\tvar$ 
  with $\trec\tvar.\SessionType$, %
  and %
  \emph{(ii)} each instance of the axiom $\& \not\in \tvar$ %
  (which, after the replacement \emph{(i)}, %
  has become $\& \not\in \trec\tvar.\SessionType$) %
  with $\someDeriv$ above. %
  We conclude $\& \not\in \SessionType \subst{\trec\tvar.\SessionType}\tvar$.

  ($\impliedby$). %
  The proof for this case is similar to %
  the proof of Proposition~\ref{lem:branch-in-t:unfolding}, %
  in the $\impliedby$ direction, %
  using $\& \not\in \SessionType$ %
  and its rules defined on page~\pageref{def:branch-not-in-t}.
  \footnote{%
    Notably, %
    the case discussed in footnote~\ref{footnote:proof:branch-in-t:unfolding} %
    is now \emph{not} absurd.%
  }%
\end{Proof}\enlargethispage{\baselineskip}

\begin{proposition}
  \label{lem:branch-not-in-t:complement}%
  $\& \in \SessionType$ holds \;if and only if\; %
  $\& \not\in \SessionType$ does \emph{not} hold.%
\end{proposition}
\begin{Proof}
  ($\implies$). %
  By induction on the derivation of %
  $\& \in \SessionType$. %
  The base case, %
  with $\SessionType = \Branch_{i\in I} \In{\Tag_i}{\Type_i}.\SessionTypeT_i$, %
  is immediate: %
  no rule yields $\& \not\in \SessionType$. %
  In the inductive case with %
  $\SessionType = \Select_{i\in I} \Out{\Tag_i}{\Type_i}.\SessionTypeT_i$, %
  for all $i \in I$ we have a premise %
  $\& \in \SessionTypeT_i$, %
  and thus (by the induction hypothesis) %
  $\& \not\in \SessionTypeT_i$ does \emph{not} hold: %
  hence, $\not\exists i \in I$ such that $\& \not\in \SessionTypeT_i$ holds, %
  and we conclude that $\& \not\in \SessionType$ does \emph{not} hold %
  by any rule. %
  In the inductive case with %
  $\SessionType = \trec\tvar.\SessionType'$, %
  we have the premise $\& \in \SessionType'$, %
  and thus (by the induction hypothesis) %
  $\& \not\in \SessionTypeT'$ does \emph{not} hold: %
  we conclude that $\& \not\in \SessionType$ does \emph{not} hold %
  by any rule. 

  ($\impliedby$). %
  We prove $\& \in \SessionType$ %
  by induction on $\SessionTypeT$, %
  examining the cases where $\& \not\in \SessionType$ does \emph{not} hold. %
  If $\SessionType = \Branch_{i\in I} \In{\Tag_i}{\Type_i}.\SessionTypeT_i$, %
  we conclude by the axiom on page~\pageref{def:branch-in-t}. %
  If $\SessionType = \Select_{i\in I} \Out{\Tag_i}{\Type_i}.\SessionTypeT_i$ %
  and $\forall i \in I: \& \not\in \SessionTypeT_i$ does \emph{not} hold, %
  by the induction hypothesis %
  we have $\forall i \in I: \& \in \SessionTypeT_i$; %
  therefore, %
  we conclude $\& \in \SessionTypeT$ %
  (By the rule for 
  selection on page~\pageref{def:branch-in-t}).
  If $\SessionType = \trec\tvar.\SessionType'$ %
  and $\& \not\in \SessionType'$ does \emph{not} hold, %
  by the induction hypothesis %
  we have $\& \in \SessionType'$: %
  therefore, %
  we conclude $\& \in \SessionTypeT$ %
  (by the rule for 
  recursion on page~\pageref{def:branch-in-t}).
\end{Proof}

\end{document}